\documentclass[12,twoside]{article}  % fleqn,
\usepackage[margin=0.9in]{geometry}  % 0.12in, 0.9in

\usepackage[square]{natbib}
\author{Conal Elliott \\[1.5ex]Target\\[1.5ex]conal@conal.net}

\usepackage[utf8]{inputenc} 
\usepackage[T1]{fontenc}
\usepackage{microtype}

\newcommand\subtitle\footnote

\usepackage{scalerel}

\usepackage{datetime}
\usdate

\newcommand\nc\newcommand
\nc\rnc\renewcommand

\usepackage{amsmath}
\usepackage{latexsym}
\usepackage{amssymb}
\usepackage{color}

\usepackage{setspace}

\usepackage{epsfig}
\usepackage{latexsym}

\nc\out[1]{}

\nc\indraft[1]{#1}

\nc\mynote[1]{\indraft{\textcolor{red}{[#1]}}}

\nc\notefoot[1]{\indraft{\footnote{\mynote{#1}}}}

\nc\todo{\mynote{To do.}}

\nc\needcite{\indraft{ \mynote{ref}}}

\nc\seclabel[1]{\label{sec:#1}}
\nc\secref[1]{Section \ref{sec:#1}}
\nc\secreftwo[2]{Sections \ref{sec:#1} and \ref{sec:#2}}
\nc\secpageref[1]{page \pageref{sec:#1}}

\nc\appref[1]{Appendix \ref{sec:#1}}

\nc\sectionl[1]{\section{#1}\seclabel{#1}}
\nc\subsectionl[1]{\subsection{#1}\seclabel{#1}}

\usepackage{subcaption}

\nc\figlabel[1]{\label{fig:#1}}
\nc\figref[1]{Figure \ref{fig:#1}}
\nc\figreftwo[2]{Figures \ref{fig:#1} and \ref{fig:#2}}
\nc\figrefthree[3]{Figures \ref{fig:#1}, \ref{fig:#2}, and \ref{fig:#3}}

\nc\incpic[1]{\includegraphics[width=\linewidth]{Figures/#1}}

\nc\figp[2]{\begin{figure}\centering #1 \hspace{-2ex} #2\end{figure}}

\nc\figoneW[3]{
{% \fbox
\begin{minipage}{#1\linewidth}
  \centering
  \setlength\mathindent{0ex}
  \incpic{#2}
  \vspace*{-4ex}
  \captionof{figure}{#3}
  \figlabel{#2}
\end{minipage}
}
}
\nc\figone{\figoneW{0.48}}

\nc\workingHere{
\vspace{1ex}
\begin{center}
\setlength{\fboxsep}{3ex}
\setlength{\fboxrule}{4pt}
\huge\textcolor{red}{\framebox{Working here}}
\end{center}
\vspace{1ex}
}

\nc\notefootsep{\indraft{\textsuperscript{,}}}

\let\oldFootnote\footnote
\nc\nextToken\relax
\rnc\footnote[1]{\oldFootnote{#1}\futurelet\nextToken\isFootnote}
\nc\isFootnote{\ifx\footnote\nextToken\notefootsep{}\fi}

\newtheorem{theorem}{Theorem}%[section]
\nc\thmLabel[1]{\label{theorem:#1}}
\nc\thmRef[1]{Theorem \ref{theorem:#1}}
\nc\thmRefTwo[2]{Theorems \ref{theorem:#1} and \ref{theorem:#2}}
\nc\thmRefs[2]{Theorems \ref{theorem:#1} through \ref{theorem:#2}}

\newtheorem{corollary}{Corollary}[theorem]
\nc\corLabel[1]{\label{corollary:#1}}
\nc\corRef[1]{Corollary \ref{corollary:#1}}
\nc\corRefTwo[2]{Corollaries \ref{corollary:#1} and \ref{corollary:#2}}
\nc\corRefs[2]{Corollaries \ref{corollary:#1} through \ref{corollary:#2}}

\newtheorem{lemma}[theorem]{Lemma}
\nc\lemLabel[1]{\label{lemma:#1}}
\nc\lemRef[1]{Lemma \ref{lemma:#1}}
\nc\lemRefTwo[2]{Lemma \ref{lemma:#1} and \ref{lemma:#2}}
\nc\lemRefs[2]{Lemma \ref{lemma:#1} through \ref{lemma:#2}}

\usepackage{hyperref}

\usepackage{fancyhdr}
\hypersetup{hidelinks} % for the long/arXiv version

\rnc\indraft[1]{}

\makeatletter
\@ifundefined{lhs2tex.lhs2tex.sty.read}%
  {\@namedef{lhs2tex.lhs2tex.sty.read}{}%
   \newcommand\SkipToFmtEnd{}%
   \newcommand\EndFmtInput{}%
   \long\def\SkipToFmtEnd#1\EndFmtInput{}%
  }\SkipToFmtEnd

\newcommand\ReadOnlyOnce[1]{\@ifundefined{#1}{\@namedef{#1}{}}\SkipToFmtEnd}
\usepackage{amstext}
\usepackage{amssymb}
\usepackage{stmaryrd}
\DeclareFontFamily{OT1}{cmtex}{}
\DeclareFontShape{OT1}{cmtex}{m}{n}
  {<5><6><7><8>cmtex8
   <9>cmtex9
   <10><10.95><12><14.4><17.28><20.74><24.88>cmtex10}{}
\DeclareFontShape{OT1}{cmtex}{m}{it}
  {<-> ssub * cmtt/m/it}{}

\DeclareFontShape{OT1}{cmtt}{bx}{n}
  {<5><6><7><8>cmtt8
   <9>cmbtt9
   <10><10.95><12><14.4><17.28><20.74><24.88>cmbtt10}{}
\DeclareFontShape{OT1}{cmtex}{bx}{n}
  {<-> ssub * cmtt/bx/n}{}
\newcommand{\Conid}[1]{\mathit{#1}}
\newcommand{\Varid}[1]{\mathit{#1}}
\newcommand{\anonymous}{\kern0.06em \vbox{\hrule\@width.5em}}

\usepackage{polytable}

\@ifundefined{mathindent}%
  {\newdimen\mathindent\mathindent\leftmargini}%
  {}%

\def\resethooks{%
  \global\let\SaveRestoreHook\empty
  \global\let\ColumnHook\empty}
\newcommand*{\savecolumns}[1][default]%
  {\g@addto@macro\SaveRestoreHook{\savecolumns[#1]}}
\newcommand*{\restorecolumns}[1][default]%
  {\g@addto@macro\SaveRestoreHook{\restorecolumns[#1]}}
\newcommand*{\aligncolumn}[2]%
  {\g@addto@macro\ColumnHook{\column{#1}{#2}}}

\resethooks

\newcommand{\onelinecommentchars}{\quad-{}- }
\newcommand{\commentbeginchars}{\enskip\{-}
\newcommand{\commentendchars}{-\}\enskip}

\newcommand{\visiblecomments}{%
  \let\onelinecomment=\onelinecommentchars
  \let\commentbegin=\commentbeginchars
  \let\commentend=\commentendchars}

\newcommand{\invisiblecomments}{%
  \let\onelinecomment=\empty
  \let\commentbegin=\empty
  \let\commentend=\empty}

\visiblecomments

\newlength{\blanklineskip}
\newcommand{\hsindent}[1]{\quad}% default is fixed indentation
\let\hspre\empty
\let\hspost\empty

\EndFmtInput
\makeatother
\ReadOnlyOnce{polycode.fmt}%
\makeatletter

\newcommand{\hsnewpar}[1]%
  {{\parskip=0pt\parindent=0pt\par\vskip #1\noindent}}

\newcommand{\hscodestyle}{}

\newcommand{\sethscode}[1]%
  {\expandafter\let\expandafter\hscode\csname #1\endcsname
   \expandafter\let\expandafter\endhscode\csname end#1\endcsname}

  {\par\noindent
   \advance\leftskip\mathindent
   \hscodestyle
   \let\\=\@normalcr
   \let\hspre\(\let\hspost\)%
   \pboxed}%
  {\endpboxed\)%
   \par\noindent
   \ignorespacesafterend}

  {\hsnewpar\abovedisplayskip
   \advance\leftskip\mathindent
   \hscodestyle
   \let\hspre\(\let\hspost\)%
   \pboxed}%
  {\endpboxed%
   \hsnewpar\belowdisplayskip
   \ignorespacesafterend}

  {\hsnewpar\abovedisplayskip
   \advance\leftskip\mathindent
   \hscodestyle
   \let\\=\@normalcr
   \(\pboxed}%
  {\endpboxed\)%
   \hsnewpar\belowdisplayskip
   \ignorespacesafterend}

\newcommand{\plainhs}{\sethscode{plainhscode}}

\plainhs

  {\hsnewpar\abovedisplayskip
   \advance\leftskip\mathindent
   \hscodestyle
   \let\\=\@normalcr
   \(\parray}%
  {\endparray\)%
   \hsnewpar\belowdisplayskip
   \ignorespacesafterend}

  {\parray}{\endparray}

  {\(\parray}{\endparray\)}

\def\codeframewidth{\arrayrulewidth}
\RequirePackage{calc}

  {\parskip=\abovedisplayskip\par\noindent
   \hscodestyle
   \arrayrulewidth=\codeframewidth
   \tabular{@{}|p{\linewidth-2\arraycolsep-2\arrayrulewidth-2pt}|@{}}%
   \hline\framedhslinecorrect\\{-1.5ex}%
   \let\endoflinesave=\\
   \let\\=\@normalcr
   \(\pboxed}%
  {\endpboxed\)%
   \framedhslinecorrect\endoflinesave{.5ex}\hline
   \endtabular
   \parskip=\belowdisplayskip\par\noindent
   \ignorespacesafterend}

\newcommand{\framedhslinecorrect}[2]%
  {#1[#2]}

  {\(\def\column##1##2{}%
   \let\>\undefined\let\<\undefined\let\\\undefined
   \newcommand\>[1][]{}\newcommand\<[1][]{}\newcommand\\[1][]{}%
   \def\fromto##1##2##3{##3}%
   }{\) }%

  {\let\orighscode=\hscode
   \let\origendhscode=\endhscode
   \def\endhscode{\def\hscode{\endgroup\def\@currenvir{hscode}\\}\begingroup}
   \orighscode\def\hscode{\endgroup\def\@currenvir{hscode}}}%
  {\origendhscode
   \global\let\hscode=\orighscode
   \global\let\endhscode=\origendhscode}%

\makeatother
\EndFmtInput
\ReadOnlyOnce{forall.fmt}%
\makeatletter

\let\HaskellResetHook\empty
\newcommand*{\AtHaskellReset}[1]{%
  \g@addto@macro\HaskellResetHook{#1}}
\newcommand*{\HaskellReset}{\HaskellResetHook}

\newcommand\hsforall{\global\let\hsdot=\hsperiodonce}
\newcommand*\hsperiodonce[2]{#2\global\let\hsdot=\hscompose}
\newcommand*\hscompose[2]{#1}

\AtHaskellReset{\global\let\hsdot=\hscompose}

\HaskellReset

\makeatother
\EndFmtInput
\newcommand{\calculationcomments}{%
  \let\onelinecomment=\onelinecommentchars
  \def\commentbegin{\quad\{ }%
  \def\commentend{\}}%
}
\calculationcomments

\nc\tit{The Simple Essence of Automatic Differentiation}
\nc\alttit{Differentiable Functional Programming Made Easy}
\date{March, 2018}

\title{\tit
\\[1ex]\emph{Extended version}\footnote{The appendices of this extended version include proofs omitted in the conference article \citep{Elliott-2018-ad-icfp}.}\ 
}

\nc\proofLabel[1]{\label{proof:#1}}
\nc\proofRef[1]{Appendix \ref{proof:#1}}
\nc\provedIn[1]{\textnormal{Proved in \proofRef{#1}}}

\newenvironment{closerCodePars}{\setlength{\blanklineskip}{1.3ex}}{}

\begin{document}

\maketitle

\begin{abstract}

Automatic differentiation (AD) in reverse mode (RAD) is a central component of deep learning and other uses of large-scale optimization. Commonly used RAD algorithms such as backpropagation, however, are complex and stateful, hindering deep understanding, improvement, and parallel execution. This paper develops a simple, generalized AD algorithm calculated from a simple, natural specification. The general algorithm is then specialized by varying the representation of derivatives. In particular, applying well-known constructions to a naive representation yields two RAD algorithms that are far simpler than previously known. In contrast to commonly used RAD implementations, the algorithms defined here involve no graphs, tapes, variables, partial derivatives, or mutation. They are inherently parallel-friendly, correct by construction, and usable directly from an existing programming language with no need for new data types or programming style, thanks to use of an AD-agnostic compiler plugin.

\end{abstract}

%%%%%%%

\sectionl{Introduction}

Accurate, efficient, and reliable computation of derivatives has become increasingly important over the last several years, thanks in large part to the successful use of \emph{backpropagation} in machine learning, including multi-layer neural networks, also known as ``deep learning'' \citep{LecunBengioHinton2015DLNature,Goodfellow2016DL}.
Backpropagation is a specialization and independent invention of the \emph{reverse mode} of automatic differentiation (AD) and is used to tune a parametric model to closely match observed data, using \emph{gradient descent} (or \emph{stochastic} gradient descent).
Machine learning and other gradient-based optimization problems typically rely on derivatives of functions with very high dimensional domains\out{---often in the hundreds of millions \citep{LecunBengioHinton2015DLNature}---} and a scalar codomain---exactly the conditions under which reverse-mode AD is much more efficient than forward-mode AD (by a factor proportional to the domain dimension).
Unfortunately, while forward-mode AD (FAD) is easily understood and implemented\needcite, reverse-mode AD (RAD) and backpropagation have had much more complicated explanations and implementations, involving mutation, graph construction and traversal, and ``tapes'' (sequences of reified, interpretable assignments, also called ``traces'' or ``Wengert lists'')\needcite.
Mutation, while motivated by efficiency concerns, makes parallel execution difficult and so undermines efficiency as well.
Construction and interpretation (or compilation) of graphs and tapes also add execution overhead.
The importance of RAD makes its current complicated and bulky implementations especially problematic.
The increasingly large machine learning (and other optimization) problems being solved with RAD (usually via backpropagation) suggest the need to find more streamlined, efficient implementations, especially with the massive hardware parallelism now readily and inexpensively available in the form of graphics processors (GPUs) and FPGAs.

Another difficulty in the practical application of AD in machine learning (ML) comes from the nature of many currently popular ML frameworks, including\out{ Theano \citep{Bergstra10theano}\notefoot{Theano doesn't seem to expose graphs.},} Caffe \citep{Jia2014Caffe}, TensorFlow \citep{Abadi2016TensorFlow}, and Keras \citep{Chollet2016KerasResources}.
These frameworks are designed around the notion of a ``graph'' (or ``network'') of interconnected nodes, each of which represents a mathematical operation---a sort of data flow graph.
Application programs construct these graphs \emph{explicitly}, creating nodes and connecting them to other nodes.
After construction, the graphs must then be processed into a representation that is more efficient to train and to evaluate.
These graphs are essentially mathematical expressions with sharing, hence directed acyclic graphs (DAGs).
This paradigm of graph construction, compilation, and execution bears a striking resemblance to what programmers and compilers do all the time:
\begin{itemize}
\item Programs are written by a human.
\item The compiler or interpreter front-end parses the program into a DAG representation.
\item The compiler back-end transforms the DAGs into a form efficient for execution.
\item A human runs the result of compilation.
\end{itemize}
When using a typical ML framework, programmers experience this sequence of steps \emph{at two levels}: working with their code \emph{and} with the graphs that their code generates.
Both levels have notions of operations, variables, information flow, values, types, and parametrization.
Both have execution models that must be understood.

\mynote{Maybe relate traditional, graph-centered ML frameworks to deep DSELs.}

A much simpler and cleaner foundation for ML would be to have just the programming language, omitting the graphs/networks altogether.
Since ML is about (mathematical) functions, one would want to choose a programming language that supports functions well, i.e., a functional language, or at least a language with strong functional features.
One might call this alternative ``differentiable functional programming''.
In this paradigm, programmers directly define their functions of interest, using the standard tools of functional programming, with the addition of a differentiation operator (a typed higher-order function, though partial since not all computable functions are differentiable).
Assuming a \emph{purely} functional language or language subset (with simple and precise mathematical denotation), the meaning of differentiation is exactly as defined in traditional calculus.

How can we realize this vision of differentiable functional programming?
One way is to create new languages, but doing so requires enormous effort to define and implement efficiently, and perhaps still more effort to evangelize.
Alternatively, we might choose a suitable purely functional language like Haskell and then add differentiation.
The present paper embodies the latter choice, augmenting the popular Haskell compiler GHC with a plugin that converts standard Haskell code into categorical form to be instantiated in any of a variety of categories, including differentiable functions \citep{Elliott-2017-compiling-to-categories}.

This paper makes the following specific contributions:
\begin{itemize}
\item
  Beginning with a simple category of derivative-augmented functions, specify AD simply and precisely by requiring this augmentation (relative to regular functions) to be homomorphic with respect to a collection of standard categorical abstractions and primitive mathematical operations.
\item
  Calculate a correct-by-construction AD implementation from the homomorphic specification.
\item
  Generalizing AD by replacing linear maps (general derivative values) with an arbitrary cartesian category \citep{Elliott-2017-compiling-to-categories}, define several AD variations, all stemming from different representations of linear maps: functions (satisfying linearity), ``generalized matrices'' (composed representable functors), continuation-based transformations of any linear map representation, and dualized versions of any linear map representation.
  The latter two variations yield correct-by-construction implementations of reverse-mode AD that are much simpler than previously known and are composed from generally useful components.
  The choice of dualized linear functions for gradient computations is particularly compelling in simplicity.
  It also appears to be quite efficient---requiring no matrix-level representations or computations---and is suitable for gradient-based optimization, e.g., for machine learning.
  In contrast to conventional reverse-mode AD algorithms, all algorithms in this paper are free of mutation and hence naturally parallel.
  A similar construction yields forward-mode AD.
\end{itemize}

\sectionl{What's a Derivative?}

\nc\set[1]{\{#1\}}

Since automatic differentiation (AD) has to do with computing derivatives, let's begin by considering what derivatives are.
If your introductory calculus class was like mine, you learned that the derivative \ensuremath{\Varid{f'}\;\Varid{x}} of a function \ensuremath{\Varid{f}\mathbin{::}\mathbb{R}\to \mathbb{R}} at a point \ensuremath{\Varid{x}} (in the domain of \ensuremath{\Varid{f}}) is a \emph{number}, defined as follows:
\begin{align} \label{eq:scalar-deriv}
\ensuremath{\Varid{f'}\;\Varid{x}\mathrel{=}\lim_{\varepsilon \to \mathrm{0}}{\frac{\Varid{f}\;(\Varid{x}\mathbin{+}\varepsilon )\mathbin{-}\Varid{f}\;\Varid{x}}{\varepsilon }}}
\end{align}
That is, \ensuremath{\Varid{f'}\;\Varid{x}} tells us how fast \ensuremath{\Varid{f}} is scaling input changes at \ensuremath{\Varid{x}}.

How well does this definition hold up beyond functions of type \ensuremath{\mathbb{R}\to \mathbb{R}}?
It will do fine with complex numbers (\ensuremath{\mathbb{C}\to \mathbb{C}}), where division is also defined.
Extending to \ensuremath{\mathbb{R}\to \mathbb{R}^n} also works if we interpret the ratio as dividing a vector (in \ensuremath{\mathbb{R}^n}) by a scalar in the usual way.
When we extend to \ensuremath{\mathbb{R}^m\to \mathbb{R}^n} (or even \ensuremath{\mathbb{R}^m\to \mathbb{R}}), however, this definition no longer makes sense, as it would rely on dividing \emph{by} a vector \ensuremath{\varepsilon \mathbin{::}\mathbb{R}^m}.

This difficulty of differentiation with non-scalar domains is usually addressed with the notion of ``partial derivatives'' with respect to the \ensuremath{\Varid{m}} scalar components of the domain \ensuremath{\mathbb{R}^m}, often written ``$\partial f / \partial x_j$'' for $j \in \set{1,\ldots,m}$.
When the codomain \ensuremath{\mathbb{R}^n} is also non-scalar (i.e., \ensuremath{\Varid{n}\mathbin{>}\mathrm{1}}), we have a \emph{matrix} $\mathbf J$ (the \emph{Jacobian}), with $\mathbf J_{ij} = \partial f_i / \partial x_j$ for $i \in \set{1,\ldots,n}$, where each $f_i$ projects out the \ensuremath{\Varid{i}^{\text{th}}} scalar value from the result of $f$.

\nc\A{\mathbf A}
\nc\B{\mathbf B}
So far, we've seen that the derivative of a function could be a single number (for \ensuremath{\mathbb{R}\to \mathbb{R}}), or a vector (for \ensuremath{\mathbb{R}\to \mathbb{R}^n}), or a matrix (for \ensuremath{\mathbb{R}^m\to \mathbb{R}^n}).
Moreover, each of these situations has an accompanying chain rule, which says how to differentiate the composition of two functions.
Where the scalar chain rule involves multiplying two scalar derivatives, the vector chain rule involves ``multiplying'' two \emph{matrices} $\A$ and $\B$ (the Jacobians), defined as follows:
$$ (\A \cdot \B)_{ij} = \sum_{k=1}^m \A_{ik} \cdot \B_{kj} $$
Since one can think of scalars as a special case of vectors, and scalar multiplication as a special case of matrix multiplication, perhaps we've reached the needed generality.
When we turn our attention to higher derivatives (which are derivatives of derivatives), however, the situation gets more complicated, and we need yet higher-dimensional representations, with correspondingly more complex chain rules.

Fortunately, there is a single, elegant generalization of differentiation with a correspondingly simple chain rule.
First, reword Definition \ref{eq:scalar-deriv} above as follows:\out{\footnote{For clarity, throughout this paper I will use ``\ensuremath{\Conid{A}\mathrel{=}\Conid{B}}'' to mean ``\ensuremath{\Conid{A}} is defined as \ensuremath{\Conid{B}}'' and ``\ensuremath{\mathrel{=}}'' to mean (more broadly) that ``\ensuremath{\Conid{A}} is equal to \ensuremath{\Conid{B}}''. The former introduces \ensuremath{\Conid{A}}, while the latter asserts that a well-defined statement of equality is in fact true.}}
$$ \ensuremath{\lim_{\varepsilon \to \mathrm{0}}{\frac{\Varid{f}\;(\Varid{x}\mathbin{+}\varepsilon )\mathbin{-}\Varid{f}\;\Varid{x}}{\varepsilon }}\mathbin{-}\Varid{f'}\;\Varid{x}\mathrel{=}\mathrm{0}} $$
Equivalently,
$$ \ensuremath{\lim_{\varepsilon \to \mathrm{0}}{\frac{\Varid{f}\;(\Varid{x}\mathbin{+}\varepsilon )\mathbin{-}(\Varid{f}\;\Varid{x}\mathbin{+}\varepsilon  \cdot \Varid{f'}\;\Varid{x})}{\varepsilon }}\mathrel{=}\mathrm{0}} $$
Notice that \ensuremath{\Varid{f'}\;\Varid{x}} is used to linearly transform \ensuremath{\varepsilon }.
Next, generalize this condition to say that \ensuremath{\Varid{f'}\;\Varid{x}} is a \emph{linear map} such that
$$\ensuremath{\lim_{\varepsilon \to \mathrm{0}}{\frac{\lVert\Varid{f}\;(\Varid{x}\mathbin{+}\varepsilon )\mathbin{-}(\Varid{f}\;\Varid{x}\mathbin{+}\Varid{f'}\;\Varid{x}\;\varepsilon )\rVert}{\lVert\varepsilon \rVert}}\mathrel{=}\mathrm{0}} .$$
In other words, \ensuremath{\Varid{f'}\;\Varid{x}} is a \emph{local linear approximation} of \ensuremath{\Varid{f}} at \ensuremath{\Varid{x}}.
When an \ensuremath{\Varid{f'}\;\Varid{x}} satisfying this condition exists, it is indeed unique \citep[chapter 2]{Spivak65}.

The derivative of a function \ensuremath{\Varid{f}\mathbin{::}\Varid{a}\to \Varid{b}} at some value in \ensuremath{\Varid{a}} is thus not a number, vector, matrix, or higher-dimensional variant, but rather a \emph{linear map} (also called ``linear transformation'') from \ensuremath{\Varid{a}} to \ensuremath{\Varid{b}}, which we will write as ``\ensuremath{\Varid{a}\multimap\Varid{b}}''.
The numbers, vectors, matrices, etc mentioned above are all different \emph{representations} of linear maps; and the various forms of ``multiplication'' appearing in their associated chain rules are all implementations of linear map \emph{composition} for those representations.
Here, \ensuremath{\Varid{a}} and \ensuremath{\Varid{b}} must be vector spaces that share a common underlying field.
Written as a Haskell-style type signature (but omitting vector space constraints),
%% %format der = "\mathop{\mathcal{D}}"
%% %format der = "\raisebox{0pt}{$\mathcal{D}$}"
\begin{hscode}\SaveRestoreHook
\column{B}{@{}>{\hspre}l<{\hspost}@{}}%
\column{E}{@{}>{\hspre}l<{\hspost}@{}}%
\>[B]{}\mathcal{D}\mathbin{::}(\Varid{a}\to \Varid{b})\to (\Varid{a}\to (\Varid{a}\multimap\Varid{b})){}\<[E]%
\ColumnHook
\end{hscode}\resethooks
From the type of \ensuremath{\mathcal{D}}, it follows that differentiating twice has the following type:\footnote{As with ``\ensuremath{\to }'', we will take ``\ensuremath{\multimap}'' to associate rightward, so \ensuremath{\Varid{u}\multimap\Varid{v}\multimap\Varid{w}} is equivalent to \ensuremath{\Varid{u}\multimap(\Varid{v}\multimap\Varid{w})}.}
\begin{hscode}\SaveRestoreHook
\column{B}{@{}>{\hspre}l<{\hspost}@{}}%
\column{E}{@{}>{\hspre}l<{\hspost}@{}}%
\>[B]{}\mathcal{D}^2\mathrel{=}\mathcal{D}\hsdot{\circ }{.}\mathcal{D}\mathbin{::}{}\;(\Varid{a}\to \Varid{b})\to (\Varid{a}\to (\Varid{a}\multimap\Varid{a}\multimap\Varid{b})){}\<[E]%
\ColumnHook
\end{hscode}\resethooks

The type \ensuremath{\Varid{a}\multimap\Varid{a}\multimap\Varid{b}} is a linear map that yields a linear map, which is the curried form of a \emph{bilinear} map.
Likewise, differentiating $k$ times yields a $k$-linear map curried $k-1$ times.
For instance, the \emph{Hessian} matrix $H$ corresponds to the second derivative of a function \ensuremath{\Varid{f}\mathbin{::}\mathbb{R}^m\to \mathbb{R}}, having $m$ rows and $m$ columns (and satisfying the symmetry condition $H_{i,j} \equiv H_{j,i}$).

\sectionl{Rules for Differentiation}

\subsectionl{Sequential Composition}

With the shift to linear maps, there is one general chain rule, having a lovely form, namely that the derivative of a composition is a \emph{composition} of the derivatives \cite[Theorem 2-2]{Spivak65}:
\begin{theorem}[compose/``chain'' rule] \thmLabel{compose}
$$\ensuremath{\mathcal{D}\;(\Varid{g}\hsdot{\circ }{.}\Varid{f})\;\Varid{a}\mathrel{=}\mathcal{D}\;\Varid{g}\;(\Varid{f}\;\Varid{a})\hsdot{\circ }{.}\mathcal{D}\;\Varid{f}\;\Varid{a}}$$
\end{theorem}
If \ensuremath{\Varid{f}\mathbin{::}\Varid{a}\to \Varid{b}} and \ensuremath{\Varid{g}\mathbin{::}\Varid{b}\to \Varid{c}}\out{, and \ensuremath{\Varid{a}\mathbin{::}\Varid{a}}}, then \ensuremath{\mathcal{D}\;\Varid{f}\;\Varid{a}\mathbin{::}\Varid{a}\multimap\Varid{b}}, and \ensuremath{\mathcal{D}\;\Varid{g}\;(\Varid{f}\;\Varid{a})\mathbin{::}\Varid{b}\multimap\Varid{c}}, so both sides of this equation have type \ensuremath{\Varid{a}\multimap\Varid{c}}.\footnote{I adopt the common, if sometimes confusing, Haskell convention of sharing names between type and value variables, e.g., with \ensuremath{\Varid{a}} (a value variable) having type \ensuremath{\Varid{a}} (a type variable).
Haskell value and type variable names live in different name spaces and are distinguished by syntactic context.}

Strictly speaking, \thmRef{compose} is not a compositional recipe for differentiating sequential compositions, i.e., it is \emph{not} the case \ensuremath{\mathcal{D}\;(\Varid{g}\hsdot{\circ }{.}\Varid{f})} can be constructed solely from \ensuremath{\mathcal{D}\;\Varid{g}} and \ensuremath{\mathcal{D}\;\Varid{f}}.
Instead, it also needs \ensuremath{\Varid{f}} itself.
Fortunately, there is a simple way to restore compositionality.
Instead of constructing just the derivative of a function \ensuremath{\Varid{f}}, suppose we \emph{augment} \ensuremath{\Varid{f}} with its derivative:

\begin{hscode}\SaveRestoreHook
\column{B}{@{}>{\hspre}l<{\hspost}@{}}%
\column{53}{@{}>{\hspre}l<{\hspost}@{}}%
\column{E}{@{}>{\hspre}l<{\hspost}@{}}%
\>[B]{}\mathcal{D}\!_{\scriptscriptstyle 0}\!\!^+\!\mathbin{::}(\Varid{a}\to \Varid{b})\to ((\Varid{a}\to \Varid{b}) \times (\Varid{a}\to (\Varid{a}\multimap\Varid{b}))){}\<[53]%
\>[53]{}\mbox{\onelinecomment  first try}{}\<[E]%
\\
\>[B]{}\mathcal{D}\!_{\scriptscriptstyle 0}\!\!^+\!\;\Varid{f}\mathrel{=}(\Varid{f},\mathcal{D}\;\Varid{f}){}\<[E]%
\ColumnHook
\end{hscode}\resethooks
As desired, this altered specification is compositional:
\begin{hscode}\SaveRestoreHook
\column{B}{@{}>{\hspre}c<{\hspost}@{}}%
\column{BE}{@{}l@{}}%
\column{5}{@{}>{\hspre}l<{\hspost}@{}}%
\column{44}{@{}>{\hspre}l<{\hspost}@{}}%
\column{E}{@{}>{\hspre}l<{\hspost}@{}}%
\>[5]{}\mathcal{D}\!_{\scriptscriptstyle 0}\!\!^+\!\;(\Varid{g}\hsdot{\circ }{.}\Varid{f}){}\<[E]%
\\
\>[B]{}\mathrel{=}{}\<[BE]%
\>[5]{}(\Varid{g}\hsdot{\circ }{.}\Varid{f},\mathcal{D}\;(\Varid{g}\hsdot{\circ }{.}\Varid{f})){}\<[44]%
\>[44]{}\mbox{\onelinecomment  definition of \ensuremath{\mathcal{D}\!_{\scriptscriptstyle 0}\!\!^+\!}}{}\<[E]%
\\
\>[B]{}\mathrel{=}{}\<[BE]%
\>[5]{}(\Varid{g}\hsdot{\circ }{.}\Varid{f},\lambda \Varid{a}\to \mathcal{D}\;\Varid{g}\;(\Varid{f}\;\Varid{a})\hsdot{\circ }{.}\mathcal{D}\;\Varid{f}\;\Varid{a}){}\<[44]%
\>[44]{}\mbox{\onelinecomment  \thmRef{compose}}{}\<[E]%
\ColumnHook
\end{hscode}\resethooks

Note that \ensuremath{\mathcal{D}\!_{\scriptscriptstyle 0}\!\!^+\!\;(\Varid{g}\hsdot{\circ }{.}\Varid{f})} is assembled entirely from components of \ensuremath{\mathcal{D}\!_{\scriptscriptstyle 0}\!\!^+\!\;\Varid{g}} and \ensuremath{\mathcal{D}\!_{\scriptscriptstyle 0}\!\!^+\!\;\Varid{f}}, which is to say from \ensuremath{\Varid{g}}, \ensuremath{\mathcal{D}\;\Varid{g}}, \ensuremath{\Varid{f}}, and \ensuremath{\mathcal{D}\;\Varid{f}}.
Writing out \ensuremath{\Varid{g}\hsdot{\circ }{.}\Varid{f}} as \ensuremath{\lambda \Varid{a}\to \Varid{g}\;(\Varid{f}\;\Varid{a})} underscores that the two parts of \ensuremath{\mathcal{D}\!_{\scriptscriptstyle 0}\!\!^+\!\;(\Varid{g}\hsdot{\circ }{.}\Varid{f})\;\Varid{a}} both involve \ensuremath{\Varid{f}\;\Varid{a}}.
Computing these parts independently thus requires redundant work.
Moreover, the chain rule itself requires applying a function and its derivative (namely \ensuremath{\Varid{f}} and \ensuremath{\mathcal{D}\;\Varid{f}}) to the same \ensuremath{\Varid{a}}.
Since the chain rule gets applied recursively to nested compositions, this redundant work multiplies greatly, resulting in an impractically expensive algorithm.

Fortunately, this efficiency problem is easily fixed.
Instead of pairing \ensuremath{\Varid{f}} and \ensuremath{\mathcal{D}\;\Varid{f}}, \emph{combine} them\out{ into a single function}:\footnote{The precedence of ``\ensuremath{ \times }'' is tighter than that of ``\ensuremath{\to }'' and ``\ensuremath{\multimap}'', so \ensuremath{\Varid{a}\to \Varid{b} \times (\Varid{a}\multimap\Varid{b})} is equivalent to \ensuremath{\Varid{a}\to (\Varid{b} \times (\Varid{a}\multimap\Varid{b}))}.}
\label{code:ad}
\begin{hscode}\SaveRestoreHook
\column{B}{@{}>{\hspre}l<{\hspost}@{}}%
\column{43}{@{}>{\hspre}l<{\hspost}@{}}%
\column{E}{@{}>{\hspre}l<{\hspost}@{}}%
\>[B]{}\mathcal{D}\!^+\!\mathbin{::}(\Varid{a}\to \Varid{b})\to (\Varid{a}\to \Varid{b} \times (\Varid{a}\multimap\Varid{b})){}\<[43]%
\>[43]{}\mbox{\onelinecomment  better!}{}\<[E]%
\\
\>[B]{}\mathcal{D}\!^+\!\;\Varid{f}\;\Varid{a}\mathrel{=}(\Varid{f}\;\Varid{a},\mathcal{D}\;\Varid{f}\;\Varid{a}){}\<[E]%
\ColumnHook
\end{hscode}\resethooks
Combining \ensuremath{\Varid{f}} and \ensuremath{\mathcal{D}\;\Varid{f}} into a single function in this way enables us to eliminate the redundant computation of \ensuremath{\Varid{f}\;\Varid{a}} in \ensuremath{\mathcal{D}\!^+\!\;(\Varid{g}\hsdot{\circ }{.}\Varid{f})\;\Varid{a}}, as follows:
\begin{corollary}[\provedIn{corollary:compose}]\corLabel{compose}
\ensuremath{\mathcal{D}\!^+\!} is (efficiently) compositional with respect to \ensuremath{(\hsdot{\circ }{.})}. Specifically,
\begin{hscode}\SaveRestoreHook
\column{B}{@{}>{\hspre}l<{\hspost}@{}}%
\column{E}{@{}>{\hspre}l<{\hspost}@{}}%
\>[B]{}\mathcal{D}\!^+\!\;(\Varid{g}\hsdot{\circ }{.}\Varid{f})\;\Varid{a}\mathrel{=}\mathbf{let}\;\{\mskip1.5mu (\Varid{b},\Varid{f'})\mathrel{=}\mathcal{D}\!^+\!\;\Varid{f}\;\Varid{a};(\Varid{c},\Varid{g'})\mathrel{=}\mathcal{D}\!^+\!\;\Varid{g}\;\Varid{b}\mskip1.5mu\}\;\mathbf{in}\;(\Varid{c},\Varid{g'}\hsdot{\circ }{.}\Varid{f'}){}\<[E]%
\ColumnHook
\end{hscode}\resethooks
\end{corollary}

\subsectionl{Parallel Composition}

The chain rule, telling how to differentiate sequential compositions, gets a lot of attention in calculus classes and in automatic and symbolic differentiation.\out{\notefoot{To do: introduce AD and SD early.}}
There are other important ways to combine functions, however, and examining them yields additional helpful tools.
Another operation (pronounced ``cross'') combines two functions in \emph{parallel} \citep{Gibbons2002Calculating}:\footnote{By ``parallel'', I mean without data dependencies. Operationally, the two functions can be applied simultaneously or not.}
\begin{hscode}\SaveRestoreHook
\column{B}{@{}>{\hspre}l<{\hspost}@{}}%
\column{E}{@{}>{\hspre}l<{\hspost}@{}}%
\>[B]{}(\times)\mathbin{::}(\Varid{a}\to \Varid{c})\to (\Varid{b}\to \Varid{d})\to (\Varid{a} \times \Varid{b}\to \Varid{c} \times \Varid{d}){}\<[E]%
\\
\>[B]{}\Varid{f}\times\Varid{g}\mathrel{=}\lambda (\Varid{a},\Varid{b})\to (\Varid{f}\;\Varid{a},\Varid{g}\;\Varid{b}){}\<[E]%
\ColumnHook
\end{hscode}\resethooks

While the derivative of a sequential composition is a sequential composition of derivatives, the derivative of a parallel composition is a parallel composition of the derivatives \citep[variant of Theorem 2-3 (3)]{Spivak65}:\notefoot{Is there a name for this rule? I've never seen it mentioned.}
\begin{theorem}[cross rule] \thmLabel{cross}
$$\ensuremath{\mathcal{D}\;(\Varid{f}\times\Varid{g})\;(\Varid{a},\Varid{b})\mathrel{=}\mathcal{D}\;\Varid{f}\;\Varid{a}\times\mathcal{D}\;\Varid{g}\;\Varid{b}}$$
\end{theorem}
If \ensuremath{\Varid{f}\mathbin{::}\Varid{a}\to \Varid{c}} and \ensuremath{\Varid{g}\mathbin{::}\Varid{b}\to \Varid{d}}, then \ensuremath{\mathcal{D}\;\Varid{f}\;\Varid{a}\mathbin{::}\Varid{a}\multimap\Varid{c}} and \ensuremath{\mathcal{D}\;\Varid{g}\;\Varid{b}\mathbin{::}\Varid{b}\multimap\Varid{d}}, so both sides of this equation have type \ensuremath{\Varid{a} \times \Varid{b}\multimap\Varid{c} \times \Varid{d}}.

\thmRef{cross} gives us what we need to construct \ensuremath{\mathcal{D}\!^+\!\;(\Varid{f}\times\Varid{g})} compositionally:
\begin{corollary}[\provedIn{corollary:cross}] \corLabel{cross}
\ensuremath{\mathcal{D}\!^+\!} is compositional with respect to \ensuremath{(\times)}. Specifically,
$$\ensuremath{\mathcal{D}\!^+\!\;(\Varid{f}\times\Varid{g})\;(\Varid{a},\Varid{b})\mathrel{=}\mathbf{let}\;\{\mskip1.5mu (\Varid{c},\Varid{f'})\mathrel{=}\mathcal{D}\!^+\!\;\Varid{f}\;\Varid{a};(\Varid{d},\Varid{g'})\mathrel{=}\mathcal{D}\!^+\!\;\Varid{g}\;\Varid{b}\mskip1.5mu\}\;\mathbf{in}\;((\Varid{c},\Varid{d}),\Varid{f'}\times\Varid{g'})}$$
\end{corollary}

An important point left implicit in the discussion above is that sequential and parallel composition preserve linearity.
This property is what makes it meaningful to use these forms to combine derivatives, i.e., linear maps, as we've done above.

\subsectionl{Linear Functions}

A function \ensuremath{\Varid{f}} is said to be \emph{linear} when it distributes over (preserves the structure of) vector addition and scalar multiplication, i.e.,
\begin{hscode}\SaveRestoreHook
\column{B}{@{}>{\hspre}l<{\hspost}@{}}%
\column{13}{@{}>{\hspre}l<{\hspost}@{}}%
\column{E}{@{}>{\hspre}l<{\hspost}@{}}%
\>[B]{}\Varid{f}\;(\Varid{a}\mathbin{+}\Varid{a'}){}\<[13]%
\>[13]{}\mathrel{=}\Varid{f}\;\Varid{a}\mathbin{+}\Varid{f}\;\Varid{a'}{}\<[E]%
\\
\>[B]{}\Varid{f}\;(\Varid{s} \cdot \Varid{a}){}\<[13]%
\>[13]{}\mathrel{=}\Varid{s} \cdot \Varid{f}\;\Varid{a}{}\<[E]%
\ColumnHook
\end{hscode}\resethooks
%% for all |a,a' :: u| and |s| taken from the scalar field underlying |u| and |v|.

\begin{samepage}
In addition to \thmRefTwo{compose}{cross}, we will want one more broadly useful rule, namely that \emph{the derivative of every linear function is itself, everywhere} \citep[Theorem 2-3 (2)]{Spivak65}:
\begin{theorem}[linear rule] \thmLabel{linear}
For all linear functions \ensuremath{\Varid{f}}, \ensuremath{\mathcal{D}\;\Varid{f}\;\Varid{a}\mathrel{=}\Varid{f}}.
\end{theorem}
\end{samepage}
This statement may sound surprising at first, but less so when we recall that the \ensuremath{\mathcal{D}\;\Varid{f}\;\Varid{a}} is a local linear approximation of \ensuremath{\Varid{f}} at \ensuremath{\Varid{a}}, so we're simply saying that linear functions are their own perfect linear approximations.

For example, consider the function \ensuremath{\Varid{id}\mathrel{=}\lambda \Varid{a}\to \Varid{a}}.
\thmRef{linear} says that \ensuremath{\mathcal{D}\;\Varid{id}\;\Varid{a}\mathrel{=}\Varid{id}}.
When expressed via typical \emph{representations} of linear maps, this property may be expressed as saying that \ensuremath{\mathcal{D}\;\Varid{id}\;\Varid{a}} is the number one or is an identity matrix (with ones on the diagonal and zeros elsewhere).
%% %format Rmn = R"^{m+n}"
Likewise, consider the (linear) function \ensuremath{\Varid{fst}\;(\Varid{a},\Varid{b})\mathrel{=}\Varid{a}}, for which \thmRef{linear} says \ensuremath{\mathcal{D}\;\Varid{fst}\;(\Varid{a},\Varid{b})\mathrel{=}\Varid{fst}}.
This property, when expressed via typical \emph{representations} of linear maps, would appear as saying that \ensuremath{\mathcal{D}\;\Varid{fst}\;\Varid{a}} comprises the partial derivatives one and zero if \ensuremath{\Varid{a},\Varid{b}\mathbin{::}\mathbb{R}}.
More generally, if \ensuremath{\Varid{a}\mathbin{::}\mathbb{R}^m} and \ensuremath{\Varid{b}\mathbin{::}\mathbb{R}^n}, then the Jacobian matrix representation has shape \ensuremath{\Varid{m} \times (\Varid{m}\mathbin{+}\Varid{n})} (i.e., \ensuremath{\Varid{m}} rows and \ensuremath{\Varid{m}\mathbin{+}\Varid{n}} columns) and is formed by the horizontal juxtaposition of an \ensuremath{\Varid{m} \times \Varid{m}} identity matrix on the left with an \ensuremath{\Varid{m} \times \Varid{n}} zero matrix on the right.
This \ensuremath{\Varid{m} \times (\Varid{m}\mathbin{+}\Varid{n})} matrix, however, represents \ensuremath{\Varid{fst}\mathbin{::}\mathbb{R}^m \times \mathbb{R}^n\multimap\mathbb{R}^m}.
Note how much simpler it is to say \ensuremath{\mathcal{D}\;\Varid{fst}\;(\Varid{a},\Varid{b})\mathrel{=}\Varid{fst}}, and with no loss of precision!

Given \thmRef{linear}, we can construct \ensuremath{\mathcal{D}\!^+\!\;\Varid{f}} for all linear \ensuremath{\Varid{f}}:
\begin{corollary} \corLabel{linear}
For all linear functions \ensuremath{\Varid{f}}, \ensuremath{\mathcal{D}\!^+\!\;\Varid{f}\mathrel{=}\lambda \Varid{a}\to (\Varid{f}\;\Varid{a},\Varid{f})}.
(Proof: immediate from the \ensuremath{\mathcal{D}\!^+\!} definition and \thmRef{linear}.)
\end{corollary}

\sectionl{Putting the Pieces Together}

The definition of \ensuremath{\mathcal{D}\!^+\!} on page \pageref{code:ad} is a precise specification; but it is not an implementation, since \ensuremath{\mathcal{D}} itself is not computable \citep{PourEl1978Diff, PourEl1983Comp}.
\corRefs{compose}{linear} provide insight into the compositional nature of \ensuremath{\mathcal{D}\!^+\!} in exactly the form we can now assemble into a correct-by-construction implementation.

Although differentiation is not computable when given just an arbitrary computable function, we can instead build up differentiable functions compositionally, using exactly the forms introduced above, (namely \ensuremath{(\hsdot{\circ }{.})}, \ensuremath{(\times)} and linear functions), together with various non-linear primitives having known derivatives.
Computations expressed in this vocabulary are differentiable by construction thanks to \corRefs{compose}{linear}.
The building blocks above are not just a random assortment, but rather a fundamental language of mathematics, logic, and computation, known as \emph{category theory} \citep{MacLane1998categories,Lawvere:2009:Conceptual,Awodey2006CT}.
While it would be unpleasant to program directly in such an austere language, its foundational nature enables instead an automatic conversion from programs written in more conventional functional languages \citep{Lambek:1980:LambdaToCCC,Lambek:1985:CCC,Elliott-2017-compiling-to-categories}.

\subsectionl{Categories}

The central notion in category theory is that of a \emph{category}, comprising \emph{objects} (generalizing sets or types) and \emph{morphisms} (generalizing functions between sets or types).
For the purpose of this paper, we will take objects to be types in our program, and morphisms to be enhanced functions.
We will introduce morphisms using Haskell-style type signatures, such as ``\ensuremath{\Varid{f}\mathbin{::}\Varid{a} \leadsto \Varid{b}\in \mathcal{U}}'', where ``\ensuremath{ \leadsto }'' refers to the morphisms for a category \ensuremath{\mathcal{U}}, with \ensuremath{\Varid{a}} and \ensuremath{\Varid{b}} being the \emph{domain} and \emph{codomain} objects/types for \ensuremath{\Varid{f}}.
In most cases, we will omit the ``\ensuremath{\in \mathcal{U}}'', where choice of category is (hopefully) clear from context.
Each category \ensuremath{\mathcal{U}} has a distinguished \emph{identity} morphism \ensuremath{\Varid{id}\mathbin{::}\Varid{a} \leadsto \Varid{a}\in \mathcal{U}} for every object/type \ensuremath{\Varid{a}} in the category.
For any two morphisms \ensuremath{\Varid{f}\mathbin{::}\Varid{a} \leadsto \Varid{b}\in \mathcal{U}} and \ensuremath{\Varid{g}\mathbin{::}\Varid{b} \leadsto \Varid{c}\in \mathcal{U}} (note same category and matching types/objects \ensuremath{\Varid{b}}), there is also the composition \ensuremath{\Varid{g}\hsdot{\circ }{.}\Varid{f}\mathbin{::}\Varid{a} \leadsto \Varid{c}\in \mathcal{U}}.
The category laws state that
(a) \ensuremath{\Varid{id}} is the left and right identity for composition, and (b) composition is associative.
You are probably already familiar with at least one example of a category, namely functions, in which \ensuremath{\Varid{id}} and \ensuremath{(\hsdot{\circ }{.})} are the identity function and function composition.

%% %format `k` = "\leadsto"
%% %format k = "(\leadsto)"

Although Haskell's type system cannot capture the category laws explicitly, we can express the two required operations as a Haskell type class, along with a familiar instance:\notefoot{Would the signatures in this paper be easier to read without using infix type variables?
For instance, ``\ensuremath{(\hsdot{\circ }{.})\mathbin{::}\Varid{k}\;\Varid{b}\;\Varid{c}\to \Varid{k}\;\Varid{a}\;\Varid{b}\to \Varid{k}\;\Varid{a}\;\Varid{c}}''.}
\\
\begin{minipage}[b]{0.48\textwidth}
\begin{hscode}\SaveRestoreHook
\column{B}{@{}>{\hspre}l<{\hspost}@{}}%
\column{3}{@{}>{\hspre}l<{\hspost}@{}}%
\column{8}{@{}>{\hspre}l<{\hspost}@{}}%
\column{E}{@{}>{\hspre}l<{\hspost}@{}}%
\>[B]{}\mathbf{class}\;\Conid{Category}\;\Varid{k}\;\mathbf{where}{}\<[E]%
\\
\>[B]{}\hsindent{3}{}\<[3]%
\>[3]{}\Varid{id}{}\<[8]%
\>[8]{}\mathbin{::}\Varid{a}\mathbin{`\Varid{k}`}\Varid{a}{}\<[E]%
\\
\>[B]{}\hsindent{3}{}\<[3]%
\>[3]{}(\hsdot{\circ }{.}){}\<[8]%
\>[8]{}\mathbin{::}(\Varid{b}\mathbin{`\Varid{k}`}\Varid{c})\to (\Varid{a}\mathbin{`\Varid{k}`}\Varid{b})\to (\Varid{a}\mathbin{`\Varid{k}`}\Varid{c}){}\<[E]%
\ColumnHook
\end{hscode}\resethooks
\end{minipage}
\begin{minipage}[b]{0ex}{\rule[1ex]{0.5pt}{0.5in}}\end{minipage}
\begin{minipage}[b]{0.45\textwidth} % \mathindent1em
\begin{hscode}\SaveRestoreHook
\column{B}{@{}>{\hspre}l<{\hspost}@{}}%
\column{3}{@{}>{\hspre}l<{\hspost}@{}}%
\column{E}{@{}>{\hspre}l<{\hspost}@{}}%
\>[B]{}\mathbf{instance}\;\Conid{Category}\;(\to )\;\mathbf{where}{}\<[E]%
\\
\>[B]{}\hsindent{3}{}\<[3]%
\>[3]{}\Varid{id}\mathrel{=}\lambda \Varid{a}\to \Varid{a}{}\<[E]%
\\
\>[B]{}\hsindent{3}{}\<[3]%
\>[3]{}\Varid{g}\hsdot{\circ }{.}\Varid{f}\mathrel{=}\lambda \Varid{a}\to \Varid{g}\;(\Varid{f}\;\Varid{a}){}\<[E]%
\ColumnHook
\end{hscode}\resethooks
\end{minipage}
\\
Another example is \emph{linear} functions, which we've written ``\ensuremath{\Varid{a}\multimap\Varid{b}}'' above.
Still another example is \emph{differentiable} functions\footnote{There are many examples of categories besides restricted forms of functions, including relations, logics, partial orders, and even matrices.}, which we can see by noting three facts:
\begin{itemize}
\item The identity function is differentiable, as witnessed by \thmRef{linear} and the linearity of \ensuremath{\Varid{id}}.
\item The composition of differentiable functions is differentiable, as \thmRef{compose} attests.
\item The category laws (identity and associativity) hold, because differentiable functions form a subset of all functions.
\end{itemize}

Each category forms its own world, with morphisms relating objects within that category.
To bridge between these worlds, there are \emph{functors}, which connect a category \ensuremath{\mathcal{U}} to a (possibly different) category \ensuremath{\mathcal{V}}.
Such a functor \ensuremath{\Conid{F}} maps objects in \ensuremath{\mathcal{U}} to objects in \ensuremath{\mathcal{V}}, \emph{and} morphisms in \ensuremath{\mathcal{U}} to morphisms in \ensuremath{\mathcal{V}}.
If \ensuremath{\Varid{f}\mathbin{::}\Varid{a} \leadsto \Varid{b}\in \mathcal{U}} is a morphism, then a \emph{functor} \ensuremath{\Conid{F}} from \ensuremath{\mathcal{U}} to \ensuremath{\mathcal{V}} transforms \ensuremath{\Varid{f}\in \mathcal{U}} to a morphism \ensuremath{\Conid{F}\;\Varid{f}\mathbin{::}\Conid{F}\;\Varid{a}\dashrightarrow\Conid{F}\;\Varid{b}\in \mathcal{V}}, i.e., the domain and codomain of the transformed morphism \ensuremath{\Conid{F}\;\Varid{f}\in \mathcal{V}} must be the transformed versions of the domain and codomain of \ensuremath{\Varid{f}\in \mathcal{U}}.
In this paper, the categories use types as objects, while the functors map these types to themselves.%
\footnote{In contrast, Haskell's functors stay within the same category and do change types.}
The functor must also preserve ``categorical'' structure:\footnote{Making the categories explicit, \ensuremath{\Conid{F}\;(\Varid{id}\in \mathcal{U})\mathrel{=}(\Varid{id}\in \mathcal{V})} and \ensuremath{\Conid{F}\;(\Varid{g}\hsdot{\circ }{.}\Varid{f}\in \mathcal{U})\mathrel{=}(\Conid{F}\;\Varid{g}\hsdot{\circ }{.}\Conid{F}\;\Varid{f}\in \mathcal{V})}.}
\begin{closerCodePars}
\begin{hscode}\SaveRestoreHook
\column{B}{@{}>{\hspre}l<{\hspost}@{}}%
\column{E}{@{}>{\hspre}l<{\hspost}@{}}%
\>[B]{}\Conid{F}\;\Varid{id}\mathrel{=}\Varid{id}{}\<[E]%
\\[\blanklineskip]%
\>[B]{}\Conid{F}\;(\Varid{g}\hsdot{\circ }{.}\Varid{f})\mathrel{=}\Conid{F}\;\Varid{g}\hsdot{\circ }{.}\Conid{F}\;\Varid{f}{}\<[E]%
\ColumnHook
\end{hscode}\resethooks
\end{closerCodePars}%

Crucially to the topic of this paper, \corRefTwo{linear}{compose} say more than that differentiable functions form a category.
They also point us to a new, easily implemented category, for which \ensuremath{\mathcal{D}\!^+\!} is in fact a functor.
This new category is simply the representation that \ensuremath{\mathcal{D}\!^+\!} produces: \ensuremath{\Varid{a}\to \Varid{b} \times (\Varid{a}\multimap\Varid{b})}, considered as having domain \ensuremath{\Varid{a}} and codomain \ensuremath{\Varid{b}}.
The functor nature of \ensuremath{\mathcal{D}\!^+\!} will be exactly what we need to in order to program in a familiar and direct way in a pleasant functional language such as Haskell and have a compiler convert to differentiable functions automatically.

To make the new category more explicit, package the result type of \ensuremath{\mathcal{D}\!^+\!} in a new data type:\out{\notefoot{Maybe format \ensuremath{\Conid{D}\;\Varid{a}\;\Varid{b}} using an infix operator. Remember that I'll need another for generalized AD (\ensuremath{\Conid{GD}}).}}
\begin{hscode}\SaveRestoreHook
\column{B}{@{}>{\hspre}l<{\hspost}@{}}%
\column{E}{@{}>{\hspre}l<{\hspost}@{}}%
\>[B]{}\mathbf{newtype}\;\Conid{D}\;\Varid{a}\;\Varid{b}\mathrel{=}\Conid{D}\;(\Varid{a}\to \Varid{b} \times (\Varid{a}\multimap\Varid{b})){}\<[E]%
\ColumnHook
\end{hscode}\resethooks
Then adapt \ensuremath{\mathcal{D}\!^+\!} to use this new data type by simply applying the \ensuremath{\Conid{D}} constructor to the result of \ensuremath{\mathcal{D}\!^+\!}:

%% Why doesn't the following definition work?

\begin{hscode}\SaveRestoreHook
\column{B}{@{}>{\hspre}l<{\hspost}@{}}%
\column{E}{@{}>{\hspre}l<{\hspost}@{}}%
\>[B]{}\hat{\mathcal{D}}\mathbin{::}(\Varid{a}\to \Varid{b})\to \Conid{D}\;\Varid{a}\;\Varid{b}{}\<[E]%
\\
\>[B]{}\hat{\mathcal{D}}\;\Varid{f}\mathrel{=}\Conid{D}\;(\mathcal{D}\!^+\!\;\Varid{f}){}\<[E]%
\ColumnHook
\end{hscode}\resethooks

\begin{closerCodePars}
Our goal is to discover a \ensuremath{\Conid{Category}} instance for \ensuremath{\Conid{D}} such that \ensuremath{\hat{\mathcal{D}}} is a functor.
This goal is essentially an algebra problem, and the desired \ensuremath{\Conid{Category}} instance is a solution to that problem.
Saying that \ensuremath{\hat{\mathcal{D}}} is a functor is equivalent to the following two conditions for all suitably typed functions \ensuremath{\Varid{f}} and \ensuremath{\Varid{g}}:\footnote{The \ensuremath{\Varid{id}} and \ensuremath{(\hsdot{\circ }{.})} on the left-hand sides are for \ensuremath{\Conid{D}}, while the ones on the right are for \ensuremath{(\to )}.}
\begin{hscode}\SaveRestoreHook
\column{B}{@{}>{\hspre}l<{\hspost}@{}}%
\column{E}{@{}>{\hspre}l<{\hspost}@{}}%
\>[B]{}\Varid{id}\mathrel{=}\hat{\mathcal{D}}\;\Varid{id}{}\<[E]%
\\[\blanklineskip]%
\>[B]{}\hat{\mathcal{D}}\;\Varid{g}\hsdot{\circ }{.}\hat{\mathcal{D}}\;\Varid{f}\mathrel{=}\hat{\mathcal{D}}\;(\Varid{g}\hsdot{\circ }{.}\Varid{f}){}\<[E]%
\ColumnHook
\end{hscode}\resethooks
Equivalently, by the definition of \ensuremath{\hat{\mathcal{D}}},
\begin{hscode}\SaveRestoreHook
\column{B}{@{}>{\hspre}l<{\hspost}@{}}%
\column{E}{@{}>{\hspre}l<{\hspost}@{}}%
\>[B]{}\Varid{id}\mathrel{=}\Conid{D}\;(\mathcal{D}\!^+\!\;\Varid{id}){}\<[E]%
\\[\blanklineskip]%
\>[B]{}\Conid{D}\;(\mathcal{D}\!^+\!\;\Varid{g})\hsdot{\circ }{.}\Conid{D}\;(\mathcal{D}\!^+\!\;\Varid{f})\mathrel{=}\Conid{D}\;(\mathcal{D}\!^+\!\;(\Varid{g}\hsdot{\circ }{.}\Varid{f})){}\<[E]%
\ColumnHook
\end{hscode}\resethooks
Now recall the following results from \corRefTwo{linear}{compose}:
\begin{hscode}\SaveRestoreHook
\column{B}{@{}>{\hspre}l<{\hspost}@{}}%
\column{E}{@{}>{\hspre}l<{\hspost}@{}}%
\>[B]{}\mathcal{D}\!^+\!\;\Varid{id}\mathrel{=}\lambda \Varid{a}\to (\Varid{id}\;\Varid{a},\Varid{id}){}\<[E]%
\\[\blanklineskip]%
\>[B]{}\mathcal{D}\!^+\!\;(\Varid{g}\hsdot{\circ }{.}\Varid{f})\mathrel{=}\lambda \Varid{a}\to \mathbf{let}\;\{\mskip1.5mu (\Varid{b},\Varid{f'})\mathrel{=}\mathcal{D}\!^+\!\;\Varid{f}\;\Varid{a};(\Varid{c},\Varid{g'})\mathrel{=}\mathcal{D}\!^+\!\;\Varid{g}\;\Varid{b}\mskip1.5mu\}\;\mathbf{in}\;(\Varid{c},\Varid{g'}\hsdot{\circ }{.}\Varid{f'}){}\<[E]%
\ColumnHook
\end{hscode}\resethooks
Then use these two facts to rewrite the right-hand sides of the functor specification for \ensuremath{\hat{\mathcal{D}}}:
\begin{hscode}\SaveRestoreHook
\column{B}{@{}>{\hspre}l<{\hspost}@{}}%
\column{E}{@{}>{\hspre}l<{\hspost}@{}}%
\>[B]{}\Varid{id}\mathrel{=}\Conid{D}\;(\lambda \Varid{a}\to (\Varid{a},\Varid{id})){}\<[E]%
\\[\blanklineskip]%
\>[B]{}\Conid{D}\;(\mathcal{D}\!^+\!\;\Varid{g})\hsdot{\circ }{.}\Conid{D}\;(\mathcal{D}\!^+\!\;\Varid{f})\mathrel{=}\Conid{D}\;(\lambda \Varid{a}\to \mathbf{let}\;\{\mskip1.5mu (\Varid{b},\Varid{f'})\mathrel{=}\mathcal{D}\!^+\!\;\Varid{f}\;\Varid{a};(\Varid{c},\Varid{g'})\mathrel{=}\mathcal{D}\!^+\!\;\Varid{g}\;\Varid{b}\mskip1.5mu\}\;\mathbf{in}\;(\Varid{c},\Varid{g'}\hsdot{\circ }{.}\Varid{f'})){}\<[E]%
\ColumnHook
\end{hscode}\resethooks
The \ensuremath{\Varid{id}} equation is trivially solvable by \emph{defining} \ensuremath{\Varid{id}\mathrel{=}\Conid{D}\;(\lambda \Varid{a}\to (\Varid{a},\Varid{id}))}.
To solve the \ensuremath{(\hsdot{\circ }{.})} equation, generalize it to a \emph{stronger} condition:\footnote{The new \ensuremath{\Varid{f}} is the old \ensuremath{\mathcal{D}\!^+\!\;\Varid{f}} and so has changed type from \ensuremath{\Varid{a}\to \Varid{b}} to \ensuremath{\Varid{a}\to \Varid{b} \times (\Varid{a}\multimap\Varid{b})}. Likewise for \ensuremath{\Varid{g}}.}
\begin{hscode}\SaveRestoreHook
\column{B}{@{}>{\hspre}l<{\hspost}@{}}%
\column{E}{@{}>{\hspre}l<{\hspost}@{}}%
\>[B]{}\Conid{D}\;\Varid{g}\hsdot{\circ }{.}\Conid{D}\;\Varid{f}\mathrel{=}\Conid{D}\;(\lambda \Varid{a}\to \mathbf{let}\;\{\mskip1.5mu (\Varid{b},\Varid{f'})\mathrel{=}\Varid{f}\;\Varid{a};(\Varid{c},\Varid{g'})\mathrel{=}\Varid{g}\;\Varid{b}\mskip1.5mu\}\;\mathbf{in}\;(\Varid{c},\Varid{g'}\hsdot{\circ }{.}\Varid{f'})){}\<[E]%
\ColumnHook
\end{hscode}\resethooks
The solution of this stronger condition is immediate, leading to the following instance as a sufficient condition for \ensuremath{\hat{\mathcal{D}}} being a functor:
\end{closerCodePars}%
\begin{hscode}\SaveRestoreHook
\column{B}{@{}>{\hspre}l<{\hspost}@{}}%
\column{3}{@{}>{\hspre}l<{\hspost}@{}}%
\column{E}{@{}>{\hspre}l<{\hspost}@{}}%
\>[B]{}\Varid{linearD}\mathbin{::}(\Varid{a}\to \Varid{b})\to \Conid{D}\;\Varid{a}\;\Varid{b}{}\<[E]%
\\
\>[B]{}\Varid{linearD}\;\Varid{f}\mathrel{=}\Conid{D}\;(\lambda \Varid{a}\to (\Varid{f}\;\Varid{a},\Varid{f})){}\<[E]%
\\[\blanklineskip]%
\>[B]{}\mathbf{instance}\;\Conid{Category}\;\Conid{D}\;\mathbf{where}{}\<[E]%
\\
\>[B]{}\hsindent{3}{}\<[3]%
\>[3]{}\Varid{id}\mathrel{=}\Varid{linearD}\;\Varid{id}{}\<[E]%
\\
\>[B]{}\hsindent{3}{}\<[3]%
\>[3]{}\Conid{D}\;\Varid{g}\hsdot{\circ }{.}\Conid{D}\;\Varid{f}\mathrel{=}\Conid{D}\;(\lambda \Varid{a}\to \mathbf{let}\;\{\mskip1.5mu (\Varid{b},\Varid{f'})\mathrel{=}\Varid{f}\;\Varid{a};(\Varid{c},\Varid{g'})\mathrel{=}\Varid{g}\;\Varid{b}\mskip1.5mu\}\;\mathbf{in}\;(\Varid{c},\Varid{g'}\hsdot{\circ }{.}\Varid{f'})){}\<[E]%
\ColumnHook
\end{hscode}\resethooks
Factoring out \ensuremath{\Varid{linearD}} will also tidy up treatment of other linear functions.

Before we get too pleased with this definition, let's remember that for \ensuremath{\Conid{D}} to be a category requires more than having definitions for \ensuremath{\Varid{id}} and \ensuremath{(\hsdot{\circ }{.})}.
These definitions must also satisfy the identity and composition laws.
How might we go about proving that they do?
Perhaps the most obvious route is take those laws, substitute our definitions of \ensuremath{\Varid{id}} and \ensuremath{(\hsdot{\circ }{.})}, and reason equationally toward the desired conclusion.
For instance, let's prove that \ensuremath{\Varid{id}\hsdot{\circ }{.}\Conid{D}\;\Varid{f}\mathrel{=}\Conid{D}\;\Varid{f}} for all \ensuremath{\Conid{D}\;\Varid{f}\mathbin{::}\Conid{D}\;\Varid{a}\;\Varid{b}}:\footnote{Note that \emph{every} morphism in \ensuremath{\Conid{D}} has the form \ensuremath{\Conid{D}\;\Varid{f}} for some \ensuremath{\Varid{f}}, so it suffices to consider this form.}\notefootsep{}\notefoot{If pinched for space, remove this proof or move it to the proof appendix.}
\begin{hscode}\SaveRestoreHook
\column{B}{@{}>{\hspre}c<{\hspost}@{}}%
\column{BE}{@{}l@{}}%
\column{5}{@{}>{\hspre}l<{\hspost}@{}}%
\column{72}{@{}>{\hspre}l<{\hspost}@{}}%
\column{E}{@{}>{\hspre}l<{\hspost}@{}}%
\>[5]{}\Varid{id}\hsdot{\circ }{.}\Conid{D}\;\Varid{f}{}\<[E]%
\\
\>[B]{}\mathrel{=}{}\<[BE]%
\>[5]{}\Conid{D}\;(\lambda \Varid{b}\to (\Varid{b},\Varid{id}))\hsdot{\circ }{.}\Conid{D}\;\Varid{f}{}\<[72]%
\>[72]{}\mbox{\onelinecomment  definition of \ensuremath{\Varid{id}} for \ensuremath{\Conid{D}}}{}\<[E]%
\\
\>[B]{}\mathrel{=}{}\<[BE]%
\>[5]{}\Conid{D}\;(\lambda \Varid{a}\to \mathbf{let}\;\{\mskip1.5mu (\Varid{b},\Varid{f'})\mathrel{=}\Varid{f}\;\Varid{a};(\Varid{c},\Varid{g'})\mathrel{=}(\Varid{b},\Varid{id})\mskip1.5mu\}\;\mathbf{in}\;(\Varid{c},\Varid{g'}\hsdot{\circ }{.}\Varid{f'})){}\<[72]%
\>[72]{}\mbox{\onelinecomment  definition of \ensuremath{(\hsdot{\circ }{.})} for \ensuremath{\Conid{D}}}{}\<[E]%
\\
\>[B]{}\mathrel{=}{}\<[BE]%
\>[5]{}\Conid{D}\;(\lambda \Varid{a}\to \mathbf{let}\;\{\mskip1.5mu (\Varid{b},\Varid{f'})\mathrel{=}\Varid{f}\;\Varid{a}\mskip1.5mu\}\;\mathbf{in}\;(\Varid{b},\Varid{id}\hsdot{\circ }{.}\Varid{f'})){}\<[72]%
\>[72]{}\mbox{\onelinecomment  substitute \ensuremath{\Varid{b}} for \ensuremath{\Varid{c}} and \ensuremath{\Varid{id}} for \ensuremath{\Varid{g'}}}{}\<[E]%
\\
\>[B]{}\mathrel{=}{}\<[BE]%
\>[5]{}\Conid{D}\;(\lambda \Varid{a}\to \mathbf{let}\;\{\mskip1.5mu (\Varid{b},\Varid{f'})\mathrel{=}\Varid{f}\;\Varid{a}\mskip1.5mu\}\;\mathbf{in}\;(\Varid{b},\Varid{f'})){}\<[72]%
\>[72]{}\mbox{\onelinecomment   \ensuremath{\Varid{id}\hsdot{\circ }{.}\Varid{f'}\mathrel{=}\Varid{f'}} (category law)}{}\<[E]%
\\
\>[B]{}\mathrel{=}{}\<[BE]%
\>[5]{}\Conid{D}\;(\lambda \Varid{a}\to \Varid{f}\;\Varid{a}){}\<[72]%
\>[72]{}\mbox{\onelinecomment  replace \ensuremath{(\Varid{b},\Varid{f'})} by its definition}{}\<[E]%
\\
\>[B]{}\mathrel{=}{}\<[BE]%
\>[5]{}\Conid{D}\;\Varid{f}{}\<[72]%
\>[72]{}\mbox{\onelinecomment  $\eta$-reduction}{}\<[E]%
\ColumnHook
\end{hscode}\resethooks

We can prove the other required properties similarly.
Fortunately, there is a way to bypass the need for these painstaking proofs, and instead rely \emph{only} on our original specification for this \ensuremath{\Conid{Category}} instance, namely that \ensuremath{\mathcal{D}\!^+\!} is a functor.
To buy this proof convenience, we have to make one concession, namely that we consider only morphisms in \ensuremath{\Conid{D}} that arise from \ensuremath{\hat{\mathcal{D}}}, i.e., only \ensuremath{\hat{\Varid{f}}\mathbin{::}\Conid{D}\;\Varid{a}\;\Varid{b}} such that \ensuremath{\hat{\Varid{f}}\mathrel{=}\hat{\mathcal{D}}\;\Varid{f}} for some \ensuremath{\Varid{f}\mathbin{::}\Varid{a}\to \Varid{b}}.
We can ensure that indeed only such \ensuremath{\hat{\Varid{f}}} do arise by making \ensuremath{\Conid{D}\;\Varid{a}\;\Varid{b}} an \emph{abstract} type, i.e., hiding its data \ensuremath{\Varid{constructor}}.\notefoot{%
For the \ensuremath{\Conid{Category}\;\Conid{D}} instance given above, the painstaking proofs appear to succeed even without this condition.
Am I missing something?}
The slightly more specialized requirement of our first identity property is then \ensuremath{\Varid{id}\hsdot{\circ }{.}\hat{\mathcal{D}}\;\Varid{f}\mathrel{=}\hat{\mathcal{D}}\;\Varid{f}} for any \ensuremath{\Varid{f}\mathbin{::}\Varid{a}\to \Varid{b}}, which follows easily:
\begin{hscode}\SaveRestoreHook
\column{B}{@{}>{\hspre}c<{\hspost}@{}}%
\column{BE}{@{}l@{}}%
\column{5}{@{}>{\hspre}l<{\hspost}@{}}%
\column{21}{@{}>{\hspre}l<{\hspost}@{}}%
\column{E}{@{}>{\hspre}l<{\hspost}@{}}%
\>[5]{}\Varid{id}\hsdot{\circ }{.}\hat{\mathcal{D}}\;\Varid{f}{}\<[E]%
\\
\>[B]{}\mathrel{=}{}\<[BE]%
\>[5]{}\hat{\mathcal{D}}\;\Varid{id}\hsdot{\circ }{.}\hat{\mathcal{D}}\;\Varid{f}{}\<[21]%
\>[21]{}\mbox{\onelinecomment  functor law for \ensuremath{\Varid{id}} (specification of \ensuremath{\hat{\mathcal{D}}})}{}\<[E]%
\\
\>[B]{}\mathrel{=}{}\<[BE]%
\>[5]{}\hat{\mathcal{D}}\;(\Varid{id}\hsdot{\circ }{.}\Varid{f}){}\<[21]%
\>[21]{}\mbox{\onelinecomment  functor law for \ensuremath{(\hsdot{\circ }{.})}}{}\<[E]%
\\
\>[B]{}\mathrel{=}{}\<[BE]%
\>[5]{}\hat{\mathcal{D}}\;\Varid{f}{}\<[21]%
\>[21]{}\mbox{\onelinecomment  category law}{}\<[E]%
\ColumnHook
\end{hscode}\resethooks
The other identity law is proved similarly.
Associativity has a similar flavor as well:
\begin{hscode}\SaveRestoreHook
\column{B}{@{}>{\hspre}c<{\hspost}@{}}%
\column{BE}{@{}l@{}}%
\column{5}{@{}>{\hspre}l<{\hspost}@{}}%
\column{30}{@{}>{\hspre}l<{\hspost}@{}}%
\column{E}{@{}>{\hspre}l<{\hspost}@{}}%
\>[5]{}\hat{\mathcal{D}}\;\Varid{h}\hsdot{\circ }{.}(\hat{\mathcal{D}}\;\Varid{g}\hsdot{\circ }{.}\hat{\mathcal{D}}\;\Varid{f}){}\<[E]%
\\
\>[B]{}\mathrel{=}{}\<[BE]%
\>[5]{}\hat{\mathcal{D}}\;\Varid{h}\hsdot{\circ }{.}\hat{\mathcal{D}}\;(\Varid{g}\hsdot{\circ }{.}\Varid{f}){}\<[30]%
\>[30]{}\mbox{\onelinecomment  functor law for \ensuremath{(\hsdot{\circ }{.})}}{}\<[E]%
\\
\>[B]{}\mathrel{=}{}\<[BE]%
\>[5]{}\hat{\mathcal{D}}\;(\Varid{h}\hsdot{\circ }{.}(\Varid{g}\hsdot{\circ }{.}\Varid{f})){}\<[30]%
\>[30]{}\mbox{\onelinecomment  functor law for \ensuremath{(\hsdot{\circ }{.})}}{}\<[E]%
\\
\>[B]{}\mathrel{=}{}\<[BE]%
\>[5]{}\hat{\mathcal{D}}\;((\Varid{h}\hsdot{\circ }{.}\Varid{g})\hsdot{\circ }{.}\Varid{f}){}\<[30]%
\>[30]{}\mbox{\onelinecomment  category law}{}\<[E]%
\\
\>[B]{}\mathrel{=}{}\<[BE]%
\>[5]{}\hat{\mathcal{D}}\;(\Varid{h}\hsdot{\circ }{.}\Varid{g})\hsdot{\circ }{.}\hat{\mathcal{D}}\;\Varid{f}{}\<[30]%
\>[30]{}\mbox{\onelinecomment  functor law for \ensuremath{(\hsdot{\circ }{.})}}{}\<[E]%
\\
\>[B]{}\mathrel{=}{}\<[BE]%
\>[5]{}(\hat{\mathcal{D}}\;\Varid{h}\hsdot{\circ }{.}\hat{\mathcal{D}}\;\Varid{g})\hsdot{\circ }{.}\hat{\mathcal{D}}\;\Varid{f}{}\<[30]%
\>[30]{}\mbox{\onelinecomment  functor law for \ensuremath{(\hsdot{\circ }{.})}}{}\<[E]%
\ColumnHook
\end{hscode}\resethooks

Note how mechanical these proofs are.
Each uses only the functor laws plus the particular category law on functions that corresponds to the one being proved for \ensuremath{\Conid{D}}.
The proofs rely on nothing about the nature of \ensuremath{\Conid{D}} or \ensuremath{\hat{\mathcal{D}}} beyond the functor laws.
The importance of this observation is that we \emph{never} need to perform these proofs when we specify category instances via a functor.

\subsectionl{Monoidal Categories}

% \nc\scrk[1]{_{\hspace{#1}\scaleto{(\leadsto)\!}{4pt}}}
\nc\scrk[1]{}

\secref{Parallel Composition} introduced parallel composition.
This operation generalizes to play an important role in category theory as part of the notion of a \emph{monoidal category}:
\\
\begin{minipage}[b]{0.59\textwidth}
\begin{hscode}\SaveRestoreHook
\column{B}{@{}>{\hspre}l<{\hspost}@{}}%
\column{3}{@{}>{\hspre}l<{\hspost}@{}}%
\column{E}{@{}>{\hspre}l<{\hspost}@{}}%
\>[B]{}\mathbf{class}\;\Conid{Category}\;\Varid{k}\Rightarrow \Conid{Monoidal}\;\Varid{k}\;\mathbf{where}{}\<[E]%
\\
\>[B]{}\hsindent{3}{}\<[3]%
\>[3]{}(\times)\mathbin{::}(\Varid{a}\mathbin{`\Varid{k}`}\Varid{c})\to (\Varid{b}\mathbin{`\Varid{k}`}\Varid{d})\to ((\Varid{a}\times\scrk{-0.4ex}\Varid{b})\mathbin{`\Varid{k}`}(\Varid{c}\times\scrk{-0.4ex}\Varid{d})){}\<[E]%
\ColumnHook
\end{hscode}\resethooks
\end{minipage}
\begin{minipage}[b]{0ex}{\rule[1ex]{0.5pt}{0.3in}}\end{minipage}
\begin{minipage}[b]{0.35\textwidth} % \mathindent1em
\begin{hscode}\SaveRestoreHook
\column{B}{@{}>{\hspre}l<{\hspost}@{}}%
\column{3}{@{}>{\hspre}l<{\hspost}@{}}%
\column{E}{@{}>{\hspre}l<{\hspost}@{}}%
\>[B]{}\mathbf{instance}\;\Conid{Monoidal}\;(\to )\;\mathbf{where}{}\<[E]%
\\
\>[B]{}\hsindent{3}{}\<[3]%
\>[3]{}\Varid{f}\times\Varid{g}\mathrel{=}\lambda (\Varid{a},\Varid{b})\to (\Varid{f}\;\Varid{a},\Varid{g}\;\Varid{b}){}\<[E]%
\ColumnHook
\end{hscode}\resethooks
\end{minipage}
\\
More generally, a category \ensuremath{\Varid{k}} can be monoidal over constructions other than products, but cartesian products (ordered pairs) suffice for this paper.

Two monoidal categories can be related by a \emph{monoidal functor}, which is a functor that also preserves the monoidal structure.
That is, a monoidal functor \ensuremath{\Conid{F}} from monoidal category \ensuremath{\mathcal{U}} to monoidal category \ensuremath{\mathcal{V}}, besides mapping objects and morphisms in \ensuremath{\mathcal{U}} to counterparts in \ensuremath{\mathcal{V}} while preserving the category structure (\ensuremath{\Varid{id}} and \ensuremath{(\hsdot{\circ }{.})}), \emph{also} preserves the monoidal structure:
\begin{hscode}\SaveRestoreHook
\column{B}{@{}>{\hspre}l<{\hspost}@{}}%
\column{E}{@{}>{\hspre}l<{\hspost}@{}}%
\>[B]{}\Conid{F}\;(\Varid{f}\times\Varid{g})\mathrel{=}\Conid{F}\;\Varid{f}\times\Conid{F}\;\Varid{g}{}\<[E]%
\ColumnHook
\end{hscode}\resethooks
Just as \corRefTwo{compose}{linear} were key to deriving a correct-by-construction \ensuremath{\Conid{Category}} instance from the specification that \ensuremath{\hat{\mathcal{D}}} is a functor, \corRef{cross} leads to a correct \ensuremath{\Conid{Monoidal}} instance from the specification that \ensuremath{\hat{\mathcal{D}}} is a monoidal functor, as we'll now see.

Let \ensuremath{\Conid{F}} be \ensuremath{\hat{\mathcal{D}}} in the reversed form of the monoidal functor equation above, and expand \ensuremath{\hat{\mathcal{D}}} to its definition as \ensuremath{\Conid{D}\hsdot{\circ }{.}\mathcal{D}\!^+\!}:
\begin{hscode}\SaveRestoreHook
\column{B}{@{}>{\hspre}l<{\hspost}@{}}%
\column{E}{@{}>{\hspre}l<{\hspost}@{}}%
\>[B]{}\Conid{D}\;(\mathcal{D}\!^+\!\;\Varid{f})\times\Conid{D}\;(\mathcal{D}\!^+\!\;\Varid{g})\mathrel{=}\Conid{D}\;(\mathcal{D}\!^+\!\;(\Varid{f}\times\Varid{g})){}\<[E]%
\ColumnHook
\end{hscode}\resethooks
By \corRef{cross},
\begin{hscode}\SaveRestoreHook
\column{B}{@{}>{\hspre}l<{\hspost}@{}}%
\column{E}{@{}>{\hspre}l<{\hspost}@{}}%
\>[B]{}\mathcal{D}\!^+\!\;(\Varid{f}\times\Varid{g})\mathrel{=}\lambda (\Varid{a},\Varid{b})\to \mathbf{let}\;\{\mskip1.5mu (\Varid{c},\Varid{f'})\mathrel{=}\mathcal{D}\!^+\!\;\Varid{f}\;\Varid{a};(\Varid{d},\Varid{g'})\mathrel{=}\mathcal{D}\!^+\!\;\Varid{g}\;\Varid{b}\mskip1.5mu\}\;\mathbf{in}\;((\Varid{c},\Varid{d}),\Varid{f'}\times\Varid{g'}){}\<[E]%
\ColumnHook
\end{hscode}\resethooks
Now substitute the left-hand side of this equation into the right-hand side of the of the monoidal functor property for \ensuremath{\hat{\mathcal{D}}}, and \emph{strengthen} the condition by generalizing from \ensuremath{\mathcal{D}\!^+\!\;\Varid{f}} and \ensuremath{\mathcal{D}\!^+\!\;\Varid{g}}:
\begin{hscode}\SaveRestoreHook
\column{B}{@{}>{\hspre}l<{\hspost}@{}}%
\column{E}{@{}>{\hspre}l<{\hspost}@{}}%
\>[B]{}\Conid{D}\;\Varid{f}\times\Conid{D}\;\Varid{g}\mathrel{=}\Conid{D}\;(\lambda (\Varid{a},\Varid{b})\to \mathbf{let}\;\{\mskip1.5mu (\Varid{c},\Varid{f'})\mathrel{=}\Varid{f}\;\Varid{a};(\Varid{d},\Varid{g'})\mathrel{=}\Varid{g}\;\Varid{b}\mskip1.5mu\}\;\mathbf{in}\;((\Varid{c},\Varid{d}),\Varid{f'}\times\Varid{g'})){}\<[E]%
\ColumnHook
\end{hscode}\resethooks
\begin{samepage}
This strengthened form of the specification can be converted directly to a sufficient definition:
\begin{hscode}\SaveRestoreHook
\column{B}{@{}>{\hspre}l<{\hspost}@{}}%
\column{3}{@{}>{\hspre}l<{\hspost}@{}}%
\column{E}{@{}>{\hspre}l<{\hspost}@{}}%
\>[B]{}\mathbf{instance}\;\Conid{Monoidal}\;\Conid{D}\;\mathbf{where}{}\<[E]%
\\
\>[B]{}\hsindent{3}{}\<[3]%
\>[3]{}\Conid{D}\;\Varid{f}\times\Conid{D}\;\Varid{g}\mathrel{=}\Conid{D}\;(\lambda (\Varid{a},\Varid{b})\to \mathbf{let}\;\{\mskip1.5mu (\Varid{c},\Varid{f'})\mathrel{=}\Varid{f}\;\Varid{a};(\Varid{d},\Varid{g'})\mathrel{=}\Varid{g}\;\Varid{b}\mskip1.5mu\}\;\mathbf{in}\;((\Varid{c},\Varid{d}),\Varid{f'}\times\Varid{g'})){}\<[E]%
\ColumnHook
\end{hscode}\resethooks
\end{samepage}

\subsectionl{Cartesian Categories}

%% %format (Exp (k) a b) = a "\Rightarrow\scrk{-0.2ex}" b

The \ensuremath{\Conid{Monoidal}} abstraction provides a way to combine two functions but not separate them.
It also gives no way to duplicate or discard information.
These additional abilities require another algebraic abstraction, namely that of \emph{cartesian category}, adding operations for projection and duplication:
\\
\begin{minipage}[b]{0.5\textwidth}
\begin{hscode}\SaveRestoreHook
\column{B}{@{}>{\hspre}l<{\hspost}@{}}%
\column{3}{@{}>{\hspre}l<{\hspost}@{}}%
\column{8}{@{}>{\hspre}l<{\hspost}@{}}%
\column{E}{@{}>{\hspre}l<{\hspost}@{}}%
\>[B]{}\mathbf{class}\;\Conid{Monoidal}\;\Varid{k}\Rightarrow \Conid{Cartesian}\;\Varid{k}\;\mathbf{where}{}\<[E]%
\\
\>[B]{}\hsindent{3}{}\<[3]%
\>[3]{}\Varid{exl}{}\<[8]%
\>[8]{}\mathbin{::}(\Varid{a}\times\scrk{-0.4ex}\Varid{b})\mathbin{`\Varid{k}`}\Varid{a}{}\<[E]%
\\
\>[B]{}\hsindent{3}{}\<[3]%
\>[3]{}\Varid{exr}{}\<[8]%
\>[8]{}\mathbin{::}(\Varid{a}\times\scrk{-0.4ex}\Varid{b})\mathbin{`\Varid{k}`}\Varid{b}{}\<[E]%
\\
\>[B]{}\hsindent{3}{}\<[3]%
\>[3]{}\Varid{dup}{}\<[8]%
\>[8]{}\mathbin{::}\Varid{a}\mathbin{`\Varid{k}`}(\Varid{a}\times\scrk{-0.4ex}\Varid{a}){}\<[E]%
\ColumnHook
\end{hscode}\resethooks
\end{minipage}
\begin{minipage}[b]{0ex}{\rule[1ex]{0.5pt}{0.67in}}\end{minipage}
\begin{minipage}[b]{0.48\textwidth} \mathindent2em
\begin{hscode}\SaveRestoreHook
\column{B}{@{}>{\hspre}l<{\hspost}@{}}%
\column{3}{@{}>{\hspre}l<{\hspost}@{}}%
\column{8}{@{}>{\hspre}l<{\hspost}@{}}%
\column{E}{@{}>{\hspre}l<{\hspost}@{}}%
\>[B]{}\mathbf{instance}\;\Conid{Cartesian}\;(\to )\;\mathbf{where}{}\<[E]%
\\
\>[B]{}\hsindent{3}{}\<[3]%
\>[3]{}\Varid{exl}{}\<[8]%
\>[8]{}\mathrel{=}\lambda (\Varid{a},\Varid{b})\to \Varid{a}{}\<[E]%
\\
\>[B]{}\hsindent{3}{}\<[3]%
\>[3]{}\Varid{exr}{}\<[8]%
\>[8]{}\mathrel{=}\lambda (\Varid{a},\Varid{b})\to \Varid{b}{}\<[E]%
\\
\>[B]{}\hsindent{3}{}\<[3]%
\>[3]{}\Varid{dup}{}\<[8]%
\>[8]{}\mathrel{=}\lambda \Varid{a}\to (\Varid{a},\Varid{a}){}\<[E]%
\ColumnHook
\end{hscode}\resethooks
\mathindent1em
\end{minipage}

\begin{closerCodePars}
Two cartesian categories can be related by a \emph{cartesian functor}, which additionally preserves the cartesian structure.
That is, a cartesian functor \ensuremath{\Conid{F}} from cartesian category \ensuremath{\mathcal{U}} to cartesian category \ensuremath{\mathcal{V}}, besides mapping objects and morphisms in \ensuremath{\mathcal{U}} to counterparts in \ensuremath{\mathcal{V}} while preserving the category and monoidal structure (\ensuremath{\Varid{id}}, \ensuremath{(\hsdot{\circ }{.})}, and \ensuremath{(\times)}), \emph{also} preserves the cartesian structure:
\begin{hscode}\SaveRestoreHook
\column{B}{@{}>{\hspre}l<{\hspost}@{}}%
\column{8}{@{}>{\hspre}l<{\hspost}@{}}%
\column{E}{@{}>{\hspre}l<{\hspost}@{}}%
\>[B]{}\Conid{F}\;\Varid{exl}{}\<[8]%
\>[8]{}\mathrel{=}\Varid{exl}{}\<[E]%
\\
\>[B]{}\Conid{F}\;\Varid{exr}{}\<[8]%
\>[8]{}\mathrel{=}\Varid{exr}{}\<[E]%
\\
\>[B]{}\Conid{F}\;\Varid{dup}{}\<[8]%
\>[8]{}\mathrel{=}\Varid{dup}{}\<[E]%
\ColumnHook
\end{hscode}\resethooks
Just as \corRefs{compose}{linear} were key to deriving a correct-by-construction \ensuremath{\Conid{Category}} and \ensuremath{\Conid{Monoidal}} instances from the specification that \ensuremath{\hat{\mathcal{D}}} is a functor and a monoidal functor respectively, \corRef{linear} enables a correct-by-construction \ensuremath{\Conid{Cartesian}} instance from the specification that \ensuremath{\hat{\mathcal{D}}} is a cartesian functor.
Let \ensuremath{\Conid{F}} be \ensuremath{\hat{\mathcal{D}}} in the reversed forms of cartesian functor equations above, and expand \ensuremath{\hat{\mathcal{D}}} to its definition as \ensuremath{\Conid{D}\hsdot{\circ }{.}\mathcal{D}\!^+\!}:
\begin{hscode}\SaveRestoreHook
\column{B}{@{}>{\hspre}l<{\hspost}@{}}%
\column{6}{@{}>{\hspre}l<{\hspost}@{}}%
\column{E}{@{}>{\hspre}l<{\hspost}@{}}%
\>[B]{}\Varid{exl}{}\<[6]%
\>[6]{}\mathrel{=}\Conid{D}\;(\mathcal{D}\!^+\!\;\Varid{exl}){}\<[E]%
\\
\>[B]{}\Varid{exr}{}\<[6]%
\>[6]{}\mathrel{=}\Conid{D}\;(\mathcal{D}\!^+\!\;\Varid{exr}){}\<[E]%
\\
\>[B]{}\Varid{dup}{}\<[6]%
\>[6]{}\mathrel{=}\Conid{D}\;(\mathcal{D}\!^+\!\;\Varid{dup}){}\<[E]%
\ColumnHook
\end{hscode}\resethooks
Next, by \corRef{linear}, together with the linearity of \ensuremath{\Varid{exl}}, \ensuremath{\Varid{exr}}, and \ensuremath{\Varid{dup}},
\begin{hscode}\SaveRestoreHook
\column{B}{@{}>{\hspre}l<{\hspost}@{}}%
\column{9}{@{}>{\hspre}l<{\hspost}@{}}%
\column{17}{@{}>{\hspre}l<{\hspost}@{}}%
\column{26}{@{}>{\hspre}l<{\hspost}@{}}%
\column{29}{@{}>{\hspre}l<{\hspost}@{}}%
\column{36}{@{}>{\hspre}c<{\hspost}@{}}%
\column{36E}{@{}l@{}}%
\column{E}{@{}>{\hspre}l<{\hspost}@{}}%
\>[B]{}\mathcal{D}\!^+\!\;\Varid{exl}{}\<[9]%
\>[9]{}\mathrel{=}\lambda \Varid{p}{}\<[17]%
\>[17]{}\to (\Varid{exl}\;{}\<[26]%
\>[26]{}\Varid{p}{}\<[29]%
\>[29]{},\Varid{exl}{}\<[36]%
\>[36]{}){}\<[36E]%
\\
\>[B]{}\mathcal{D}\!^+\!\;\Varid{exr}{}\<[9]%
\>[9]{}\mathrel{=}\lambda \Varid{p}{}\<[17]%
\>[17]{}\to (\Varid{exr}\;{}\<[26]%
\>[26]{}\Varid{p}{}\<[29]%
\>[29]{},\Varid{exr}{}\<[36]%
\>[36]{}){}\<[36E]%
\\
\>[B]{}\mathcal{D}\!^+\!\;\Varid{dup}{}\<[9]%
\>[9]{}\mathrel{=}\lambda \Varid{a}{}\<[17]%
\>[17]{}\to (\Varid{dup}\;{}\<[26]%
\>[26]{}\Varid{a}{}\<[29]%
\>[29]{},\Varid{dup}{}\<[36]%
\>[36]{}){}\<[36E]%
\ColumnHook
\end{hscode}\resethooks
Now substitute the left-hand sides of these three properties into the right-hand sides of the of the cartesian functor properties for \ensuremath{\hat{\mathcal{D}}}, and recall the definition of \ensuremath{\Varid{linearD}}:
\begin{hscode}\SaveRestoreHook
\column{B}{@{}>{\hspre}l<{\hspost}@{}}%
\column{6}{@{}>{\hspre}l<{\hspost}@{}}%
\column{E}{@{}>{\hspre}l<{\hspost}@{}}%
\>[B]{}\Varid{exl}{}\<[6]%
\>[6]{}\mathrel{=}\Varid{linearD}\;\Varid{exl}{}\<[E]%
\\
\>[B]{}\Varid{exr}{}\<[6]%
\>[6]{}\mathrel{=}\Varid{linearD}\;\Varid{exr}{}\<[E]%
\\
\>[B]{}\Varid{dup}{}\<[6]%
\>[6]{}\mathrel{=}\Varid{linearD}\;\Varid{dup}{}\<[E]%
\ColumnHook
\end{hscode}\resethooks
\end{closerCodePars}%
This form of the specification can be turned directly into a sufficient definition:
\begin{hscode}\SaveRestoreHook
\column{B}{@{}>{\hspre}l<{\hspost}@{}}%
\column{3}{@{}>{\hspre}l<{\hspost}@{}}%
\column{8}{@{}>{\hspre}l<{\hspost}@{}}%
\column{E}{@{}>{\hspre}l<{\hspost}@{}}%
\>[B]{}\mathbf{instance}\;\Conid{Cartesian}\;\Conid{D}\;\mathbf{where}{}\<[E]%
\\
\>[B]{}\hsindent{3}{}\<[3]%
\>[3]{}\Varid{exl}{}\<[8]%
\>[8]{}\mathrel{=}\Varid{linearD}\;\Varid{exl}{}\<[E]%
\\
\>[B]{}\hsindent{3}{}\<[3]%
\>[3]{}\Varid{exr}{}\<[8]%
\>[8]{}\mathrel{=}\Varid{linearD}\;\Varid{exr}{}\<[E]%
\\
\>[B]{}\hsindent{3}{}\<[3]%
\>[3]{}\Varid{dup}{}\<[8]%
\>[8]{}\mathrel{=}\Varid{linearD}\;\Varid{dup}{}\<[E]%
\ColumnHook
\end{hscode}\resethooks

\subsectionl{Cocartesian Categories}

%% %format -+> = ->

%% Define and use a type of additive functions

Cartesian categories have a dual, known as \emph{cocartesian categories}, in which each cartesian operation has a mirror image with morphisms reversed (swapping domain and codomain) and coproducts replacing products.
In general, each category can have its own notion of coproduct, e.g., sum (disjoint union) types for the \ensuremath{(\to )} category.
In this paper, however, coproducts will coincide with categorical products\out{ (both being ordered pairs)}, i.e., we'll be using biproduct categories \citep{MacedoOliveira2013Typing}:
\begin{hscode}\SaveRestoreHook
\column{B}{@{}>{\hspre}l<{\hspost}@{}}%
\column{3}{@{}>{\hspre}l<{\hspost}@{}}%
\column{8}{@{}>{\hspre}c<{\hspost}@{}}%
\column{8E}{@{}l@{}}%
\column{12}{@{}>{\hspre}l<{\hspost}@{}}%
\column{E}{@{}>{\hspre}l<{\hspost}@{}}%
\>[B]{}\mathbf{class}\;\Conid{Category}\;\Varid{k}\Rightarrow \Conid{Cocartesian}\;\Varid{k}\;\mathbf{where}{}\<[E]%
\\
\>[B]{}\hsindent{3}{}\<[3]%
\>[3]{}\Varid{inl}{}\<[8]%
\>[8]{}\mathbin{::}{}\<[8E]%
\>[12]{}\Varid{a}\mathbin{`\Varid{k}`}(\Varid{a}\times\scrk{-0.4ex}\Varid{b}){}\<[E]%
\\
\>[B]{}\hsindent{3}{}\<[3]%
\>[3]{}\Varid{inr}{}\<[8]%
\>[8]{}\mathbin{::}{}\<[8E]%
\>[12]{}\Varid{b}\mathbin{`\Varid{k}`}(\Varid{a}\times\scrk{-0.4ex}\Varid{b}){}\<[E]%
\\
\>[B]{}\hsindent{3}{}\<[3]%
\>[3]{}\Varid{jam}{}\<[8]%
\>[8]{}\mathbin{::}{}\<[8E]%
\>[12]{}(\Varid{a}\times\scrk{-0.4ex}\Varid{a})\mathbin{`\Varid{k}`}\Varid{a}{}\<[E]%
\ColumnHook
\end{hscode}\resethooks
Unlike the other classes, there is no \ensuremath{\Conid{Cocartesian}\;(\to )} instance, and fortunately we will not need such an instance below.
(There is an instance when using sums instead of cartesian products for coproducts.)
Instead, we can define a category \ensuremath{(\mathbin{\rightarrow^{\!\!+}\!})} of \emph{additive functions} that will have a \ensuremath{\Conid{Cocartesian}} instance and that we can use to represent derivatives, as shown in \figref{AddFun}.
These instances rely on one more feature of the \ensuremath{\Conid{Category}} class not yet mentioned, namely an associated constraint \citep{Bolingbroke2011CK} \ensuremath{\Conid{Obj}_{\Varid{k}}}.
In the actual class definitions, \ensuremath{\Conid{Obj}_{\Varid{k}}} constrains the types involved in all categorical operations.
\begin{figure}
\begin{center}
\begin{minipage}[b]{0.50\textwidth}
\begin{hscode}\SaveRestoreHook
\column{B}{@{}>{\hspre}l<{\hspost}@{}}%
\column{3}{@{}>{\hspre}l<{\hspost}@{}}%
\column{8}{@{}>{\hspre}l<{\hspost}@{}}%
\column{E}{@{}>{\hspre}l<{\hspost}@{}}%
\>[B]{}\mathbf{newtype}\;\Varid{a}\mathbin{\rightarrow^{\!\!+}\!}\Varid{b}\mathrel{=}\Conid{AddFun}\;(\Varid{a}\to \Varid{b}){}\<[E]%
\\[\blanklineskip]%
\>[B]{}\mathbf{instance}\;\Conid{Category}\;(\mathbin{\rightarrow^{\!\!+}\!})\;\mathbf{where}{}\<[E]%
\\
\>[B]{}\hsindent{3}{}\<[3]%
\>[3]{}\mathbf{type}\;\Conid{Obj}_{(\mathbin{\rightarrow^{\!\!+}\!})}\mathrel{=}\Conid{Additive}{}\<[E]%
\\
\>[B]{}\hsindent{3}{}\<[3]%
\>[3]{}\Varid{id}\mathrel{=}\Conid{AddFun}\;\Varid{id}{}\<[E]%
\\
\>[B]{}\hsindent{3}{}\<[3]%
\>[3]{}\Conid{AddFun}\;\Varid{g}\hsdot{\circ }{.}\Conid{AddFun}\;\Varid{f}\mathrel{=}\Conid{AddFun}\;(\Varid{g}\hsdot{\circ }{.}\Varid{f}){}\<[E]%
\\[\blanklineskip]%
\>[B]{}\mathbf{instance}\;\Conid{Monoidal}\;(\mathbin{\rightarrow^{\!\!+}\!})\;\mathbf{where}{}\<[E]%
\\
\>[B]{}\hsindent{3}{}\<[3]%
\>[3]{}\Conid{AddFun}\;\Varid{f}\times\Conid{AddFun}\;\Varid{g}\mathrel{=}\Conid{AddFun}\;(\Varid{f}\times\Varid{g}){}\<[E]%
\\[\blanklineskip]%
\>[B]{}\mathbf{instance}\;\Conid{Cartesian}\;(\mathbin{\rightarrow^{\!\!+}\!})\;\mathbf{where}{}\<[E]%
\\
\>[B]{}\hsindent{3}{}\<[3]%
\>[3]{}\Varid{exl}{}\<[8]%
\>[8]{}\mathrel{=}\Conid{AddFun}\;\Varid{exl}{}\<[E]%
\\
\>[B]{}\hsindent{3}{}\<[3]%
\>[3]{}\Varid{exr}{}\<[8]%
\>[8]{}\mathrel{=}\Conid{AddFun}\;\Varid{exr}{}\<[E]%
\\
\>[B]{}\hsindent{3}{}\<[3]%
\>[3]{}\Varid{dup}{}\<[8]%
\>[8]{}\mathrel{=}\Conid{AddFun}\;\Varid{dup}{}\<[E]%
\ColumnHook
\end{hscode}\resethooks
\end{minipage}
\begin{minipage}[b]{0ex}{\rule[1ex]{0.5pt}{2.3in}}\end{minipage}
\begin{minipage}[b]{0.48\textwidth} \mathindent2em
\begin{hscode}\SaveRestoreHook
\column{B}{@{}>{\hspre}l<{\hspost}@{}}%
\column{3}{@{}>{\hspre}l<{\hspost}@{}}%
\column{7}{@{}>{\hspre}l<{\hspost}@{}}%
\column{9}{@{}>{\hspre}l<{\hspost}@{}}%
\column{E}{@{}>{\hspre}l<{\hspost}@{}}%
\>[B]{}\mathbf{instance}\;\Conid{Cocartesian}\;(\mathbin{\rightarrow^{\!\!+}\!})\;\mathbf{where}{}\<[E]%
\\
\>[B]{}\hsindent{3}{}\<[3]%
\>[3]{}\Varid{inl}{}\<[9]%
\>[9]{}\mathrel{=}\Conid{AddFun}\;\Varid{inlF}{}\<[E]%
\\
\>[B]{}\hsindent{3}{}\<[3]%
\>[3]{}\Varid{inr}{}\<[9]%
\>[9]{}\mathrel{=}\Conid{AddFun}\;\Varid{inrF}{}\<[E]%
\\
\>[B]{}\hsindent{3}{}\<[3]%
\>[3]{}\Varid{jam}{}\<[9]%
\>[9]{}\mathrel{=}\Conid{AddFun}\;\Varid{jamF}{}\<[E]%
\\[\blanklineskip]%
\>[B]{}\Varid{inlF}{}\<[7]%
\>[7]{}\mathbin{::}\Conid{Additive}\;\Varid{b}\Rightarrow \Varid{a}\to \Varid{a} \times \Varid{b}{}\<[E]%
\\
\>[B]{}\Varid{inrF}{}\<[7]%
\>[7]{}\mathbin{::}\Conid{Additive}\;\Varid{a}\Rightarrow \Varid{b}\to \Varid{a} \times \Varid{b}{}\<[E]%
\\
\>[B]{}\Varid{jamF}{}\<[7]%
\>[7]{}\mathbin{::}\Conid{Additive}\;\Varid{a}\Rightarrow \Varid{a} \times \Varid{a}\to \Varid{a}{}\<[E]%
\\[\blanklineskip]%
\>[B]{}\Varid{inlF}{}\<[7]%
\>[7]{}\mathrel{=}\lambda \Varid{a}\to (\Varid{a},\mathrm{0}){}\<[E]%
\\
\>[B]{}\Varid{inrF}{}\<[7]%
\>[7]{}\mathrel{=}\lambda \Varid{b}\to (\mathrm{0},\Varid{b}){}\<[E]%
\\
\>[B]{}\Varid{jamF}{}\<[7]%
\>[7]{}\mathrel{=}\lambda (\Varid{a},\Varid{b})\to \Varid{a}\mathbin{+}\Varid{b}{}\<[E]%
\ColumnHook
\end{hscode}\resethooks
\end{minipage}
\caption{Additive functions}
\figlabel{AddFun}
\end{center}
\end{figure}

Unsurprisingly, there is a notion of \emph{cocartesian functor}, saying that the cocartesian structure is preserved, i.e.,
\begin{closerCodePars}
\begin{hscode}\SaveRestoreHook
\column{B}{@{}>{\hspre}l<{\hspost}@{}}%
\column{8}{@{}>{\hspre}l<{\hspost}@{}}%
\column{E}{@{}>{\hspre}l<{\hspost}@{}}%
\>[B]{}\Conid{F}\;\Varid{inl}{}\<[8]%
\>[8]{}\mathrel{=}\Varid{inl}{}\<[E]%
\\
\>[B]{}\Conid{F}\;\Varid{inr}{}\<[8]%
\>[8]{}\mathrel{=}\Varid{inr}{}\<[E]%
\\
\>[B]{}\Conid{F}\;\Varid{jam}{}\<[8]%
\>[8]{}\mathrel{=}\Varid{jam}{}\<[E]%
\ColumnHook
\end{hscode}\resethooks
\end{closerCodePars}%

\subsectionl{Derived Operations}

With \ensuremath{\Varid{dup}}, we can define an alternative to \ensuremath{(\times)} that takes two morphisms sharing a domain:
\begin{hscode}\SaveRestoreHook
\column{B}{@{}>{\hspre}l<{\hspost}@{}}%
\column{E}{@{}>{\hspre}l<{\hspost}@{}}%
\>[B]{}(\mathbin{\vartriangle})\mathbin{::}\Conid{Cartesian}\;\Varid{k}\Rightarrow (\Varid{a}\mathbin{`\Varid{k}`}\Varid{c})\to (\Varid{a}\mathbin{`\Varid{k}`}\Varid{d})\to (\Varid{a}\mathbin{`\Varid{k}`}(\Varid{c}\times\scrk{-0.4ex}\Varid{d})){}\<[E]%
\\
\>[B]{}\Varid{f}\mathbin{\vartriangle}\Varid{g}\mathrel{=}(\Varid{f}\times\Varid{g})\hsdot{\circ }{.}\Varid{dup}{}\<[E]%
\ColumnHook
\end{hscode}\resethooks
The \ensuremath{(\mathbin{\vartriangle})} operation is particularly useful for translating the $\lambda$-calculus to categorical form \citep[Section 3]{Elliott-2017-compiling-to-categories}.

Dually, \ensuremath{\Varid{jam}} lets us define a second alternative to \ensuremath{(\times)} for two morphisms sharing a \emph{codomain}:
\begin{hscode}\SaveRestoreHook
\column{B}{@{}>{\hspre}l<{\hspost}@{}}%
\column{E}{@{}>{\hspre}l<{\hspost}@{}}%
\>[B]{}(\mathbin{\triangledown})\mathbin{::}\Conid{Cocartesian}\;\Varid{k}\Rightarrow (\Varid{c}\mathbin{`\Varid{k}`}\Varid{a})\to (\Varid{d}\mathbin{`\Varid{k}`}\Varid{a})\to ((\Varid{c}\times\scrk{-0.4ex}\Varid{d})\mathbin{`\Varid{k}`}\Varid{a}){}\<[E]%
\\
\>[B]{}\Varid{f}\mathbin{\triangledown}\Varid{g}\mathrel{=}\Varid{jam}\hsdot{\circ }{.}(\Varid{f}\times\Varid{g}){}\<[E]%
\ColumnHook
\end{hscode}\resethooks
The \ensuremath{(\mathbin{\vartriangle})} and \ensuremath{(\mathbin{\triangledown})} operations\out{ (sometimes called ``fork'' and ``join'')} are invertible in uncurried form \citep{Gibbons2002Calculating}:
\begin{hscode}\SaveRestoreHook
\column{B}{@{}>{\hspre}l<{\hspost}@{}}%
\column{9}{@{}>{\hspre}l<{\hspost}@{}}%
\column{25}{@{}>{\hspre}l<{\hspost}@{}}%
\column{E}{@{}>{\hspre}l<{\hspost}@{}}%
\>[B]{}\Varid{fork}{}\<[9]%
\>[9]{}\mathbin{::}\Conid{Cartesian}\;{}\<[25]%
\>[25]{}\Varid{k}\Rightarrow (\Varid{a}\mathbin{`\Varid{k}`}\Varid{c}) \times (\Varid{a}\mathbin{`\Varid{k}`}\Varid{d})\to (\Varid{a}\mathbin{`\Varid{k}`}(\Varid{c}\times\scrk{-0.4ex}\Varid{d})){}\<[E]%
\\
\>[B]{}\Varid{unfork}{}\<[9]%
\>[9]{}\mathbin{::}\Conid{Cartesian}\;{}\<[25]%
\>[25]{}\Varid{k}\Rightarrow (\Varid{a}\mathbin{`\Varid{k}`}(\Varid{c}\times\scrk{-0.4ex}\Varid{d}))\to (\Varid{a}\mathbin{`\Varid{k}`}\Varid{c}) \times (\Varid{a}\mathbin{`\Varid{k}`}\Varid{d}){}\<[E]%
\\[\blanklineskip]%
\>[B]{}\Varid{join}{}\<[9]%
\>[9]{}\mathbin{::}\Conid{Cocartesian}\;{}\<[25]%
\>[25]{}\Varid{k}\Rightarrow (\Varid{c}\mathbin{`\Varid{k}`}\Varid{a}) \times (\Varid{d}\mathbin{`\Varid{k}`}\Varid{a})\to ((\Varid{c}\times\scrk{-0.4ex}\Varid{d})\mathbin{`\Varid{k}`}\Varid{a}){}\<[E]%
\\
\>[B]{}\Varid{unjoin}{}\<[9]%
\>[9]{}\mathbin{::}\Conid{Cocartesian}\;{}\<[25]%
\>[25]{}\Varid{k}\Rightarrow ((\Varid{c}\times\scrk{-0.4ex}\Varid{d})\mathbin{`\Varid{k}`}\Varid{a})\to (\Varid{c}\mathbin{`\Varid{k}`}\Varid{a}) \times (\Varid{d}\mathbin{`\Varid{k}`}\Varid{a}){}\<[E]%
\ColumnHook
\end{hscode}\resethooks
where
\\[1.5ex]
\begin{minipage}[b]{0.38\textwidth} % \mathindent1em
\begin{hscode}\SaveRestoreHook
\column{B}{@{}>{\hspre}l<{\hspost}@{}}%
\column{E}{@{}>{\hspre}l<{\hspost}@{}}%
\>[B]{}\Varid{fork}\;(\Varid{f},\Varid{g})\mathrel{=}\Varid{f}\mathbin{\vartriangle}\Varid{g}{}\<[E]%
\\
\>[B]{}\Varid{unfork}\;\Varid{h}\mathrel{=}(\Varid{exl}\hsdot{\circ }{.}\Varid{h},\Varid{exr}\hsdot{\circ }{.}\Varid{h}){}\<[E]%
\ColumnHook
\end{hscode}\resethooks
\end{minipage}
\begin{minipage}[b]{0ex}{\rule[1ex]{0.5pt}{0.32in}}\end{minipage}
\begin{minipage}[b]{0.48\textwidth} \mathindent2.5em
\begin{hscode}\SaveRestoreHook
\column{B}{@{}>{\hspre}l<{\hspost}@{}}%
\column{E}{@{}>{\hspre}l<{\hspost}@{}}%
\>[B]{}\Varid{join}\;(\Varid{f},\Varid{g})\mathrel{=}\Varid{f}\mathbin{\triangledown}\Varid{g}{}\<[E]%
\\
\>[B]{}\Varid{unjoin}\;\Varid{h}\mathrel{=}(\Varid{h}\hsdot{\circ }{.}\Varid{inl},\Varid{h}\hsdot{\circ }{.}\Varid{inr}){}\<[E]%
\ColumnHook
\end{hscode}\resethooks
\end{minipage}

\subsectionl{Numeric Operations}

So far, the vocabulary we've considered comprises linear functions and combining forms (\ensuremath{(\hsdot{\circ }{.})} and \ensuremath{(\times)}) that preserve linearity.
To make differentiation interesting, we'll need some non-linear primitives as well.
Let's now add these primitives, while continuing to derive correct implementations from simple, regular specifications in terms of homomorphisms (structure-preserving transformations).
We'll define a collection of interfaces for numeric operations, roughly imitating Haskell's numeric type class hierarchy.

Haskell provides the following basic class:
\begin{hscode}\SaveRestoreHook
\column{B}{@{}>{\hspre}l<{\hspost}@{}}%
\column{3}{@{}>{\hspre}l<{\hspost}@{}}%
\column{E}{@{}>{\hspre}l<{\hspost}@{}}%
\>[B]{}\mathbf{class}\;\Conid{Num}\;\Varid{a}\;\mathbf{where}{}\<[E]%
\\
\>[B]{}\hsindent{3}{}\<[3]%
\>[3]{}\Varid{negate}\mathbin{::}\Varid{a}\to \Varid{a}{}\<[E]%
\\
\>[B]{}\hsindent{3}{}\<[3]%
\>[3]{}(\mathbin{+}),(\mathbin{*})\mathbin{::}\Varid{a}\to \Varid{a}\to \Varid{a}{}\<[E]%
\\
\>[B]{}\hsindent{3}{}\<[3]%
\>[3]{}\mathbin{...}{}\<[E]%
\ColumnHook
\end{hscode}\resethooks
Although this class can accommodate many different types of ``numbers'', the class operations are all committed to being functions.
A more flexible alternative allows operations to be non-functions\out{ as well}:
\\
\begin{minipage}[b]{0.4\textwidth}
\begin{hscode}\SaveRestoreHook
\column{B}{@{}>{\hspre}l<{\hspost}@{}}%
\column{3}{@{}>{\hspre}l<{\hspost}@{}}%
\column{E}{@{}>{\hspre}l<{\hspost}@{}}%
\>[B]{}\mathbf{class}\;\Conid{NumCat}\;\Varid{k}\;\Varid{a}\;\mathbf{where}{}\<[E]%
\\
\>[B]{}\hsindent{3}{}\<[3]%
\>[3]{}\Varid{negateC}\mathbin{::}\Varid{a}\mathbin{`\Varid{k}`}\Varid{a}{}\<[E]%
\\
\>[B]{}\hsindent{3}{}\<[3]%
\>[3]{}\Varid{addC}\mathbin{::}(\Varid{a} \times \Varid{a})\mathbin{`\Varid{k}`}\Varid{a}{}\<[E]%
\\
\>[B]{}\hsindent{3}{}\<[3]%
\>[3]{}\Varid{mulC}\mathbin{::}(\Varid{a} \times \Varid{a})\mathbin{`\Varid{k}`}\Varid{a}{}\<[E]%
\\
\>[B]{}\hsindent{3}{}\<[3]%
\>[3]{}\mathbin{...}{}\<[E]%
\ColumnHook
\end{hscode}\resethooks
\end{minipage}
\begin{minipage}[b]{0ex}{\rule[1.4ex]{0.5pt}{0.83in}}\end{minipage}
\begin{minipage}[b]{0.48\textwidth} \mathindent2em
\begin{hscode}\SaveRestoreHook
\column{B}{@{}>{\hspre}l<{\hspost}@{}}%
\column{3}{@{}>{\hspre}l<{\hspost}@{}}%
\column{9}{@{}>{\hspre}l<{\hspost}@{}}%
\column{E}{@{}>{\hspre}l<{\hspost}@{}}%
\>[B]{}\mathbf{instance}\;\Conid{Num}\;\Varid{a}\Rightarrow \Conid{NumCat}\;(\to )\;\Varid{a}\;\mathbf{where}{}\<[E]%
\\
\>[B]{}\hsindent{3}{}\<[3]%
\>[3]{}\Varid{negateC}\mathrel{=}\Varid{negate}{}\<[E]%
\\
\>[B]{}\hsindent{3}{}\<[3]%
\>[3]{}\Varid{addC}{}\<[9]%
\>[9]{}\mathrel{=}\Varid{uncurry}\;(\mathbin{+}){}\<[E]%
\\
\>[B]{}\hsindent{3}{}\<[3]%
\>[3]{}\Varid{mulC}{}\<[9]%
\>[9]{}\mathrel{=}\Varid{uncurry}\;(\cdot){}\<[E]%
\\
\>[B]{}\hsindent{3}{}\<[3]%
\>[3]{}\mathbin{...}{}\<[E]%
\ColumnHook
\end{hscode}\resethooks
\end{minipage}
\\
Besides generalizing from \ensuremath{(\to )} to \ensuremath{\Varid{k}}, we've also uncurried the operations, so as to demand less of supporting categories \ensuremath{\Varid{k}}.
There are similar classes for other operations, such as division, powers and roots, and transcendental functions (\ensuremath{\Varid{sin}}, \ensuremath{\Varid{cos}}, \ensuremath{\Varid{exp}} etc).
Note that the \ensuremath{(\to )} instance uses the operations from the standard numeric classes (\ensuremath{\Conid{Num}} etc).

Differentiation rules for these operations are part of basic differential calculus:\footnote{The conventional differentiation rules shown here treat derivatives as numbers rather than linear maps.}
\begin{hscode}\SaveRestoreHook
\column{B}{@{}>{\hspre}l<{\hspost}@{}}%
\column{9}{@{}>{\hspre}c<{\hspost}@{}}%
\column{9E}{@{}l@{}}%
\column{12}{@{}>{\hspre}l<{\hspost}@{}}%
\column{E}{@{}>{\hspre}l<{\hspost}@{}}%
\>[B]{}\mathcal{D}\;(\Varid{negate}\;\Varid{u})\mathrel{=}\Varid{negate}\;(\mathcal{D}\;\Varid{u}){}\<[E]%
\\
\>[B]{}\mathcal{D}\;(\Varid{u}{}\<[9]%
\>[9]{}\mathbin{+}{}\<[9E]%
\>[12]{}\Varid{v})\mathrel{=}\mathcal{D}\;\Varid{u}\mathbin{+}\mathcal{D}\;\Varid{v}{}\<[E]%
\\
\>[B]{}\mathcal{D}\;(\Varid{u}\cdot\Varid{v})\mathrel{=}\Varid{u}\cdot\mathcal{D}\;\Varid{v}\mathbin{+}\Varid{v}\cdot\mathcal{D}\;\Varid{u}{}\<[E]%
\ColumnHook
\end{hscode}\resethooks
This conventional form is unnecessarily complex, as each of these rules implicitly involves not just a numeric operation, but also an application of the chain rule.
This form is also imprecise about the nature of \ensuremath{\Varid{u}} and \ensuremath{\Varid{v}}.
If they are functions, then one needs to explain arithmetic on functions; and if they are not functions, then differentiation of non-functions needs explanation.\out{\footnote{Arithmetic on functions is usually defined pointwise, e.g., $u + v = \ t -> u t + v t$.}}

A precise and simpler presentation is to remove the arguments and talk about differentiating the primitive operations in isolation.
We have the chain rule to account for context, so we do not need to involve it in every numeric operation.
Since negation and (uncurried) addition are linear, we already know how to differentiate them.
Multiplication is a little more involved \citep[Theorem 2-3 (2)]{Spivak65}:
\begin{hscode}\SaveRestoreHook
\column{B}{@{}>{\hspre}l<{\hspost}@{}}%
\column{E}{@{}>{\hspre}l<{\hspost}@{}}%
\>[B]{}\mathcal{D}\;\Varid{mulC}\;(\Varid{a},\Varid{b})\mathrel{=}\lambda (\Varid{da},\Varid{db})\to \Varid{da}\cdot\Varid{b}\mathbin{+}\Varid{a}\cdot\Varid{db}{}\<[E]%
\ColumnHook
\end{hscode}\resethooks
Note the linearity of the right-hand side, so that the derivative of \ensuremath{\Varid{mulC}} at \ensuremath{(\Varid{a},\Varid{b})} for real values has the expected type: \ensuremath{\mathbb{R} \times \mathbb{R}\multimap\mathbb{R}}.\footnote{The derivative of uncurried multiplication generalizes to an arbitrary \emph{bilinear} function \ensuremath{\Varid{f}\mathbin{::}\Varid{a} \times \Varid{b}\to \Varid{c}} \citep[Problem 2-12]{Spivak65}: \ensuremath{\mathcal{D}\;\Varid{f}\;(\Varid{a},\Varid{b})\mathrel{=}\lambda (\Varid{da},\Varid{db})\to \Varid{f}\;(\Varid{da},\Varid{b})\mathbin{+}\Varid{f}\;(\Varid{a},\Varid{db})}.}
To make the linearity more apparent, and to prepare for variations later in this paper, let's now rephrase \ensuremath{\mathcal{D}\;\Varid{mulC}} without using lambda directly.
Just as \ensuremath{\Conid{Category}}, \ensuremath{\Conid{Monoidal}}, \ensuremath{\Conid{Cartesian}}\out{, \ensuremath{\Conid{Cocartesian}}}, \ensuremath{\Conid{NumCat}}, etc generalize operations beyond functions, it will also be handy to generalize scalar multiplication as well:
\\
\begin{minipage}[b]{0.35\textwidth} % \mathindent1em
\begin{hscode}\SaveRestoreHook
\column{B}{@{}>{\hspre}l<{\hspost}@{}}%
\column{3}{@{}>{\hspre}l<{\hspost}@{}}%
\column{E}{@{}>{\hspre}l<{\hspost}@{}}%
\>[B]{}\mathbf{class}\;\Conid{Scalable}\;\Varid{k}\;\Varid{a}\;\mathbf{where}{}\<[E]%
\\
\>[B]{}\hsindent{3}{}\<[3]%
\>[3]{}\Varid{scale}\mathbin{::}\Varid{a}\to (\Varid{a}\mathbin{`\Varid{k}`}\Varid{a}){}\<[E]%
\ColumnHook
\end{hscode}\resethooks
\end{minipage}
\begin{minipage}[b]{0ex}{\rule[1ex]{0.5pt}{0.32in}}\end{minipage}
\begin{minipage}[b]{0.48\textwidth} \mathindent2em
\begin{hscode}\SaveRestoreHook
\column{B}{@{}>{\hspre}l<{\hspost}@{}}%
\column{3}{@{}>{\hspre}l<{\hspost}@{}}%
\column{E}{@{}>{\hspre}l<{\hspost}@{}}%
\>[B]{}\mathbf{instance}\;\Conid{Num}\;\Varid{a}\Rightarrow \Conid{Scalable}\;(\mathbin{\rightarrow^{\!\!+}\!})\;\Varid{a}\;\mathbf{where}{}\<[E]%
\\
\>[B]{}\hsindent{3}{}\<[3]%
\>[3]{}\Varid{scale}\;\Varid{a}\mathrel{=}\Conid{AddFun}\;(\lambda \Varid{da}\to \Varid{a}\cdot\Varid{da}){}\<[E]%
\ColumnHook
\end{hscode}\resethooks
\end{minipage}
\\
Since uncurried multiplication is bilinear, its partial application as \ensuremath{\Varid{scale}\;\Varid{a}} (for functions) is linear for all \ensuremath{\Varid{a}}.
Now we can rephrase the product rule in terms of more general, linear language, using the derived \ensuremath{(\mathbin{\triangledown})} operation defined in \secref{Derived Operations}:
\begin{hscode}\SaveRestoreHook
\column{B}{@{}>{\hspre}l<{\hspost}@{}}%
\column{E}{@{}>{\hspre}l<{\hspost}@{}}%
\>[B]{}\mathcal{D}\;\Varid{mulC}\;(\Varid{a},\Varid{b})\mathrel{=}\Varid{scale}\;\Varid{b}\mathbin{\triangledown}\Varid{scale}\;\Varid{a}{}\<[E]%
\ColumnHook
\end{hscode}\resethooks

This product rule, along with the linearity of negation and uncurried addition, enables using the same style of derivation as with operations from \ensuremath{\Conid{Category}}, \ensuremath{\Conid{Monoidal}},
and \ensuremath{\Conid{Cartesian}}
above.
As usual, specify the \ensuremath{\Conid{NumCat}} instance for differentiable functions by saying that \ensuremath{\hat{\mathcal{D}}} preserves (\ensuremath{\Conid{NumCat}}) structure, i.e., \ensuremath{\hat{\mathcal{D}}\;\Varid{negateC}\mathrel{=}\Varid{negateC}}, \ensuremath{\hat{\mathcal{D}}\;\Varid{addC}\mathrel{=}\Varid{addC}}, and \ensuremath{\hat{\mathcal{D}}\;\Varid{mulC}\mathrel{=}\Varid{mulC}}.
Reasoning as before, we get another correct-by-construction instance for differentiable functions:
\begin{hscode}\SaveRestoreHook
\column{B}{@{}>{\hspre}l<{\hspost}@{}}%
\column{3}{@{}>{\hspre}l<{\hspost}@{}}%
\column{9}{@{}>{\hspre}l<{\hspost}@{}}%
\column{E}{@{}>{\hspre}l<{\hspost}@{}}%
\>[B]{}\mathbf{instance}\;\Conid{NumCat}\;\Conid{D}\;\mathbf{where}{}\<[E]%
\\
\>[B]{}\hsindent{3}{}\<[3]%
\>[3]{}\Varid{negateC}\mathrel{=}\Varid{linearD}\;\Varid{negateC}{}\<[E]%
\\
\>[B]{}\hsindent{3}{}\<[3]%
\>[3]{}\Varid{addC}{}\<[9]%
\>[9]{}\mathrel{=}\Varid{linearD}\;\Varid{addC}{}\<[E]%
\\
\>[B]{}\hsindent{3}{}\<[3]%
\>[3]{}\Varid{mulC}{}\<[9]%
\>[9]{}\mathrel{=}\Conid{D}\;(\lambda (\Varid{a},\Varid{b})\to (\Varid{a}\cdot\Varid{b},\Varid{scale}\;\Varid{b}\mathbin{\triangledown}\Varid{scale}\;\Varid{a})){}\<[E]%
\ColumnHook
\end{hscode}\resethooks
Similar reasoning applies to other numeric operations, e.g.,
\begin{hscode}\SaveRestoreHook
\column{B}{@{}>{\hspre}l<{\hspost}@{}}%
\column{3}{@{}>{\hspre}l<{\hspost}@{}}%
\column{9}{@{}>{\hspre}l<{\hspost}@{}}%
\column{E}{@{}>{\hspre}l<{\hspost}@{}}%
\>[B]{}\mathbf{instance}\;\Conid{FloatingCat}\;\Conid{D}\;\mathbf{where}{}\<[E]%
\\
\>[B]{}\hsindent{3}{}\<[3]%
\>[3]{}\Varid{sinC}{}\<[9]%
\>[9]{}\mathrel{=}\Conid{D}\;(\lambda \Varid{a}\to (\Varid{sin}\;\Varid{a},\Varid{scale}\;(\Varid{cos}\;\Varid{a}))){}\<[E]%
\\
\>[B]{}\hsindent{3}{}\<[3]%
\>[3]{}\Varid{cosC}{}\<[9]%
\>[9]{}\mathrel{=}\Conid{D}\;(\lambda \Varid{a}\to (\Varid{cos}\;\Varid{a},\Varid{scale}\;(\mathbin{-}\Varid{sin}\;\Varid{a}))){}\<[E]%
\\
\>[B]{}\hsindent{3}{}\<[3]%
\>[3]{}\Varid{expC}{}\<[9]%
\>[9]{}\mathrel{=}\Conid{D}\;(\lambda \Varid{a}\to \mathbf{let}\;\Varid{e}\mathrel{=}\Varid{exp}\;\Varid{a}\;\mathbf{in}\;(\Varid{e},\Varid{scale}\;\Varid{e})){}\<[E]%
\\
\>[9]{}\mathbin{...}{}\<[E]%
\ColumnHook
\end{hscode}\resethooks
In what follows, the \ensuremath{\Varid{scale}} operation will play a more important role than merely tidying definitions.

\sectionl{Examples}

Let's now look at some AD examples, to which we will return later in the paper:
% In this section and later ones, we will use a few running examples:
\begin{hscode}\SaveRestoreHook
\column{B}{@{}>{\hspre}l<{\hspost}@{}}%
\column{E}{@{}>{\hspre}l<{\hspost}@{}}%
\>[B]{}\Varid{sqr}\mathbin{::}\Conid{Num}\;\Varid{a}\Rightarrow \Varid{a}\to \Varid{a}{}\<[E]%
\\
\>[B]{}\Varid{sqr}\;\Varid{a}\mathrel{=}\Varid{a}\cdot\Varid{a}{}\<[E]%
\\[\blanklineskip]%
\>[B]{}\Varid{magSqr}\mathbin{::}\Conid{Num}\;\Varid{a}\Rightarrow \Varid{a} \times \Varid{a}\to \Varid{a}{}\<[E]%
\\
\>[B]{}\Varid{magSqr}\;(\Varid{a},\Varid{b})\mathrel{=}\Varid{sqr}\;\Varid{a}\mathbin{+}\Varid{sqr}\;\Varid{b}{}\<[E]%
\\[\blanklineskip]%
\>[B]{}\Varid{cosSinProd}\mathbin{::}\Conid{Floating}\;\Varid{a}\Rightarrow \Varid{a} \times \Varid{a}\to \Varid{a} \times \Varid{a}{}\<[E]%
\\
\>[B]{}\Varid{cosSinProd}\;(\Varid{x},\Varid{y})\mathrel{=}(\Varid{cos}\;\Varid{z},\Varid{sin}\;\Varid{z})\;\mathbf{where}\;\Varid{z}\mathrel{=}\Varid{x}\cdot\Varid{y}{}\<[E]%
\ColumnHook
\end{hscode}\resethooks
A compiler plugin converts these definitions to categorical vocabulary \citep{Elliott-2017-compiling-to-categories}:
\begin{hscode}\SaveRestoreHook
\column{B}{@{}>{\hspre}l<{\hspost}@{}}%
\column{E}{@{}>{\hspre}l<{\hspost}@{}}%
\>[B]{}\Varid{sqr}\mathrel{=}\Varid{mulC}\hsdot{\circ }{.}(\Varid{id}\mathbin{\vartriangle}\Varid{id}){}\<[E]%
\\[\blanklineskip]%
\>[B]{}\Varid{magSqr}\mathrel{=}\Varid{addC}\hsdot{\circ }{.}(\Varid{mulC}\hsdot{\circ }{.}(\Varid{exl}\mathbin{\vartriangle}\Varid{exl})\mathbin{\vartriangle}\Varid{mulC}\hsdot{\circ }{.}(\Varid{exr}\mathbin{\vartriangle}\Varid{exr})){}\<[E]%
\\[\blanklineskip]%
\>[B]{}\Varid{cosSinProd}\mathrel{=}(\Varid{cosC}\mathbin{\vartriangle}\Varid{sinC})\hsdot{\circ }{.}\Varid{mulC}{}\<[E]%
\ColumnHook
\end{hscode}\resethooks
To visualize computations before differentiation, we can interpret these categorical expressions in a category of graphs \citep[Section 7]{Elliott-2017-compiling-to-categories}, with the results rendered in \figreftwo{magSqr}{cosSinProd}.
\figp{
\figone{magSqr}{\ensuremath{\Varid{magSqr}}}}{
\figone{cosSinProd}{\ensuremath{\Varid{cosSinProd}}}}
To see the differentiable versions, interpret these same expressions in the category of differentiable functions (\ensuremath{\Conid{D}} from \secref{Categories}), remove the \ensuremath{\Conid{D}} constructors to reveal the function representation, convert these functions to categorical form as well, and finally interpret the result in the graph category.
The results are rendered in \figreftwo{magSqr-adf}{cosSinProd-adf}.
\figp{
\figone{magSqr-adf}{\ensuremath{\hat{\mathcal{D}}\;\Varid{magSqr}}}}{
\figone{cosSinProd-adf}{\ensuremath{\hat{\mathcal{D}}\;\Varid{cosSinProd}}}}
Some remarks:
\begin{itemize}
\item The derivatives are (linear) functions, as depicted in boxes.
\item Work is shared between the function's result (sometimes called the ``primal'') and its derivative in \figref{cosSinProd-adf}.
\item The graphs shown here are used \emph{solely} for visualizing functions before and after differentiation, playing no role in the programming interface or in the implementation of differentiation.
\end{itemize}

\sectionl{Programming as Defining and Solving Algebra Problems}

Stepping back to consider what we've done, a general recipe emerges:\out{\notefoot{Go over the wording of this section to make as clear as I can.}}
\begin{itemize}
\item Start with an expensive or even non-computable specification (here involving differentiation).
\item Build the desired result into the representation of a new data type (here as the combination of a function and its derivative).
\item Try to show that conversion from a simpler form (here regular functions) to the new data type---even if not computable---is \emph{compositional} with respect to a well-understood collection of algebraic abstractions (here \ensuremath{\Conid{Category}} etc).
\item If compositionality fails (as with \ensuremath{\mathcal{D}}, unadorned differentiation, in \secref{Sequential Composition}), examine the failure to find an augmented specification, iterating as needed until converging on a representation and corresponding specification that \emph{is} compositional.
\item Set up an algebra problem whose solution will be an instance of the well-understood algebraic abstraction for the chosen representation.
These algebra problems always have a particular stylized form, namely that the operation being solved for is a \emph{homomorphism} for the chosen abstractions (here including a category homomorphism, also called a ``functor'').
\item Solve the algebra problem by using the compositionality properties.
\item Rest assured that the solution satisfies the required laws, at least when the new data type is kept abstract, thanks to the homomorphic specification.
\end{itemize}
The result of this recipe is not quite an implementation of our homomorphic specification, which may after all be non-computable.
Rather, it gives a computable alternative that is nearly as useful: if the input to the specified conversion is expressed in the vocabulary of the chosen algebraic abstraction, then a re-interpretation of that vocabulary in the new data type is the result of the (possibly non-computable) specification.
Furthermore, if we can \emph{automatically} convert conventionally written functional programs into the chosen algebraic vocabulary \citep{Elliott-2017-compiling-to-categories}, then those programs can be re-interpreted to compute the desired specification.

\sectionl{Generalizing Automatic Differentiation}

\corRefs{compose}{linear} all have the same form: an operation on \ensuremath{\Conid{D}} (differentiable functions) is defined entirely via the same operation on \ensuremath{(\multimap)} (linear maps).
Specifically, the sequential and parallel composition of differentiable functions rely (respectively) on sequential and parallel composition of linear maps, and likewise for each other operation.
These corollaries follow closely from \thmRefs{compose}{linear}, which relate derivatives for these operations to the corresponding operations on linear maps.
These properties make for a pleasantly poetic theory, but they also have a powerful, tangible benefit, which is that we can replace linear maps by any of a much broader variety of underlying categories to arrive at a greatly generalized notion of AD.

%% %format GD (k) a b = a "\leadsto_{"k"}" b

A few small changes to the non-generalized definitions derived in \secref{Putting the Pieces Together} result in the generalized AD definitions shown in \figref{GAD}:
\begin{itemize}
\item The new category takes as parameter a category \ensuremath{\Varid{k}} that replaces \ensuremath{(\multimap)} in \ensuremath{\Conid{D}}.
\item The \ensuremath{\Varid{linearD}} function takes two arrows, previously identified.\notefoot{Alternatively, posit an embedding function \ensuremath{\Varid{lin}\mathbin{::}(\Varid{a}\multimap\Varid{b})\to (\Varid{a}\to \Varid{b})}, write \thmRef{linear} as \ensuremath{\mathcal{D}\;(\Varid{lin}\;\Varid{f})\;\Varid{a}\mathrel{=}\Varid{f}}, and change to \ensuremath{\Varid{linearD}\mathbin{::}(\Varid{a}\multimap\Varid{b})\to \Conid{D}\;\Varid{a}\;\Varid{b}}.
Then retroactively make \ensuremath{\Varid{lin}} a method of a new class.
Could incremental computation implement \ensuremath{\Varid{lin}}?}
\item The functionality needed of the underlying category becomes explicit.
\item The constraint \ensuremath{\Conid{Obj}_{\Conid{D}_{\Varid{k}}}} is defined to be the conjunction of \ensuremath{\Conid{Additive}} (needed for the \ensuremath{\Conid{Cocartesian}} instance) and \ensuremath{\Conid{Obj}_{\Varid{k}}} (needed for all instances).
\end{itemize}
\begin{figure}
\begin{center}
\begin{hscode}\SaveRestoreHook
\column{B}{@{}>{\hspre}l<{\hspost}@{}}%
\column{3}{@{}>{\hspre}l<{\hspost}@{}}%
\column{8}{@{}>{\hspre}l<{\hspost}@{}}%
\column{12}{@{}>{\hspre}l<{\hspost}@{}}%
\column{23}{@{}>{\hspre}l<{\hspost}@{}}%
\column{24}{@{}>{\hspre}l<{\hspost}@{}}%
\column{32}{@{}>{\hspre}l<{\hspost}@{}}%
\column{E}{@{}>{\hspre}l<{\hspost}@{}}%
\>[B]{}\mathbf{newtype}\;\Conid{D}_{\Varid{k}}\;\Varid{a}\;\Varid{b}\mathrel{=}\Conid{D}\;(\Varid{a}\to \Varid{b} \times (\Varid{a}\mathbin{`\Varid{k}`}\Varid{b})){}\<[E]%
\\[\blanklineskip]%
\>[B]{}\Varid{linearD}\mathbin{::}(\Varid{a}\to \Varid{b})\to (\Varid{a}\mathbin{`\Varid{k}`}\Varid{b})\to \Conid{D}_{\Varid{k}}\;\Varid{a}\;\Varid{b}{}\<[E]%
\\
\>[B]{}\Varid{linearD}\;\Varid{f}\;\Varid{f'}\mathrel{=}\Conid{D}\;(\lambda \Varid{a}\to (\Varid{f}\;\Varid{a},\Varid{f'})){}\<[E]%
\\[\blanklineskip]%
\>[B]{}\mathbf{instance}\;\Conid{Category}\;\Varid{k}\Rightarrow \Conid{Category}\;\Conid{D}_{\Varid{k}}\;\mathbf{where}{}\<[E]%
\\
\>[B]{}\hsindent{3}{}\<[3]%
\>[3]{}\mathbf{type}\;\Conid{Obj}_{\Conid{D}_{\Varid{k}}}\mathrel{=}\Conid{Additive}\mathrel{\wedge}\Conid{Obj}_{\Varid{k}}{}\<[E]%
\\
\>[B]{}\hsindent{3}{}\<[3]%
\>[3]{}\Varid{id}\mathrel{=}\Varid{linearD}\;\Varid{id}\;\Varid{id}{}\<[E]%
\\
\>[B]{}\hsindent{3}{}\<[3]%
\>[3]{}\Conid{D}\;\Varid{g}\hsdot{\circ }{.}\Conid{D}\;\Varid{f}\mathrel{=}\Conid{D}\;(\lambda \Varid{a}\to \mathbf{let}\;\{\mskip1.5mu (\Varid{b},\Varid{f'})\mathrel{=}\Varid{f}\;\Varid{a};(\Varid{c},\Varid{g'})\mathrel{=}\Varid{g}\;\Varid{b}\mskip1.5mu\}\;\mathbf{in}\;(\Varid{c},\Varid{g'}\hsdot{\circ }{.}\Varid{f'})){}\<[E]%
\\[\blanklineskip]%
\>[B]{}\mathbf{instance}\;\Conid{Monoidal}\;\Varid{k}\Rightarrow \Conid{Monoidal}\;\Conid{D}_{\Varid{k}}\;\mathbf{where}{}\<[E]%
\\
\>[B]{}\hsindent{3}{}\<[3]%
\>[3]{}\Conid{D}\;\Varid{f}\times\Conid{D}\;\Varid{g}\mathrel{=}\Conid{D}\;(\lambda (\Varid{a},\Varid{b})\to \mathbf{let}\;\{\mskip1.5mu (\Varid{c},\Varid{f'})\mathrel{=}\Varid{f}\;\Varid{a};(\Varid{d},\Varid{g'})\mathrel{=}\Varid{g}\;\Varid{b}\mskip1.5mu\}\;\mathbf{in}\;((\Varid{c},\Varid{d}),\Varid{f'}\times\Varid{g'})){}\<[E]%
\\[\blanklineskip]%
\>[B]{}\mathbf{instance}\;\Conid{Cartesian}\;\Varid{k}\Rightarrow \Conid{Cartesian}\;\Conid{D}_{\Varid{k}}\;\mathbf{where}{}\<[E]%
\\
\>[B]{}\hsindent{3}{}\<[3]%
\>[3]{}\Varid{exl}{}\<[8]%
\>[8]{}\mathrel{=}\Varid{linearD}\;\Varid{exl}\;{}\<[23]%
\>[23]{}\Varid{exl}{}\<[E]%
\\
\>[B]{}\hsindent{3}{}\<[3]%
\>[3]{}\Varid{exr}{}\<[8]%
\>[8]{}\mathrel{=}\Varid{linearD}\;\Varid{exr}\;{}\<[23]%
\>[23]{}\Varid{exr}{}\<[E]%
\\
\>[B]{}\hsindent{3}{}\<[3]%
\>[3]{}\Varid{dup}{}\<[8]%
\>[8]{}\mathrel{=}\Varid{linearD}\;\Varid{dup}\;{}\<[23]%
\>[23]{}\Varid{dup}{}\<[E]%
\\[\blanklineskip]%
\>[B]{}\mathbf{instance}\;\Conid{Cocartesian}\;\Varid{k}\Rightarrow \Conid{Cocartesian}\;\Conid{D}_{\Varid{k}}\;\mathbf{where}{}\<[E]%
\\
\>[B]{}\hsindent{3}{}\<[3]%
\>[3]{}\Varid{inl}{}\<[8]%
\>[8]{}\mathrel{=}\Varid{linearD}\;\Varid{inlF}\;{}\<[24]%
\>[24]{}\Varid{inl}{}\<[E]%
\\
\>[B]{}\hsindent{3}{}\<[3]%
\>[3]{}\Varid{inr}{}\<[8]%
\>[8]{}\mathrel{=}\Varid{linearD}\;\Varid{inrF}\;{}\<[24]%
\>[24]{}\Varid{inr}{}\<[E]%
\\
\>[B]{}\hsindent{3}{}\<[3]%
\>[3]{}\Varid{jam}{}\<[8]%
\>[8]{}\mathrel{=}\Varid{linearD}\;\Varid{jamF}\;{}\<[24]%
\>[24]{}\Varid{jam}{}\<[E]%
\\[\blanklineskip]%
\>[B]{}\mathbf{instance}\;\Conid{Scalable}\;\Varid{k}\;\Varid{s}\Rightarrow \Conid{NumCat}\;\Conid{D}_{\Varid{k}}\;\Varid{s}\;\mathbf{where}{}\<[E]%
\\
\>[B]{}\hsindent{3}{}\<[3]%
\>[3]{}\Varid{negateC}{}\<[12]%
\>[12]{}\mathrel{=}\Varid{linearD}\;{}\<[23]%
\>[23]{}\Varid{negateC}\;{}\<[32]%
\>[32]{}\Varid{negateC}{}\<[E]%
\\
\>[B]{}\hsindent{3}{}\<[3]%
\>[3]{}\Varid{addC}{}\<[12]%
\>[12]{}\mathrel{=}\Varid{linearD}\;{}\<[23]%
\>[23]{}\Varid{addC}\;{}\<[32]%
\>[32]{}\Varid{addC}{}\<[E]%
\\
\>[B]{}\hsindent{3}{}\<[3]%
\>[3]{}\Varid{mulC}{}\<[12]%
\>[12]{}\mathrel{=}\Conid{D}\;(\lambda (\Varid{a},\Varid{b})\to (\Varid{a}\cdot\Varid{b},\Varid{scale}\;\Varid{b}\mathbin{\triangledown}\Varid{scale}\;\Varid{a})){}\<[E]%
\ColumnHook
\end{hscode}\resethooks
\caption{Generalized automatic differentiation}
\figlabel{GAD}
\end{center}
\end{figure}

\sectionl{Matrices}

As an aside, let's consider matrices---the representation typically used in linear algebra---and especially the property of rectangularity.
There are three (non-exclusive) possibilities for a nonempty matrix \ensuremath{\Conid{W}}:
\begin{itemize}
\item \ensuremath{\Varid{width}\;\Conid{W}\mathrel{=}\Varid{height}\;\Conid{W}\mathrel{=}\mathrm{1}};
\item \ensuremath{\Conid{W}} is the horizontal juxtaposition of two matrices \ensuremath{\Conid{U}} and \ensuremath{\Conid{V}} with \ensuremath{\Varid{height}\;\Conid{W}\mathrel{=}\Varid{height}\;\Conid{U}\mathrel{=}\Varid{height}\;\Conid{V}}, and \ensuremath{\Varid{width}\;\Conid{W}\mathrel{=}\Varid{width}\;\Conid{U}\mathbin{+}\Varid{width}\;\Conid{V}}; or
\item \ensuremath{\Conid{W}} is the vertical juxtaposition of two matrices \ensuremath{\Conid{U}} and \ensuremath{\Conid{V}} with \ensuremath{\Varid{width}\;\Conid{W}\mathrel{=}\Varid{width}\;\Conid{U}\mathrel{=}\Varid{width}\;\Conid{V}}, and \ensuremath{\Varid{height}\;\Conid{W}\mathrel{=}\Varid{height}\;\Conid{U}\mathbin{+}\Varid{height}\;\Conid{V}}.
\end{itemize}
These three shape constraints establish and preserve rectangularity.

The vocabulary we have needed from generalized linear maps so far is exactly that of \ensuremath{\Conid{Category}}, \ensuremath{\Conid{Cartesian}}, \ensuremath{\Conid{Cocartesian}}, and \ensuremath{\Conid{Scalable}}.
Let's now extract just three operations from this vocabulary:
\begin{closerCodePars}
\begin{hscode}\SaveRestoreHook
\column{B}{@{}>{\hspre}l<{\hspost}@{}}%
\column{3}{@{}>{\hspre}l<{\hspost}@{}}%
\column{10}{@{}>{\hspre}l<{\hspost}@{}}%
\column{E}{@{}>{\hspre}l<{\hspost}@{}}%
\>[3]{}\Varid{scale}{}\<[10]%
\>[10]{}\mathbin{::}\Varid{a}\to (\Varid{a}\mathbin{`\Varid{k}`}\Varid{a}){}\<[E]%
\\
\>[3]{}(\mathbin{\triangledown}){}\<[10]%
\>[10]{}\mathbin{::}(\Varid{a}\mathbin{`\Varid{k}`}\Varid{c})\to (\Varid{b}\mathbin{`\Varid{k}`}\Varid{c})\to ((\Varid{a} \times \Varid{b})\mathbin{`\Varid{k}`}\Varid{c}){}\<[E]%
\\
\>[3]{}(\mathbin{\vartriangle}){}\<[10]%
\>[10]{}\mathbin{::}(\Varid{a}\mathbin{`\Varid{k}`}\Varid{c})\to (\Varid{a}\mathbin{`\Varid{k}`}\Varid{d})\to (\Varid{a}\mathbin{`\Varid{k}`}(\Varid{c} \times \Varid{d})){}\<[E]%
\ColumnHook
\end{hscode}\resethooks
\end{closerCodePars}%
These operations exactly correspond to the three possibilities above for a nonempty matrix \ensuremath{\Conid{W}}, with the width and height constraints captured neatly by types.
When matrices are used to represent linear maps, the domain and codomain types for the corresponding linear map are determined by the width and height of the matrix, respectively (assuming the convention of matrix on the left multiplied by a column vector on the right), together with the type of the matrix elements.

\out{\mynote{Maybe say something about block matrices and their use in efficient matrix computations, citing \citet{MacedoOliveira2013Typing}.}}

\sectionl{Extracting a Data Representation}

The generalized form of AD in \secref{Generalizing Automatic Differentiation} allows for different representations of linear maps (as well as alternatives to linear maps).
One simple choice is to use functions, as in \figreftwo{magSqr-adf}{cosSinProd-adf}.
Although this choice is simple and reliable, sometimes we need a \emph{data} representation.
For instance,
\begin{itemize}
\item Gradient-based optimization (including in machine learning) works by searching for local minima in the domain of a differentiable function \ensuremath{\Varid{f}\mathbin{::}\Varid{a}\to \Varid{s}}, where \ensuremath{\Varid{a}} is a vector space over the scalar field \ensuremath{\Varid{s}}.
      Each step in the search is in the direction opposite of the gradient of \ensuremath{\Varid{f}}, which is a vector form of \ensuremath{\mathcal{D}\;\Varid{f}}.
\item Computer graphics shading models rely on normal vectors.
      For surfaces represented in parametric form, i.e., as \ensuremath{\Varid{f}\mathbin{::}\mathbb{R}_{2}\to \mathbb{R}_{3}}, normal vectors are calculated from the partial derivatives of \ensuremath{\Varid{f}} as vectors, which are the rows of the $3 \times 2$ Jacobian matrix that represents the derivative of \ensuremath{\Varid{f}} at any given point \ensuremath{\Varid{p}\mathbin{::}\mathbb{R}_{2}}.
\end{itemize}

Given a linear map \ensuremath{\Varid{f'}\mathbin{::}\Conid{U}\multimap\Conid{V}} represented as a function, it is possible to extract a Jacobian matrix (including the special case of a gradient vector) by applying \ensuremath{\Varid{f'}} to every vector in a basis of \ensuremath{\Conid{U}}.
A particularly convenient basis is the sequence of column vectors of an identity matrix, where the \ensuremath{\Varid{i}^{\text{th}}} such vector has a one in the \ensuremath{\Varid{i}^{\text{th}}} position and zeros elsewhere.
If \ensuremath{\Conid{U}} has dimension \ensuremath{\Varid{m}} (e.g., \ensuremath{\Conid{U}\mathrel{=}\mathbb{R}^m}), this sampling requires \ensuremath{\Varid{m}} passes.
\out{Considering the nature of the sparse vectors used as arguments, each pass likely computes inefficiently.
Alternatively, the computations can be done using a sparse vector representation, but such an implementation involves considerable complexity and poses difficulties for efficient, massively parallel, SIMD implementations, such as graphics processors\needcite.}
If \ensuremath{\Varid{m}} is small, then this method of extracting a Jacobian is tolerably efficient, but as dimension grows, it becomes quite expensive.
In particular, many useful problems involve gradient-based optimization over very high-dimensional spaces, which is the worst case for this technique.

\sectionl{Generalized Matrices}

Rather than representing derivatives as functions and then extracting a (Jacobian) matrix, a more conventional alternative is to construct and combine matrices in the first place.
These matrices are usually rectangular arrays, representing \ensuremath{\mathbb{R}^m\multimap\mathbb{R}^n}, which interferes with the composability we get from  organizing around binary cartesian products, as in the \ensuremath{\Conid{Monoidal}}, \ensuremath{\Conid{Cartesian}}, and \ensuremath{\Conid{Cocartesian}} categorical interfaces.

There is, however, an especially convenient perspective on linear algebra, known as \emph{free vector spaces}\needcite\out{FreeVectorSpaceOverASet}.
Given a scalar field \ensuremath{\Varid{s}}, any free vector space has the form \ensuremath{\Varid{p}\to \Varid{s}} for some \ensuremath{\Varid{p}}, where the cardinality of \ensuremath{\Varid{p}} is the dimension of the vector space (and only finitely many \ensuremath{\Varid{p}} values can have non-zero images).
Scaling a vector \ensuremath{\Varid{v}\mathbin{::}\Varid{p}\to \Varid{s}} or adding two such vectors is defined in the usual way for functions.
Rather than using functions directly as a representation, one can instead use any representation isomorphic to such a function.
In particular, we can represent vector spaces over a given field as a \emph{representable functor}, i.e., a functor \ensuremath{\Conid{F}} such that $\exists p \, \forall s$ \ensuremath{\Conid{F}\;\Varid{s} \cong \Varid{p}\to \Varid{s}} (where ``\ensuremath{ \cong }'' denotes isomorphism)\out{\notefoot{Relate this notion of \emph{functor} to the one used for specifying \ensuremath{\hat{\mathcal{D}}}.}}
This method is convenient in a richly typed functional language like Haskell, which comes with libraries of functor-level building blocks.
Four such building blocks are functor product, functor composition, and their corresponding identities, which are the unit functor (containing no elements) and the identity functor (containing one element) \citep{Magalhaes:2010,HaskellWikiGhcGenerics}.
One must then define the standard functionality for linear maps in the form of instances of \ensuremath{\Conid{Category}}, \ensuremath{\Conid{Monoidal}}, \ensuremath{\Conid{Cartesian}}, \ensuremath{\Conid{Cocartesian}}, and \ensuremath{\Conid{Scalable}}.
Details are worked out by \citet[Section 7.4 and Appendix A]{Elliott-2017-compiling-to-categories}.
One can use other representable functors as well, including length-typed vectors \citep{vector-sized}.

All of these functors give data representations of functions that save recomputation over a native function representation, as a form of functional memoization \cite{Hinze00memofunctions}.
They also provide a composable, type-safe alternative to the more commonly used multi-dimensional arrays (often called ``tensors'') in machine learning libraries.

\sectionl{Efficiency of Composition}

With the function representation of linear maps, composition is simple and efficient, but extracting a matrix can be quite expensive, as described in \secref{Extracting a Data Representation}.
The generalized matrix representation of \secref{Generalized Matrices} eliminates the need for this expensive extraction step but at the cost of more expensive construction operations used throughout.

One particularly important efficiency concern is that of (generalized) matrix multiplication.
Although matrix multiplication is associative (because it correctly implements composition of linear maps represented as matrices), different associations can result in very different computational cost.
The problem of optimally associating a chain of matrix multiplications can be solved via dynamic programming in $O(n^3)$ time \citep[Section 15.2]{CLRS} or in $O(n \log n)$ time with a more subtle algorithm \citep{Hu:Shing:1981}.
Solving this problem requires knowing only the sizes (heights and widths) of the matrices involved, and those sizes depend only on the types involved for a strongly typed linear map representation.
One can thus choose an optimal association at compile time rather than waiting for run-time and then solving the problem repeatedly.
A more sophisticated version of this question is known as the ``optimal Jacobian accumulation'' problem and is NP-complete \citep {Naumann2008OptimalJA}.

Alternatively, for some kinds of problems we might want to choose a particular association for sequential composition.
For instance, gradient-based optimization (including its use in machine learning) uses ``reverse-mode'' automatic differentiation (RAD), which is to say fully left-associated compositions.\notefoot{Is RAD always optimal for gradient problems?}
(Dually, ``foward-mode'' AD fully right-associates.)
Reverse mode (including its specialization, backpropagation) is much more efficient for these problems, but is also typically given much more complicated explanations and implementations, involving mutation, graph construction, and ``tapes''\needcite.
One of the main purposes of this paper is to demonstrate that these complications are inessential and that RAD can instead be specified and implemented quite simply.

\sectionl{Reverse-Mode Automatic Differentiation}

The AD algorithm derived in \secref{Putting the Pieces Together} and generalized in \figref{GAD} can be thought of as a family of algorithms.
For fully right-associated compositions, it becomes forward mode AD; for fully left-associated compositions, reverse-mode AD; and for all other associations, various mixed modes.

Let's now look at how to separate the associations used in formulating a differentiable function from the associations used to compose its derivatives.
A practical reason for making this separation is that we want to do gradient-based optimization (calling for left association), while modular program organization results in a mixture of compositions.
Fortunately, a fairly simple technique removes the tension between efficient execution and modular program organization.

Given any category \ensuremath{\Varid{k}}, we can represent its morphisms by the intent to left-compose with some to-be-given morphism \ensuremath{\Varid{h}}.
That is, represent \ensuremath{\Varid{f}\mathbin{::}\Varid{a}\mathbin{`\Varid{k}`}\Varid{b}} by the function \ensuremath{(\hsdot{\circ }{.}\:{}\Varid{f})\mathbin{::}(\Varid{b}\mathbin{`\Varid{k}`}\Varid{r})\to (\Varid{a}\mathbin{`\Varid{k}`}\Varid{r})}, where \ensuremath{\Varid{r}} is any object in \ensuremath{\Varid{k}}.\footnote{Following Haskell notation for \emph{right sections}, ``\ensuremath{(\hsdot{\circ }{.}\:{}\Varid{f})}'' is shorthand for \ensuremath{\lambda \Varid{h}\to \Varid{h}\hsdot{\circ }{.}\Varid{f}}.}
The morphism \ensuremath{\Varid{h}} will be a \emph{continuation}, finishing the journey from \ensuremath{\Varid{f}} all the way to the codomain of the overall function being assembled.
Building a category around this idea results in converting \emph{all} composition patterns into fully left-associated form.
This trick is akin to conversion to continuation-passing style \citep{Reynolds72definitionalinterpreters,Appel2007CC,Kennedy2007ContCont}.
%% Give each computation a continuation saying how the result will ultimately be consumed.
Compositions in the computation become compositions in the continuation\out{ which is post-/left-composed with the main computation}.
For instance, \ensuremath{\Varid{g}\hsdot{\circ }{.}\Varid{f}} with a continuation \ensuremath{\Varid{k}} (i.e., \ensuremath{\Varid{k}\hsdot{\circ }{.}(\Varid{g}\hsdot{\circ }{.}\Varid{f})}) becomes \ensuremath{\Varid{f}} with a continuation \ensuremath{\Varid{k}\hsdot{\circ }{.}\Varid{g}} (i.e., \ensuremath{(\Varid{k}\hsdot{\circ }{.}\Varid{g})\hsdot{\circ }{.}\Varid{f}}).
The initial continuation is \ensuremath{\Varid{id}} (because \ensuremath{\Varid{id}\hsdot{\circ }{.}\Varid{f}\mathrel{=}\Varid{f}}).

Now package up the continuation representation as a transformation from category \ensuremath{\Varid{k}} and codomain \ensuremath{\Varid{r}} to a new category, \ensuremath{\Conid{Cont}_{\Varid{k}}^{\Varid{r}}}:
\begin{hscode}\SaveRestoreHook
\column{B}{@{}>{\hspre}l<{\hspost}@{}}%
\column{E}{@{}>{\hspre}l<{\hspost}@{}}%
\>[B]{}\mathbf{newtype}\;\Conid{Cont}_{\Varid{k}}^{\Varid{r}}\;\Varid{a}\;\Varid{b}\mathrel{=}\Conid{Cont}\;((\Varid{b}\mathbin{`\Varid{k}`}\Varid{r})\to (\Varid{a}\mathbin{`\Varid{k}`}\Varid{r})){}\<[E]%
\\[\blanklineskip]%
\>[B]{}\Varid{cont}\mathbin{::}\Conid{Category}\;\Varid{k}\Rightarrow (\Varid{a}\mathbin{`\Varid{k}`}\Varid{b})\to \Conid{Cont}_{\Varid{k}}^{\Varid{r}}\;\Varid{a}\;\Varid{b}{}\<[E]%
\\
\>[B]{}\Varid{cont}\;\Varid{f}\mathrel{=}\Conid{Cont}\;(\hsdot{\circ }{.}\:{}\Varid{f}){}\<[E]%
\ColumnHook
\end{hscode}\resethooks
As usual, we can derive instances for our new category by homomorphic specification:
\begin{theorem}[\provedIn{theorem:cont}]\thmLabel{cont}
Given the definitions in \figref{cont}, \ensuremath{\Varid{cont}} is a homomorphism with respect to each instantiated class.
\end{theorem}
Note the pleasant symmetries in \figref{cont}.
Each \ensuremath{\Conid{Cartesian}} or \ensuremath{\Conid{Cocartesian}} operation on \ensuremath{\Conid{Cont}_{\Varid{k}}^{\Varid{r}}} is defined via the dual \ensuremath{\Conid{Cocartesian}} or \ensuremath{\Conid{Cartesian}} operation, together with the \ensuremath{\Varid{join}}/\ensuremath{\Varid{unjoin}} isomorphism.

\begin{figure}
\begin{center}
\begin{hscode}\SaveRestoreHook
\column{B}{@{}>{\hspre}l<{\hspost}@{}}%
\column{3}{@{}>{\hspre}l<{\hspost}@{}}%
\column{4}{@{}>{\hspre}l<{\hspost}@{}}%
\column{8}{@{}>{\hspre}l<{\hspost}@{}}%
\column{E}{@{}>{\hspre}l<{\hspost}@{}}%
\>[B]{}\mathbf{newtype}\;\Conid{Cont}_{\Varid{k}}^{\Varid{r}}\;\Varid{a}\;\Varid{b}\mathrel{=}\Conid{Cont}\;((\Varid{b}\mathbin{`\Varid{k}`}\Varid{r})\to (\Varid{a}\mathbin{`\Varid{k}`}\Varid{r})){}\<[E]%
\\[\blanklineskip]%
\>[B]{}\mathbf{instance}\;\Conid{Category}\;\Varid{k}\Rightarrow \Conid{Category}\;\Conid{Cont}_{\Varid{k}}^{\Varid{r}}\;\mathbf{where}{}\<[E]%
\\
\>[B]{}\hsindent{3}{}\<[3]%
\>[3]{}\Varid{id}\mathrel{=}\Conid{Cont}\;\Varid{id}{}\<[E]%
\\
\>[B]{}\hsindent{3}{}\<[3]%
\>[3]{}\Conid{Cont}\;\Varid{g}\hsdot{\circ }{.}\Conid{Cont}\;\Varid{f}\mathrel{=}\Conid{Cont}\;(\Varid{f}\hsdot{\circ }{.}\Varid{g}){}\<[E]%
\\[\blanklineskip]%
\>[B]{}\mathbf{instance}\;\Conid{Monoidal}\;\Varid{k}\Rightarrow \Conid{Monoidal}\;\Conid{Cont}_{\Varid{k}}^{\Varid{r}}\;\mathbf{where}{}\<[E]%
\\
\>[B]{}\hsindent{3}{}\<[3]%
\>[3]{}\Conid{Cont}\;\Varid{f}\times\Conid{Cont}\;\Varid{g}\mathrel{=}\Conid{Cont}\;(\Varid{join}\hsdot{\circ }{.}(\Varid{f}\times\Varid{g})\hsdot{\circ }{.}\Varid{unjoin}){}\<[E]%
\\[\blanklineskip]%
\>[B]{}\mathbf{instance}\;\Conid{Cartesian}\;\Varid{k}\Rightarrow \Conid{Cartesian}\;\Conid{Cont}_{\Varid{k}}^{\Varid{r}}\;\mathbf{where}{}\<[E]%
\\
\>[B]{}\hsindent{3}{}\<[3]%
\>[3]{}\Varid{exl}{}\<[8]%
\>[8]{}\mathrel{=}\Conid{Cont}\;(\Varid{join}\hsdot{\circ }{.}\Varid{inl}){}\<[E]%
\\
\>[B]{}\hsindent{3}{}\<[3]%
\>[3]{}\Varid{exr}{}\<[8]%
\>[8]{}\mathrel{=}\Conid{Cont}\;(\Varid{join}\hsdot{\circ }{.}\Varid{inr}){}\<[E]%
\\
\>[B]{}\hsindent{3}{}\<[3]%
\>[3]{}\Varid{dup}{}\<[8]%
\>[8]{}\mathrel{=}\Conid{Cont}\;(\Varid{jam}\hsdot{\circ }{.}\Varid{unjoin}){}\<[E]%
\\[\blanklineskip]%
\>[B]{}\mathbf{instance}\;\Conid{Cocartesian}\;\Varid{k}\Rightarrow \Conid{Cocartesian}\;\Conid{Cont}_{\Varid{k}}^{\Varid{r}}\;\mathbf{where}{}\<[E]%
\\
\>[B]{}\hsindent{3}{}\<[3]%
\>[3]{}\Varid{inl}{}\<[8]%
\>[8]{}\mathrel{=}\Conid{Cont}\;(\Varid{exl}\hsdot{\circ }{.}\Varid{unjoin}){}\<[E]%
\\
\>[B]{}\hsindent{3}{}\<[3]%
\>[3]{}\Varid{inr}{}\<[8]%
\>[8]{}\mathrel{=}\Conid{Cont}\;(\Varid{exr}\hsdot{\circ }{.}\Varid{unjoin}){}\<[E]%
\\
\>[B]{}\hsindent{3}{}\<[3]%
\>[3]{}\Varid{jam}{}\<[8]%
\>[8]{}\mathrel{=}\Conid{Cont}\;(\Varid{join}\hsdot{\circ }{.}\Varid{dup}){}\<[E]%
\\[\blanklineskip]%
\>[B]{}\mathbf{instance}\;\Conid{Scalable}\;\Varid{k}\;\Varid{a}\Rightarrow \Conid{Scalable}\;\Conid{Cont}_{\Varid{k}}^{\Varid{r}}\;\Varid{a}\;\mathbf{where}{}\<[E]%
\\
\>[B]{}\hsindent{4}{}\<[4]%
\>[4]{}\Varid{scale}\;\Varid{s}\mathrel{=}\Conid{Cont}\;(\Varid{scale}\;\Varid{s}){}\<[E]%
\ColumnHook
\end{hscode}\resethooks
\caption{Continuation category transformer (specified by functoriality of \ensuremath{\Varid{cont}})}
\figlabel{cont}
\end{center}
\end{figure}

\out{\mynote{Mention Cayley's Theorem: that any monoid is equivalent to a monoid of functions under composition.
I think \ensuremath{\Conid{Cont}_{\Varid{k}}^{\Varid{r}}} is a generalization from \ensuremath{\Conid{Monoid}} to \ensuremath{\Conid{Category}}.
Also generalizes to the contravariant Yoneda lemma.}}

The instances for \ensuremath{\Conid{Cont}_{\Varid{k}}^{\Varid{r}}} constitute a simple algorithm for reverse-mode AD.
\out{\mynote{Contrast with other presentations.}}
\figreftwo{magSqr-adr}{cosSinProd-adr} show the results of \ensuremath{\Conid{Cont}_{\Varid{k}}^{\Varid{r}}} corresponding to \figreftwo{magSqr}{cosSinProd} and \figreftwo{magSqr-adf}{cosSinProd-adf}.
\figp{
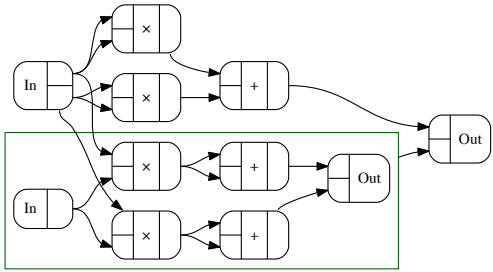
\figoneW{0.40}{magSqr-adr}{\ensuremath{\Varid{magSqr}} in \ensuremath{\Conid{D}_{\Conid{Cont}_{(\mathbin{\rightarrow^{\!\!+}\!})}^{\mathbb{R}}}}}}{
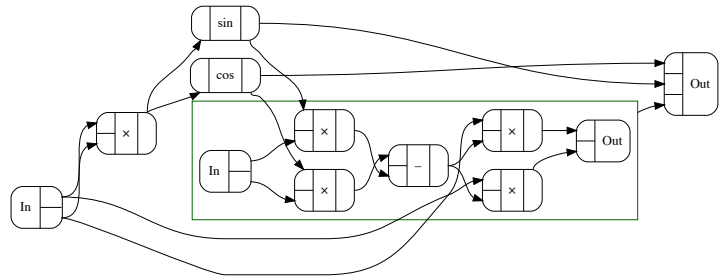
\figoneW{0.57}{cosSinProd-adr}{\ensuremath{\Varid{cosSinProd}} in \ensuremath{\Conid{D}_{\Conid{Cont}_{(\mathbin{\rightarrow^{\!\!+}\!})}^{\mathbb{R}}}}}}
The derivatives are represented as (linear) functions again, but reversed (mapping from codomain to domain).

\sectionl{Gradients and Duality}

As a special case of reverse-mode automatic differentiation, let's consider its use to compute \emph{gradients}, i.e., derivatives of functions with a scalar codomain, as with gradient-based optimization.
%% This case is very important for gradient-based optimization.

Given a vector space \ensuremath{\Conid{A}} over a scalar field \ensuremath{\Varid{s}}, the \emph{dual} of \ensuremath{\Conid{A}} is \ensuremath{\Conid{A}\multimap\Varid{s}}, i.e., the linear maps to the underlying field \citep[]{Lang1987LinearAlgebra}.\footnote{These linear maps are variously known as ``linear functionals'', ``linear forms'', ``one-forms'', and ``covectors''.}
This dual space is also a vector space, and when \ensuremath{\Conid{A}} has finite dimension, it is isomorphic to its dual.
In particular, every linear map in \ensuremath{\Conid{A}\multimap\Varid{s}} has the form \ensuremath{\Varid{dot}\;\Varid{u}} for some \ensuremath{\Varid{u}\mathbin{::}\Conid{A}}, where \ensuremath{\Varid{dot}} is the curried dot product:\notefoot{Maybe I don't need this isomorphism, and it suffices to consider those linear maps that do correspond to \ensuremath{\Varid{dot}\;\Varid{u}} for some \ensuremath{\Varid{u}}.}
\begin{hscode}\SaveRestoreHook
\column{B}{@{}>{\hspre}l<{\hspost}@{}}%
\column{E}{@{}>{\hspre}l<{\hspost}@{}}%
\>[B]{}\mathbf{class}\;\Conid{HasDot}^{\Varid{s}}\;\Varid{u}\;\mathbf{where}\;\Varid{dot}\mathbin{::}\Varid{u}\to (\Varid{u}\multimap\Varid{s}){}\<[E]%
\\[\blanklineskip]%
\>[B]{}\mathbf{instance}\;\Conid{HasDot}^{\mathbb{R}}\;\mathbb{R}\;\mathbf{where}\;\Varid{dot}\mathrel{=}\Varid{scale}{}\<[E]%
\\[\blanklineskip]%
\>[B]{}\mathbf{instance}\;(\Conid{HasDot}^{\Varid{s}}\;\Varid{a},\Conid{HasDot}^{\Varid{s}}\;\Varid{b})\Rightarrow \Conid{HasDot}^{\Varid{s}}\;(\Varid{a} \times \Varid{b})\;\mathbf{where}\;\Varid{dot}\;(\Varid{u},\Varid{v})\mathrel{=}\Varid{dot}\;\Varid{u}\mathbin{\triangledown}\Varid{dot}\;\Varid{v}{}\<[E]%
\ColumnHook
\end{hscode}\resethooks

The \ensuremath{\Conid{Cont}_{\Varid{k}}^{\Varid{r}}} construction from \secref{Reverse-Mode Automatic Differentiation} works for \emph{any} type/object \ensuremath{\Varid{r}}, so let's take \ensuremath{\Varid{r}} to be the scalar field \ensuremath{\Varid{s}}.
The internal representation of \ensuremath{\Conid{Cont}_{(\multimap)}^{\Varid{s}}\;\Varid{a}\;\Varid{b}} is \ensuremath{(\Varid{b}\multimap\Varid{s})\to (\Varid{a}\multimap\Varid{s})}, which is isomorphic to \ensuremath{\Varid{b}\to \Varid{a}}.
Call this representation the \emph{dual} (or ``opposite'') of \ensuremath{\Varid{k}}:
%% %format Dual = Op
\begin{hscode}\SaveRestoreHook
\column{B}{@{}>{\hspre}l<{\hspost}@{}}%
\column{E}{@{}>{\hspre}l<{\hspost}@{}}%
\>[B]{}\mathbf{newtype}\;\Conid{Dual}_{\Varid{k}}\;\Varid{a}\;\Varid{b}\mathrel{=}\Conid{Dual}\;(\Varid{b}\mathbin{`\Varid{k}`}\Varid{a}){}\<[E]%
\ColumnHook
\end{hscode}\resethooks
To construct dual representations of (generalized) linear maps, it suffices to convert from \ensuremath{\Conid{Cont}_{\Varid{k}}^{\Varid{s}}} to \ensuremath{\Conid{Dual}_{\Varid{k}}} by a functor we will now derive.
Composing this new functor with \ensuremath{\Varid{cont}\mathbin{::}(\Varid{a}\mathbin{`\Varid{k}`}\Varid{b})\to \Conid{Cont}_{\Varid{k}}^{\Varid{s}}\;\Varid{a}\;\Varid{b}} will give us a functor from \ensuremath{\Varid{k}} to \ensuremath{\Conid{Dual}_{\Varid{k}}}.
The new to-be-derived functor:
\begin{hscode}\SaveRestoreHook
\column{B}{@{}>{\hspre}l<{\hspost}@{}}%
\column{E}{@{}>{\hspre}l<{\hspost}@{}}%
\>[B]{}\Varid{asDual}\mathbin{::}(\Conid{HasDot}^{\Varid{s}}\;\Varid{a},\Conid{HasDot}^{\Varid{s}}\;\Varid{b})\Rightarrow \Conid{Cont}_{\Varid{k}}^{\Varid{s}}\;\Varid{a}\;\Varid{b}\to \Conid{Dual}_{\Varid{k}}\;\Varid{a}\;\Varid{b}{}\<[E]%
\\
\>[B]{}\Varid{asDual}\;(\Conid{Cont}\;\Varid{f})\mathrel{=}\Conid{Dual}\;(\Varid{onDot}\;\Varid{f}){}\<[E]%
\ColumnHook
\end{hscode}\resethooks
where \ensuremath{\Varid{onDot}} uses both halves of the isomorphism between \ensuremath{\Varid{a}\multimap\Varid{s}} and \ensuremath{\Varid{a}}:\out{\notefoot{Maybe drop \ensuremath{\Varid{onDot}} in favor of its definition.}}
%% %format unDot = dot"^{\scriptscriptstyle -\!1}"
\begin{hscode}\SaveRestoreHook
\column{B}{@{}>{\hspre}l<{\hspost}@{}}%
\column{E}{@{}>{\hspre}l<{\hspost}@{}}%
\>[B]{}\Varid{onDot}\mathbin{::}(\Conid{HasDot}^{\Varid{s}}\;\Varid{a},\Conid{HasDot}^{\Varid{s}}\;\Varid{b})\Rightarrow ((\Varid{b}\multimap\Varid{s})\to (\Varid{a}\multimap\Varid{s}))\to (\Varid{b}\multimap\Varid{a}){}\<[E]%
\\
\>[B]{}\Varid{onDot}\;\Varid{f}\mathrel{=}\Varid{dot}^{-1}\hsdot{\circ }{.}\Varid{f}\hsdot{\circ }{.}\Varid{dot}{}\<[E]%
\ColumnHook
\end{hscode}\resethooks

As usual, we can derive instances for our new category by homomorphic specification:
\begin{theorem}[\provedIn{theorem:asDual}]\thmLabel{asDual}
Given the definitions in \figref{asDual}, \ensuremath{\Varid{asDual}} is a homomorphism with respect to each instantiated class.
\end{theorem}
\begin{figure}
\begin{center}
\begin{hscode}\SaveRestoreHook
\column{B}{@{}>{\hspre}l<{\hspost}@{}}%
\column{4}{@{}>{\hspre}l<{\hspost}@{}}%
\column{9}{@{}>{\hspre}l<{\hspost}@{}}%
\column{10}{@{}>{\hspre}l<{\hspost}@{}}%
\column{E}{@{}>{\hspre}l<{\hspost}@{}}%
\>[B]{}\mathbf{instance}\;\Conid{Category}\;\Varid{k}\Rightarrow \Conid{Category}\;\Conid{Dual}_{\Varid{k}}\;\mathbf{where}{}\<[E]%
\\
\>[B]{}\hsindent{4}{}\<[4]%
\>[4]{}\Varid{id}\mathrel{=}\Conid{Dual}\;\Varid{id}{}\<[E]%
\\
\>[B]{}\hsindent{4}{}\<[4]%
\>[4]{}\Conid{Dual}\;\Varid{g}\hsdot{\circ }{.}\Conid{Dual}\;\Varid{f}\mathrel{=}\Conid{Dual}\;(\Varid{f}\hsdot{\circ }{.}\Varid{g}){}\<[E]%
\\[\blanklineskip]%
\>[B]{}\mathbf{instance}\;\Conid{Monoidal}\;\Varid{k}\Rightarrow \Conid{Monoidal}\;\Conid{Dual}_{\Varid{k}}\;\mathbf{where}{}\<[E]%
\\
\>[B]{}\hsindent{4}{}\<[4]%
\>[4]{}\Conid{Dual}\;\Varid{f}\times\Conid{Dual}\;\Varid{g}\mathrel{=}\Conid{Dual}\;(\Varid{f}\times\Varid{g}){}\<[E]%
\\[\blanklineskip]%
\>[B]{}\mathbf{instance}\;\Conid{Cartesian}\;\Varid{k}\Rightarrow \Conid{Cartesian}\;\Conid{Dual}_{\Varid{k}}\;\mathbf{where}{}\<[E]%
\\
\>[B]{}\hsindent{4}{}\<[4]%
\>[4]{}\Varid{exl}{}\<[9]%
\>[9]{}\mathrel{=}\Conid{Dual}\;\Varid{inl}{}\<[E]%
\\
\>[B]{}\hsindent{4}{}\<[4]%
\>[4]{}\Varid{exr}{}\<[9]%
\>[9]{}\mathrel{=}\Conid{Dual}\;\Varid{inr}{}\<[E]%
\\
\>[B]{}\hsindent{4}{}\<[4]%
\>[4]{}\Varid{dup}{}\<[9]%
\>[9]{}\mathrel{=}\Conid{Dual}\;\Varid{jam}{}\<[E]%
\\[\blanklineskip]%
\>[B]{}\mathbf{instance}\;\Conid{Cocartesian}\;\Varid{k}\Rightarrow \Conid{Cocartesian}\;\Conid{Dual}_{\Varid{k}}\;\mathbf{where}{}\<[E]%
\\
\>[B]{}\hsindent{4}{}\<[4]%
\>[4]{}\Varid{inl}{}\<[10]%
\>[10]{}\mathrel{=}\Conid{Dual}\;\Varid{exl}{}\<[E]%
\\
\>[B]{}\hsindent{4}{}\<[4]%
\>[4]{}\Varid{inr}{}\<[10]%
\>[10]{}\mathrel{=}\Conid{Dual}\;\Varid{exr}{}\<[E]%
\\
\>[B]{}\hsindent{4}{}\<[4]%
\>[4]{}\Varid{jam}{}\<[10]%
\>[10]{}\mathrel{=}\Conid{Dual}\;\Varid{dup}{}\<[E]%
\\[\blanklineskip]%
\>[B]{}\mathbf{instance}\;\Conid{Scalable}\;\Varid{k}\Rightarrow \Conid{Scalable}\;\Conid{Dual}_{\Varid{k}}\;\mathbf{where}{}\<[E]%
\\
\>[B]{}\hsindent{4}{}\<[4]%
\>[4]{}\Varid{scale}\;\Varid{s}\mathrel{=}\Conid{Dual}\;(\Varid{scale}\;\Varid{s}){}\<[E]%
\ColumnHook
\end{hscode}\resethooks
\caption{Dual category transformer (specified by functoriality of \ensuremath{\Varid{asDual}})}
\figlabel{asDual}
\end{center}
\end{figure}

Note that the instances in \figref{asDual} exactly dualize a computation, reversing sequential compositions and swapping corresponding \ensuremath{\Conid{Cartesian}} and \ensuremath{\Conid{Cocartesian}} operations.
Likewise for the derived operations:
\begin{corollary}[\provedIn{corollary:dual-derived}]\corLabel{dual-derived}
%% |Dual f &&& Dual g == Dual (f ### g)|, and |Dual f ### Dual g == Dual (f &&& g)|.
The \ensuremath{(\mathbin{\vartriangle})} and \ensuremath{(\mathbin{\triangledown})} operations mutually dualize:
$$\ensuremath{\Conid{Dual}\;\Varid{f}\mathbin{\vartriangle}\Conid{Dual}\;\Varid{g}\mathrel{=}\Conid{Dual}\;(\Varid{f}\mathbin{\triangledown}\Varid{g})}$$
$$\ensuremath{\Conid{Dual}\;\Varid{f}\mathbin{\triangledown}\Conid{Dual}\;\Varid{g}\mathrel{=}\Conid{Dual}\;(\Varid{f}\mathbin{\vartriangle}\Varid{g})}$$
\end{corollary}
Recall from \secref{Matrices}, that \ensuremath{\Varid{scale}} forms $1 \times 1$ matrices, while \ensuremath{(\mathbin{\triangledown})} and \ensuremath{(\mathbin{\vartriangle})} correspond to horizontal and vertical juxtaposition, respectively.
Thus, from a matrix perspective, duality is \emph{transposition}, turning an $m \times n$ matrix into an $n \times m$ matrix.
Note, however, that \ensuremath{\Conid{Dual}_{\Varid{k}}} involves no actual matrix computations unless \ensuremath{\Varid{k}} does.
In particular, we can simply use the category of linear functions \ensuremath{(\mathbin{\rightarrow^{\!\!+}\!})}.

\figreftwo{magSqr-gradr}{cos-xpytz-gradr} show the results of reverse-mode AD via \ensuremath{\Conid{D}_{\Conid{Dual}_{(\mathbin{\rightarrow^{\!\!+}\!})}}}.
Compare \figref{magSqr-gradr} with\out{ the same example in} \figreftwo{magSqr-adf}{magSqr-adr}.
\figp{
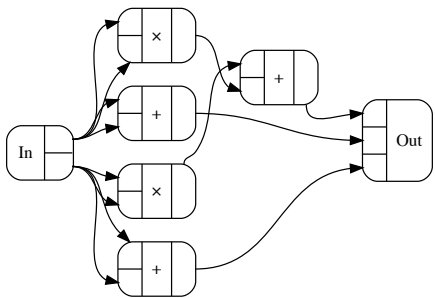
\figoneW{0.34}{magSqr-gradr}{\ensuremath{\Varid{magSqr}} in \ensuremath{\Conid{D}_{\Conid{Dual}_{(\mathbin{\rightarrow^{\!\!+}\!})}}}}}{
\figoneW{0.62}{cos-xpytz-gradr}{\ensuremath{\lambda ((\Varid{x},\Varid{y}),\Varid{z})\to \Varid{cos}\;(\Varid{x}\mathbin{+}\Varid{y}\cdot\Varid{z})} in \ensuremath{\Conid{D}_{\Conid{Dual}_{(\mathbin{\rightarrow^{\!\!+}\!})}}}}
}

\sectionl{Forward-Mode Automatic Differentiation}

It may be interesting to note that we can turn the \ensuremath{\Conid{Cont}} and \ensuremath{\Conid{Dual}} techniques around to yield category transformers that perform full \emph{right-} instead of left-association, converting the general, mode-independent algorithm into forward mode, thus yielding an algorithm preferable for low-dimensional domains (rather than codomains):
\begin{hscode}\SaveRestoreHook
\column{B}{@{}>{\hspre}l<{\hspost}@{}}%
\column{E}{@{}>{\hspre}l<{\hspost}@{}}%
\>[B]{}\mathbf{newtype}\;\Conid{Begin}_{\Varid{k}}^{\Varid{r}}\;\Varid{a}\;\Varid{b}\mathrel{=}\Conid{Begin}\;((\Varid{r}\mathbin{`\Varid{k}`}\Varid{a})\to (\Varid{r}\mathbin{`\Varid{k}`}\Varid{b})){}\<[E]%
\\[\blanklineskip]%
\>[B]{}\Varid{begin}\mathbin{::}\Conid{Category}\;\Varid{k}\Rightarrow (\Varid{a}\mathbin{`\Varid{k}`}\Varid{b})\to \Conid{Begin}_{\Varid{k}}^{\Varid{r}}\;\Varid{a}\;\Varid{b}{}\<[E]%
\\
\>[B]{}\Varid{begin}\;\Varid{f}\mathrel{=}\Conid{Begin}\;(\Varid{f}\:{}\hsdot{\circ }{.}){}\<[E]%
\ColumnHook
\end{hscode}\resethooks
As usual, we can derive instances for our new category by homomorphic specification (for \ensuremath{\Varid{begin}}).
Then choose \ensuremath{\Varid{r}} to be the scalar field \ensuremath{\Varid{s}}, as in \secref{Gradients and Duality}, noting that \ensuremath{(\Varid{s}\multimap\Varid{a}) \cong \Varid{a}}.

\sectionl{Scaling Up}

So far, we have considered binary products.
Practical applications, including machine learning and other optimization problems, often involve very high-dimensional spaces.
While those spaces can be encoded as nested binary products, doing so would result in unwieldy representations and prohibitively long compilation and execution times.
A practical alternative is to consider $n$-ary products, for which we can again use representable functors.
To construct and consume these ``indexed'' (bi)products, we'll need an indexed variant of \ensuremath{\Conid{Monoidal}}, replacing the two arguments to \ensuremath{(\times)} by a (representable) functor \ensuremath{\Varid{h}} of morphisms:
\\
\begin{minipage}[b]{0.49\textwidth} % \mathindent1em
\begin{hscode}\SaveRestoreHook
\column{B}{@{}>{\hspre}l<{\hspost}@{}}%
\column{3}{@{}>{\hspre}l<{\hspost}@{}}%
\column{E}{@{}>{\hspre}l<{\hspost}@{}}%
\>[B]{}\mathbf{class}\;\Conid{Category}\;\Varid{k}\Rightarrow \Conid{MonoidalI}\;\Varid{k}\;\Varid{h}\;\mathbf{where}{}\<[E]%
\\
\>[B]{}\hsindent{3}{}\<[3]%
\>[3]{}\Varid{crossI}\mathbin{::}\Varid{h}\;(\Varid{a}\mathbin{`\Varid{k}`}\Varid{b})\to (\Varid{h}\;\Varid{a}\mathbin{`\Varid{k}`}\Varid{h}\;\Varid{b}){}\<[E]%
\ColumnHook
\end{hscode}\resethooks
\end{minipage}
\begin{minipage}[b]{0ex}{\rule[1ex]{0.5pt}{0.3in}}\end{minipage}
\begin{minipage}[b]{0.48\textwidth} % \mathindent1em
\begin{hscode}\SaveRestoreHook
\column{B}{@{}>{\hspre}l<{\hspost}@{}}%
\column{3}{@{}>{\hspre}l<{\hspost}@{}}%
\column{E}{@{}>{\hspre}l<{\hspost}@{}}%
\>[B]{}\mathbf{instance}\;\Conid{Zip}\;\Varid{h}\Rightarrow \Conid{MonoidalI}\;(\to )\;\Varid{h}\;\mathbf{where}{}\<[E]%
\\
\>[B]{}\hsindent{3}{}\<[3]%
\>[3]{}\Varid{crossI}\mathrel{=}\Varid{zipWith}\;\Varid{id}{}\<[E]%
\ColumnHook
\end{hscode}\resethooks
\end{minipage}
\\
Note that the collected morphisms must all agree in domain and codomain.
While not required for the general categorical notion of products, this restriction accommodates Haskell's type system and seems adequate in practice so far.

Where the \ensuremath{\Conid{Cartesian}} class has two projection methods and a duplication method, the indexed counterpart has a collection of projections and one replication method:%
\footnote{The \ensuremath{\Varid{index}} and \ensuremath{\Varid{tabulate}} functions convert from functor to function and back \citep{Kmett2011Adj}.}

\begin{hscode}\SaveRestoreHook
\column{B}{@{}>{\hspre}l<{\hspost}@{}}%
\column{3}{@{}>{\hspre}l<{\hspost}@{}}%
\column{10}{@{}>{\hspre}l<{\hspost}@{}}%
\column{E}{@{}>{\hspre}l<{\hspost}@{}}%
\>[B]{}\mathbf{class}\;\Conid{MonoidalI}\;\Varid{k}\;\Varid{h}\Rightarrow \Conid{CartesianI}\;\Varid{k}\;\Varid{h}\;\mathbf{where}{}\<[E]%
\\
\>[B]{}\hsindent{3}{}\<[3]%
\>[3]{}\Varid{exI}{}\<[10]%
\>[10]{}\mathbin{::}\Varid{h}\;(\Varid{h}\;\Varid{a}\mathbin{`\Varid{k}`}\Varid{a}){}\<[E]%
\\
\>[B]{}\hsindent{3}{}\<[3]%
\>[3]{}\Varid{replI}{}\<[10]%
\>[10]{}\mathbin{::}\Varid{a}\mathbin{`\Varid{k}`}\Varid{h}\;\Varid{a}{}\<[E]%
\\[\blanklineskip]%
\>[B]{}\mathbf{instance}\;(\Conid{Representable}\;\Varid{h},\Conid{Zip}\;\Varid{h},\Conid{Pointed}\;\Varid{h})\Rightarrow \Conid{CartesianI}\;(\to )\;\Varid{h}\;\mathbf{where}{}\<[E]%
\\
\>[B]{}\hsindent{3}{}\<[3]%
\>[3]{}\Varid{exI}{}\<[10]%
\>[10]{}\mathrel{=}\Varid{tabulate}\;(\Varid{flip}\;\Varid{index}){}\<[E]%
\\
\>[B]{}\hsindent{3}{}\<[3]%
\>[3]{}\Varid{replI}{}\<[10]%
\>[10]{}\mathrel{=}\Varid{point}{}\<[E]%
\ColumnHook
\end{hscode}\resethooks
Dually, where the \ensuremath{\Conid{Cocartesian}} class has two injection methods and a binary combination method, the indexed counterpart has a collection of injections and one collection-combining method:
\begin{hscode}\SaveRestoreHook
\column{B}{@{}>{\hspre}l<{\hspost}@{}}%
\column{3}{@{}>{\hspre}l<{\hspost}@{}}%
\column{10}{@{}>{\hspre}l<{\hspost}@{}}%
\column{E}{@{}>{\hspre}l<{\hspost}@{}}%
\>[B]{}\mathbf{class}\;\Conid{MonoidalI}\;\Varid{k}\;\Varid{h}\Rightarrow \Conid{CocartesianI}\;\Varid{k}\;\Varid{h}\;\mathbf{where}{}\<[E]%
\\
\>[B]{}\hsindent{3}{}\<[3]%
\>[3]{}\Varid{inI}{}\<[10]%
\>[10]{}\mathbin{::}\Varid{h}\;(\Varid{a}\mathbin{`\Varid{k}`}\Varid{h}\;\Varid{a}){}\<[E]%
\\
\>[B]{}\hsindent{3}{}\<[3]%
\>[3]{}\Varid{jamI}{}\<[10]%
\>[10]{}\mathbin{::}\Varid{h}\;\Varid{a}\mathbin{`\Varid{k}`}\Varid{a}{}\<[E]%
\ColumnHook
\end{hscode}\resethooks

\noindent
There are also indexed variants of the derived operations \ensuremath{(\mathbin{\vartriangle})} and \ensuremath{(\mathbin{\triangledown})} from \secref{Derived Operations}:
\begin{hscode}\SaveRestoreHook
\column{B}{@{}>{\hspre}l<{\hspost}@{}}%
\column{E}{@{}>{\hspre}l<{\hspost}@{}}%
\>[B]{}\Varid{forkI}\mathbin{::}\Conid{CartesianI}\;\Varid{k}\;\Varid{h}\Rightarrow \Varid{h}\;(\Varid{a}\mathbin{`\Varid{k}`}\Varid{b})\to (\Varid{a}\mathbin{`\Varid{k}`}\Varid{h}\;\Varid{b}){}\<[E]%
\\
\>[B]{}\Varid{forkI}\;\Varid{fs}\mathrel{=}\Varid{crossI}\;\Varid{fs}\hsdot{\circ }{.}\Varid{replI}{}\<[E]%
\\[\blanklineskip]%
\>[B]{}\Varid{unforkF}\mathbin{::}\Conid{CartesianI}\;\Varid{k}\;\Varid{h}\Rightarrow (\Varid{a}\mathbin{`\Varid{k}`}\Varid{h}\;\Varid{b})\to \Varid{h}\;(\Varid{a}\mathbin{`\Varid{k}`}\Varid{b}){}\<[E]%
\\
\>[B]{}\Varid{unforkF}\;\Varid{f}\mathrel{=}\Varid{fmap}\;(\hsdot{\circ }{.}\:{}\Varid{f})\;\Varid{exI}{}\<[E]%
\\[\blanklineskip]%
\>[B]{}\Varid{joinI}\mathbin{::}\Conid{CartesianI}\;\Varid{k}\;\Varid{h}\Rightarrow \Varid{h}\;(\Varid{b}\mathbin{`\Varid{k}`}\Varid{a})\to (\Varid{h}\;\Varid{b}\mathbin{`\Varid{k}`}\Varid{a}){}\<[E]%
\\
\>[B]{}\Varid{joinI}\;\Varid{fs}\mathrel{=}\Varid{jamI}\hsdot{\circ }{.}\Varid{crossI}\;\Varid{fs}{}\<[E]%
\\[\blanklineskip]%
\>[B]{}\Varid{unjoinPF}\mathbin{::}\Conid{CocartesianI}\;\Varid{k}\;\Varid{h}\Rightarrow (\Varid{h}\;\Varid{b}\mathbin{`\Varid{k}`}\Varid{a})\to \Varid{h}\;(\Varid{b}\mathbin{`\Varid{k}`}\Varid{a}){}\<[E]%
\\
\>[B]{}\Varid{unjoinPF}\;\Varid{f}\mathrel{=}\Varid{fmap}\;(\Varid{f}\:{}\hsdot{\circ }{.})\;\Varid{inI}{}\<[E]%
\ColumnHook
\end{hscode}\resethooks

As usual, we can derive instances by homomorphic specification:
\begin{theorem}[\provedIn{theorem:indexed}]\thmLabel{indexed}
Given the definitions in \figref{indexed}, \ensuremath{\hat{\mathcal{D}}} is a homomorphism with respect to each instantiated class.
\end{theorem}
%% \figref{indexed} assumes the following definitions:
%% \begin{code}
%% \end{code}
\begin{figure}
\begin{center}
\begin{hscode}\SaveRestoreHook
\column{B}{@{}>{\hspre}l<{\hspost}@{}}%
\column{3}{@{}>{\hspre}l<{\hspost}@{}}%
\column{10}{@{}>{\hspre}l<{\hspost}@{}}%
\column{E}{@{}>{\hspre}l<{\hspost}@{}}%
\>[B]{}\mathbf{instance}\;(\Conid{MonoidalI}\;\Varid{k}\;\Varid{h},\Conid{Zip}\;\Varid{h})\Rightarrow \Conid{MonoidalI}\;\Conid{D}_{\Varid{k}}\;\Varid{h}\;\mathbf{where}{}\<[E]%
\\
\>[B]{}\hsindent{3}{}\<[3]%
\>[3]{}\Varid{crossI}\;\Varid{fs}\mathrel{=}\Conid{D}\;(\Varid{second}\;\Varid{crossI}\hsdot{\circ }{.}\Varid{unzip}\hsdot{\circ }{.}\Varid{crossI}\;(\Varid{fmap}\;\Varid{unD}\;\Varid{fs})){}\<[E]%
\\[\blanklineskip]%
\>[B]{}\mathbf{instance}\;(\Conid{CartesianI}\;(\to )\;\Varid{h},\Conid{CartesianI}\;\Varid{k}\;\Varid{h},\Conid{Zip}\;\Varid{h})\Rightarrow \Conid{CartesianI}\;\Conid{D}_{\Varid{k}}\;\Varid{h}\;\mathbf{where}{}\<[E]%
\\
\>[B]{}\hsindent{3}{}\<[3]%
\>[3]{}\Varid{exI}{}\<[10]%
\>[10]{}\mathrel{=}\Varid{linearD}\;\Varid{exI}\;\Varid{exI}{}\<[E]%
\\
\>[B]{}\hsindent{3}{}\<[3]%
\>[3]{}\Varid{replI}{}\<[10]%
\>[10]{}\mathrel{=}\Varid{zipWith}\;\Varid{linearD}\;\Varid{replI}\;\Varid{replI}{}\<[E]%
\\[\blanklineskip]%
\>[B]{}\mathbf{instance}\;(\Conid{CocartesianI}\;\Varid{k}\;\Varid{h},\Conid{Zip}\;\Varid{h})\Rightarrow \Conid{CocartesianI}\;\Conid{D}_{\Varid{k}}\;\Varid{h}\;\mathbf{where}{}\<[E]%
\\
\>[B]{}\hsindent{3}{}\<[3]%
\>[3]{}\Varid{inI}{}\<[10]%
\>[10]{}\mathrel{=}\Varid{zipWith}\;\Varid{linearD}\;\Varid{inIF}\;\Varid{inI}{}\<[E]%
\\
\>[B]{}\hsindent{3}{}\<[3]%
\>[3]{}\Varid{jamI}{}\<[10]%
\>[10]{}\mathrel{=}\Varid{linearD}\;\Varid{sum}\;\Varid{jamI}{}\<[E]%
\\[\blanklineskip]%
\>[B]{}\Varid{unD}\mathbin{::}\Conid{D}\;\Varid{a}\;\Varid{b}\to (\Varid{a}\to (\Varid{b} \times (\Varid{a}\multimap\Varid{b}))){}\<[E]%
\\
\>[B]{}\Varid{unD}\;(\Conid{D}\;\Varid{f})\mathrel{=}\Varid{f}{}\<[E]%
\\[\blanklineskip]%
\>[B]{}\Varid{unzip}\mathbin{::}\Conid{Functor}\;\Varid{h}\Rightarrow \Varid{h}\;(\Varid{a} \times \Varid{b})\to \Varid{h}\;\Varid{a} \times \Varid{h}\;\Varid{b}{}\<[E]%
\\
\>[B]{}\Varid{unzip}\mathrel{=}\Varid{fmap}\;\Varid{exl}\mathbin{\vartriangle}\Varid{fmap}\;\Varid{exr}{}\<[E]%
\\[\blanklineskip]%
\>[B]{}\Varid{second}\mathbin{::}\Conid{Monoidal}\;\Varid{k}\Rightarrow (\Varid{b}\mathbin{`\Varid{k}`}\Varid{d})\to ((\Varid{a} \times \Varid{b})\mathbin{`\Varid{k}`}(\Varid{a} \times \Varid{d})){}\<[E]%
\\
\>[B]{}\Varid{second}\;\Varid{g}\mathrel{=}\Varid{id}\times\Varid{g}{}\<[E]%
\\[\blanklineskip]%
\>[B]{}\Varid{inIF}\mathbin{::}(\Conid{Additive}\;\Varid{a},\Conid{Foldable}\;\Varid{h})\Rightarrow \Varid{h}\;(\Varid{a}\to \Varid{h}\;\Varid{a}){}\<[E]%
\\
\>[B]{}\Varid{inIF}\mathrel{=}\Varid{tabulate}\;(\lambda \Varid{i}\;\Varid{a}\to \Varid{tabulate}\;(\lambda \Varid{j}\to \mathbf{if}\;\Varid{i}\mathrel{=}\Varid{j}\;\mathbf{then}\;\Varid{a}\;\mathbf{else}\;\mathrm{0})){}\<[E]%
\\[\blanklineskip]%
\>[B]{}\mathbf{class}\;\Conid{Zip}\;\Varid{h}\;\mathbf{where}\;\Varid{zipWith}\mathbin{::}(\Varid{a}\to \Varid{b}\to \Varid{c})\to \Varid{h}\;\Varid{a}\to \Varid{h}\;\Varid{b}\to \Varid{h}\;\Varid{c}{}\<[E]%
\ColumnHook
\end{hscode}\resethooks
\caption{AD for indexed biproducts}
\figlabel{indexed}
\end{center}
\end{figure}

These indexed operations are useful in themselves but can be used to derive other operations.
For instance, note the similarity between the types of \ensuremath{\Varid{crossI}} and \ensuremath{\Varid{fmap}}:
\begin{hscode}\SaveRestoreHook
\column{B}{@{}>{\hspre}l<{\hspost}@{}}%
\column{9}{@{}>{\hspre}l<{\hspost}@{}}%
\column{26}{@{}>{\hspre}l<{\hspost}@{}}%
\column{34}{@{}>{\hspre}l<{\hspost}@{}}%
\column{E}{@{}>{\hspre}l<{\hspost}@{}}%
\>[B]{}\Varid{crossI}{}\<[9]%
\>[9]{}\mathbin{::}\Conid{Monoidal}\;{}\<[26]%
\>[26]{}\Varid{h}\Rightarrow \Varid{h}\;{}\<[34]%
\>[34]{}(\Varid{a}\to \Varid{b})\to (\Varid{h}\;\Varid{a}\to \Varid{h}\;\Varid{b}){}\<[E]%
\\
\>[B]{}\Varid{fmap}{}\<[9]%
\>[9]{}\mathbin{::}\Conid{Functor}\;{}\<[26]%
\>[26]{}\Varid{h}\Rightarrow {}\<[34]%
\>[34]{}(\Varid{a}\to \Varid{b})\to (\Varid{h}\;\Varid{a}\to \Varid{h}\;\Varid{b}){}\<[E]%
\ColumnHook
\end{hscode}\resethooks
In fact, the following relationship holds: \ensuremath{\Varid{fmap}\mathrel{=}\Varid{crossI}\hsdot{\circ }{.}\Varid{replI}}.
This equation, together with the differentiation rules for \ensuremath{\Varid{crossI}}, \ensuremath{\Varid{replI}}, and \ensuremath{(\hsdot{\circ }{.})} determines differentiation for \ensuremath{\Varid{fmap}\;\Varid{f}}.

As with \figref{GAD}, the operations defined in \figref{indexed} rely on corresponding operations for the category parameter \ensuremath{\Varid{k}}.
Fortunately, all of those operations are linear or preserve linearity, so they can all be defined on the various representations of derivatives (linear maps) used for AD in this paper.
\figreftwo{ixAdditive}{ixDual} show instances for two of the linear map representations defined in this paper, with the third left as an exercise for the reader.
({The constraint \ensuremath{\Conid{Additive}_{1}\;\Varid{h}} means that \ensuremath{\forall \Varid{a}\hsforall \hsdot{\circ }{.}\Conid{Additive}\;\Varid{a}\Rightarrow \Conid{Additive}\;(\Varid{h}\;\Varid{a})}.})
\begin{figure}
\begin{center}
\begin{hscode}\SaveRestoreHook
\column{B}{@{}>{\hspre}l<{\hspost}@{}}%
\column{3}{@{}>{\hspre}l<{\hspost}@{}}%
\column{9}{@{}>{\hspre}l<{\hspost}@{}}%
\column{10}{@{}>{\hspre}l<{\hspost}@{}}%
\column{E}{@{}>{\hspre}l<{\hspost}@{}}%
\>[B]{}\mathbf{instance}\;(\Conid{Zip}\;\Varid{h},\Conid{Additive}_{1}\;\Varid{h})\Rightarrow \Conid{MonoidalI}\;(\mathbin{\rightarrow^{\!\!+}\!})\;\Varid{h}\;\mathbf{where}{}\<[E]%
\\
\>[B]{}\hsindent{3}{}\<[3]%
\>[3]{}\Varid{crossI}\mathrel{=}\Conid{AddFun}\hsdot{\circ }{.}\Varid{crossI}\hsdot{\circ }{.}\Varid{fmap}\;\Varid{unAddFun}{}\<[E]%
\\[\blanklineskip]%
\>[B]{}\mathbf{instance}\;(\Conid{Representable}\;\Varid{h},\Conid{Zip}\;\Varid{h},\Conid{Pointed}\;\Varid{h},\Conid{Additive}_{1}\;\Varid{h})\Rightarrow \Conid{CartesianI}\;(\mathbin{\rightarrow^{\!\!+}\!})\;\Varid{h}\;\mathbf{where}{}\<[E]%
\\
\>[B]{}\hsindent{3}{}\<[3]%
\>[3]{}\Varid{exI}{}\<[9]%
\>[9]{}\mathrel{=}\Varid{fmap}\;\Conid{AddFun}\;\Varid{exI}{}\<[E]%
\\
\>[B]{}\hsindent{3}{}\<[3]%
\>[3]{}\Varid{replI}\mathrel{=}\Conid{AddFun}\;\Varid{replI}{}\<[E]%
\\[\blanklineskip]%
\>[B]{}\mathbf{instance}\;(\Conid{Foldable}\;\Varid{h},\Conid{Additive}_{1}\;\Varid{h})\Rightarrow \Conid{CocartesianI}\;(\mathbin{\rightarrow^{\!\!+}\!})\;\Varid{h}\;\mathbf{where}{}\<[E]%
\\
\>[B]{}\hsindent{3}{}\<[3]%
\>[3]{}\Varid{inI}{}\<[10]%
\>[10]{}\mathrel{=}\Varid{fmap}\;\Conid{AddFun}\;\Varid{inIF}{}\<[E]%
\\
\>[B]{}\hsindent{3}{}\<[3]%
\>[3]{}\Varid{jamI}{}\<[10]%
\>[10]{}\mathrel{=}\Conid{AddFun}\;\Varid{sum}{}\<[E]%
\ColumnHook
\end{hscode}\resethooks
\vspace{-3ex}
\caption{Indexed instances for \ensuremath{(\mathbin{\rightarrow^{\!\!+}\!})}}
\figlabel{ixAdditive}
\end{center}
\end{figure}
\begin{figure}
\begin{center}
\begin{hscode}\SaveRestoreHook
\column{B}{@{}>{\hspre}l<{\hspost}@{}}%
\column{3}{@{}>{\hspre}l<{\hspost}@{}}%
\column{10}{@{}>{\hspre}l<{\hspost}@{}}%
\column{E}{@{}>{\hspre}l<{\hspost}@{}}%
\>[B]{}\mathbf{instance}\;(\Conid{MonoidalI}\;\Varid{k}\;\Varid{h},\Conid{Functor}\;\Varid{h},\Conid{Additive}_{1}\;\Varid{h})\Rightarrow \Conid{MonoidalI}\;(\Conid{Dual}\;\Varid{k})\;\Varid{h}\;\mathbf{where}{}\<[E]%
\\
\>[B]{}\hsindent{3}{}\<[3]%
\>[3]{}\Varid{crossI}\mathrel{=}\Conid{Dual}\hsdot{\circ }{.}\Varid{crossI}\hsdot{\circ }{.}\Varid{fmap}\;\Varid{unDual}{}\<[E]%
\\[\blanklineskip]%
\>[B]{}\mathbf{instance}\;(\Conid{CocartesianI}\;\Varid{k}\;\Varid{h},\Conid{Functor}\;\Varid{h},\Conid{Additive}_{1}\;\Varid{h})\Rightarrow \Conid{CartesianI}\;(\Conid{Dual}\;\Varid{k})\;\Varid{h}\;\mathbf{where}{}\<[E]%
\\
\>[B]{}\hsindent{3}{}\<[3]%
\>[3]{}\Varid{exI}{}\<[10]%
\>[10]{}\mathrel{=}\Varid{fmap}\;\Conid{Dual}\;\Varid{inI}{}\<[E]%
\\
\>[B]{}\hsindent{3}{}\<[3]%
\>[3]{}\Varid{replI}{}\<[10]%
\>[10]{}\mathrel{=}\Conid{Dual}\;\Varid{jamI}{}\<[E]%
\\[\blanklineskip]%
\>[B]{}\mathbf{instance}\;(\Conid{CartesianI}\;\Varid{k}\;\Varid{h},\Conid{Functor}\;\Varid{h},\Conid{Additive}_{1}\;\Varid{h})\Rightarrow \Conid{CocartesianI}\;(\Conid{Dual}\;\Varid{k})\;\Varid{h}\;\mathbf{where}{}\<[E]%
\\
\>[B]{}\hsindent{3}{}\<[3]%
\>[3]{}\Varid{inI}{}\<[10]%
\>[10]{}\mathrel{=}\Varid{fmap}\;\Conid{Dual}\;\Varid{exI}{}\<[E]%
\\
\>[B]{}\hsindent{3}{}\<[3]%
\>[3]{}\Varid{jamI}{}\<[10]%
\>[10]{}\mathrel{=}\Conid{Dual}\;\Varid{replI}{}\<[E]%
\ColumnHook
\end{hscode}\resethooks
\vspace{-3ex}
\caption{Indexed instances for \ensuremath{\Conid{Dual}_{\Varid{k}}}}
\figlabel{ixDual}
\end{center}
\end{figure}

\mynote{Discuss other bulk operations (\ensuremath{\Varid{zipWith}}, etc)?}

\sectionl{Related Work}

The literature on automatic differentiation is vast, beginning with forward mode \citep{Wengert64} and later reverse mode \citep{Speelpenning:1980:CFP,Rall1981Automatic}, with many developments since \citep{Griewank89onAD,GriewankWalther2008EvalDerivs}.
While most techniques and uses of AD have been directed at imperative programming, there are also variations for functional programs \citep{Karczmarczuk1999FunCoding,Karczmarczuk00adjointcodes,Karczmarczuk2001FunDif,Pearlmutter2007LMH,Pearlmutter2008RAF,Elliott2009-beautiful-differentiation}.
The work in this paper differs in being phrased at the level of functions/morphisms and specified by functoriality, without any mention or manipulation of graphs or other syntactic representations.\footnote{Of course the Haskell compiler itself manipulates syntax trees, and the compiler plugin that converts Haskell code to categorical form helps do so, but both are entirely domain-independent, with no knowledge of or special support for differentiation or linear algebra \citep{Elliott-2017-compiling-to-categories}.}
Moreover, the specifications in this paper are simple enough that the various forms of AD presented can be calculated into being \citep{BirddeMoor96AOP,Oliveira2018Calc}, and so are correct by construction%
.

\citet{Pearlmutter2008RAF} make the following observation:
\begin{quotation}\noindent
In this context, reverse-mode AD refers to a particular construction in which the primal data-flow graph is transformed to construct an adjoint graph that computes the sensitivity values. In the adjoint, the direction of the data-flow edges are reversed; addition nodes are replaced by fanout nodes; fanout nodes are replaced by addition nodes; and other nodes are replaced by multiplication by their linearizations. The main constructions of this paper can, in this context, be viewed as a method for constructing scaffolding that supports this adjoint computation.
\end{quotation}
The \ensuremath{\Conid{Cont}} and \ensuremath{\Conid{Dual}} category transformers described in \secreftwo{Reverse-Mode Automatic Differentiation}{Gradients and Duality} (shown in \figreftwo{cont}{asDual}) above explain this ``adjoint graph'' construction without involving graphs.
Data-flow edge reversal corresponds to the reversal of \ensuremath{(\hsdot{\circ }{.})} (from \ensuremath{\Conid{Category}}), while fanout and addition correspond to \ensuremath{\Varid{dup}} and \ensuremath{\Varid{jam}} (from \ensuremath{\Conid{Cartesian}} and \ensuremath{\Conid{Cocartesian}} respectively), which are mutually dual.
\citet{Pearlmutter2008RAF} further remark:
\begin{quotation}\noindent
The main technical difficulty to be faced is that reverse-mode AD must convert fanout (multiple use of a variable) in the untransformed code into addition in the reverse phase of the transformed code. We address this by expressing all straight-line code segments in A-normal form, which makes fanout lexically apparent. 
\end{quotation}
The categorical approach in this paper also makes fanout easily apparent, as occurrences of \ensuremath{\Varid{dup}}, which are produced during translation from Haskell to categorical form \citep{Elliott-2017-compiling-to-categories} (via \ensuremath{(\mathbin{\vartriangle})} as defined in \secref{Derived Operations} above).
This translation is specified and implemented independently of AD and so presents no additional complexity.

Closely related to our choice of derivatives as linear maps and their categorical generalizations is the work of \citet{MacedoOliveira2013Typing}, also based on biproducts (though not addressing differentiation).
That work uses natural numbers as categorical objects to capture the dimensions of vectors and matrices, while the current paper uses vector spaces themselves.
The difference is perhaps minor, however, since natural numbers can be thought of as representing finite sets (of corresponding cardinality), which are \emph{bases} of finite-dimensional free vector spaces (as in \secref{Generalized Matrices}).
On the other hand, the duality-based gradient algorithm of \secref{Gradients and Duality} involves no matrices at all in their traditional representation (arrays of numbers) or generalized sense of \secref{Generalized Matrices} (representable functors).

Also sharing a categorical style is the work of \citet{Fong2017BackpropAF}, formulating the backpropropagation algorithm as a functor.
That work, which also uses biproducts (in monoidal but not cartesian form), does not appear to be separable from the application to machine learning, and so would seem to complement this paper.
Backpropagation is a specialization of reverse-mode AD to the context of machine learning, discovered by \citet{Linnainmaa1970MS} and popularized by \citet{Rumelhart1988backprop}.

The continuation transformation of \secref{Reverse-Mode Automatic Differentiation} was inspired by Mitch Wand's work on continuation-based program transformation \citep{Wand80continuation-basedprogram}.
He derived a variety of algorithms based on a single elegant technique: transform a simple recursive program into continuation-passing form, examine the continuations that arise, and find a data (rather than function) representation for them.
Each such representation is a monoid, with its identity and associative operation corresponding to identity and composition of the continuations.
Monoids are categories with only one object, but the technique extends to general categories.
Cayley's theorem for groups (or monoids) captures this same insight and is a corollary (in retrospect) of the Yoneda lemma \cite[Section 2.2]{Riehl2016category}.
The idea of using data representations for functions (``defunctionalization'') was pioneered by \citet{Reynolds72definitionalinterpreters} and further explored by \citet{Danvy2001DW}.

The notion of derivatives as linear maps is the basis of calculus on manifolds \cite{Spivak65} and was also used for AD by \citet{Elliott2009-beautiful-differentiation}.
The latter addressed only forward-mode AD but also included all orders of derivatives.

While there are many forward-mode AD libraries for Haskell, reverse mode (RAD) has been much more difficult.
The most successful implementation appears to be in the \emph{ad} library \citep{Kmett2010AD}.
One RAD implementation in that library uses stable names \citep{PeytonJones99Stretching} and reification \citep{Gill2009TOS} to recover sharing information.
Another maintains a Wengert list (or ``tape'') with the help of a reflection library \citep{Kiselyov2004FPI}.
Both implementations rely on hidden, carefully crafted use of side effects.

Chris \citet{Olah2015NNTFP} shared a vision for ``differentiable functional programming'' similar to that in \secref{Introduction}.
He pointed out that most of the patterns now used in machine learning are already found in functional programming:
\begin{quotation}
These neural network patterns are just higher order functions---that is, functions which take functions as arguments. Things like that have been studied extensively in functional programming. In fact, many of these network patterns correspond to extremely common functions, like fold. The only unusual thing is that, instead of receiving normal functions as arguments, they receive chunks of neural network.
\end{quotation}
The current paper carries this perspective further, suggesting that the essence is \emph{differentiable functions}, with ``networks'' (graphs) being an unnecessary (and apparently unwise) operational choice.

This paper builds on a compiler plugin that translates Haskell programs into categorical form to be specialized to various specific categories, including differentiable functions \citep{Elliott-2017-compiling-to-categories}.
(The plugin knows nothing about any specific category, including differentiable functions.)
Another instance of generalized AD given there is automatic incremental evaluation of functional programs.
Relative to that work, the new contributions are the \ensuremath{\Conid{Cont}_{\Varid{k}}^{\Varid{r}}} and \ensuremath{\Conid{Dual}_{\Varid{k}}} categories, their use to succinctly implement reverse-mode AD (by instantiating the generalized differentiation category), the precise specification of instances for \ensuremath{\Conid{D}}, \ensuremath{\Conid{Cont}_{\Varid{k}}^{\Varid{r}}}, and \ensuremath{\Conid{Dual}_{\Varid{k}}} via functoriality, and the calculation of implementations from these specifications.

The implementations in this paper are quite simple and appear to be efficient as well.
For instance, the duality-based version (\secref{Gradients and Duality}) involves no matrices.
Moreover, typical reverse-mode AD (RAD) implementations use mutation to incrementally update derivative contributions from each \emph{use} of a variable or intermediate computation, holding onto all of these accumulators until the very end of the derivative computation.
For this reason, such implementations tend to use considerable memory\needcite.
In contrast, the implementations in this paper (\secreftwo{Reverse-Mode Automatic Differentiation}{Gradients and Duality}) are free of mutation and can easily free (reuse) memory as they run, keeping memory use low.
Given the prominent use of AD, particularly with large data, performance is crucial, so it will be worthwhile to examine and compare time and space use in detail.
Lack of mutation also makes the algorithms in this paper naturally parallel, potentially leading to considerable speed improvement, especially when using the functor-level (bulk) vocabulary in \secref{Scaling Up}.

\sectionl{Conclusions}

This paper develops a simple, mode-independent algorithm for automatic differentiation (AD) (\secref{Putting the Pieces Together}), calculated from a simple, natural specification in terms of elementary category theory (functoriality).
It then generalizes the algorithm, replacing linear maps (as derivatives) by an arbitrary biproduct category (\figref{GAD}).
Specializing this general algorithm to two well-known categorical constructions (\figreftwo{cont}{asDual})---also calculated---yields reverse-mode AD (RAD) for general derivatives and for gradients.
These RAD implementations are far simpler than previously known.
In contrast to common approaches to AD, the new algorithms involve no graphs, tapes, variables, partial derivatives, or mutation, and are usable directly from an existing programming language with no need for new data types or programming style (thanks to use of an AD-agnostic compiler plugin).
Only the simple essence remains.

Future work includes detailed performance analysis (compared with backpropagation and other conventional AD algorithms); efficient higher-order differentiation; and applying generalized AD to derivative-like notions, including subdifferentiation \citep{Rockafellar1966} and automatic incrementalization (continuing previous work \citep{Elliott-2017-compiling-to-categories}).

AD is typically said to be about the chain rule for sequential composition (\thmRef{compose})\needcite.
This paper rounds out the story with two more rules: one for parallel composition and one for all linear operations (\thmRefTwo{cross}{linear}).
Parallel composition is usually left implicit in the special-case treatment of a collection of non-unary operations, such as addition, multiplication, division, and dot products.
With explicit, general support for parallel composition, all operations come to be on equal footing, regardless of arity (as illustrated in \figref{GAD}).

AD is also typically presented in opposition to symbolic differentiation (SD), with the latter described as applying differentiation rules symbolically.
The main criticism of SD is that it can blow up expressions, resulting a great deal of redundant work\needcite.
Secondly, SD requires implementation of symbolic manipulation as in a computer algebra system.
In contrast, AD is described as a numeric method and can retain the complexity of the original function (within a small constant factor) if carefully implemented, as in reverse mode.
The approach explored in this paper suggests a different perspective: automatic differentiation \emph{is} symbolic differentiation performed by a compiler.
Compilers already work symbolically and already take care to preserve sharing in computations, addressing both criticisms of SD.

The specification and implementation of AD in a simple, correct-by-construction, and apparently efficient manner, together with its use from a typed functional language (here via a compiler plugin), make a step toward the vision of differentiable functional programming for machine learning and other uses, as outlined in \secref{Introduction}.
Programmers then define their functions just as they are accustomed, differentiating where desired, without the intrusion of operational notions such as graphs with questionably defined, extralinguistic semantics.

In retrospect, two key principles enable the results in this paper:
\begin{enumerate}
\item
  Focus on abstract notions (specified denotationally and/or axiomatically) rather than particular representations (here, derivatives as linear maps rather than as matrices).
  Then transform a correct, naive representation into subtler, more efficient representations.
\item
  Capture the main concepts of interest directly, as first-class values (here, differentiable functions).
\end{enumerate}
The second principle leads us into a quandary, because most programming languages (including Haskell) are much better suited to expressing regular computable functions than other function-like things, and yet the main AD concept is exactly a function-like thing (differentiable functions).
This imbalance in suitability stems from built-in language support for functions---such as lambda, application, and variables---and would seem to explain two common strategies in AD: use of explicit graph representations (complicating use and implementation), and overloading numeric operators (abandoning the second principle, encouraging \emph{forward} mode AD, and leading to incorrect nested differentiation \citep{Siskind2008nesting-forward-mode-AD}).
Fortunately, we can instead extend the notational convenience of functions to other function-like things by
writing in a conventional functional language and automatically translating to other categories \citep{Elliott-2017-compiling-to-categories}.

\sectionl{Acknowledgments}

The investigation of reverse-mode AD and its specialization to scalar-valued functions (as in backpropagation) were inspired by a conversation with Wang Ruikang.

\appendix

\vspace{2ex}

\mynote{The appendices that follow appear in the extended version of this paper and are cited from the shorter, conference version.}

\sectionl{Terminal and Initial Objects}

In the biproduct setting of this paper, terminal and initial objects coincide and may be taken to be any singleton type.
We may as well choose the unit type, having exactly one element, representing a canonical zero-dimensional vector space, and written ``\ensuremath{()}'' in Haskell:\footnote{In a more general categorical setting, terminal and initial objects need not coincide and are defined per category.}\footnote{As with \ensuremath{\Conid{Cocartesian}},  in the actual implementation, the \ensuremath{\Conid{Initial}} definition has no \ensuremath{\Conid{Additive}} constraint or \ensuremath{\Conid{Initial}\;(\to )} instance, and instead has a \ensuremath{\Conid{Initial}} instance for additive functions.}
\begin{hscode}\SaveRestoreHook
\column{B}{@{}>{\hspre}l<{\hspost}@{}}%
\column{24}{@{}>{\hspre}l<{\hspost}@{}}%
\column{30}{@{}>{\hspre}l<{\hspost}@{}}%
\column{46}{@{}>{\hspre}l<{\hspost}@{}}%
\column{52}{@{}>{\hspre}l<{\hspost}@{}}%
\column{E}{@{}>{\hspre}l<{\hspost}@{}}%
\>[B]{}\mathbf{class}\;\Conid{Terminal}\;\Varid{k}\;{}\<[24]%
\>[24]{}\mathbf{where}\;\Varid{it}\mathbin{::}{}\<[52]%
\>[52]{}\Varid{a}\mathbin{`\Varid{k}`}(){}\<[E]%
\\
\>[B]{}\mathbf{class}\;\Conid{Initial}\;\Varid{k}\;{}\<[24]%
\>[24]{}\mathbf{where}\;\Varid{ti}\mathbin{::}\Conid{Additive}\;{}\<[46]%
\>[46]{}\Varid{a}\Rightarrow {}\<[52]%
\>[52]{}()\mathbin{`\Varid{k}`}\Varid{a}{}\<[E]%
\\[\blanklineskip]%
\>[B]{}\mathbf{instance}\;\Conid{Terminal}\;(\to )\;{}\<[30]%
\>[30]{}\mathbf{where}\;\Varid{it}\mathrel{=}\lambda \anonymous \to (){}\<[E]%
\\
\>[B]{}\mathbf{instance}\;\Conid{Initial}\;(\to )\;{}\<[30]%
\>[30]{}\mathbf{where}\;\Varid{ti}\mathrel{=}\lambda ()\to \mathrm{0}{}\<[E]%
\ColumnHook
\end{hscode}\resethooks
Differentiation is trivial, since \ensuremath{\Varid{it}} and \ensuremath{\Varid{ti}} on functions are both linear.

\sectionl{Abelian Categories}

Another perspective on the operations we've considered is that morphisms sharing any particular domain and codomain (i.e., hom-sets) form an abelian group.
The zero for \ensuremath{\Varid{a}\mathbin{`\Varid{k}`}\Varid{b}} results from the composition of initial and terminal morphisms:
\begin{hscode}\SaveRestoreHook
\column{B}{@{}>{\hspre}l<{\hspost}@{}}%
\column{3}{@{}>{\hspre}l<{\hspost}@{}}%
\column{E}{@{}>{\hspre}l<{\hspost}@{}}%
\>[B]{}\mathbf{instance}\;(\Conid{Cartesian}\;\Varid{k},\Conid{Cocartesian}\;\Varid{k},\Conid{Terminal}\;\Varid{k},\Conid{InitialCat}\;\Varid{k})\Rightarrow \Conid{Additive}\;(\Varid{a}\mathbin{`\Varid{k}`}\Varid{b})\;\mathbf{where}{}\<[E]%
\\
\>[B]{}\hsindent{3}{}\<[3]%
\>[3]{}\mathrm{0}\mathrel{=}\Varid{ti}\hsdot{\circ }{.}\Varid{it}{}\<[E]%
\\
\>[B]{}\hsindent{3}{}\<[3]%
\>[3]{}\Varid{f}\mathbin{+}\Varid{g}\mathrel{=}\Varid{jam}\hsdot{\circ }{.}(\Varid{f}\times\Varid{g})\hsdot{\circ }{.}\Varid{dup}\mbox{\onelinecomment  \ensuremath{\mathrel{=}\Varid{jam}\hsdot{\circ }{.}(\Varid{f}\mathbin{\vartriangle}\Varid{g})\mathrel{=}(\Varid{f}\mathbin{\triangledown}\Varid{g})\hsdot{\circ }{.}\Varid{dup}}.}{}\<[E]%
\ColumnHook
\end{hscode}\resethooks
%% TODO: replace uses of |zero| and |(^+^)| by |zero| and |(+)|
The following identities hold (with ``\ensuremath{\hsdot{\circ }{.}}'' binding more tightly than ``\ensuremath{\mathbin{+}}'') \cite[Equations 16 and 17]{MacedoOliveira2013Typing}:
\begin{hscode}\SaveRestoreHook
\column{B}{@{}>{\hspre}l<{\hspost}@{}}%
\column{E}{@{}>{\hspre}l<{\hspost}@{}}%
\>[B]{}\Varid{u}\mathbin{\vartriangle}\Varid{v}\mathrel{=}\Varid{u}\hsdot{\circ }{.}\Varid{exl}\mathbin{+}\Varid{v}\hsdot{\circ }{.}\Varid{exr}{}\<[E]%
\\
\>[B]{}\Varid{u}\mathbin{\triangledown}\Varid{v}\mathrel{=}\Varid{inl}\hsdot{\circ }{.}\Varid{u}\mathbin{+}\Varid{inr}\hsdot{\circ }{.}\Varid{v}{}\<[E]%
\ColumnHook
\end{hscode}\resethooks
In particular,
\begin{hscode}\SaveRestoreHook
\column{B}{@{}>{\hspre}l<{\hspost}@{}}%
\column{E}{@{}>{\hspre}l<{\hspost}@{}}%
\>[B]{}\Varid{u}\mathbin{\vartriangle}\mathrm{0}\mathrel{=}\Varid{u}\hsdot{\circ }{.}\Varid{exl}{}\<[E]%
\\
\>[B]{}\mathrm{0}\mathbin{\vartriangle}\Varid{v}\mathrel{=}\Varid{v}\hsdot{\circ }{.}\Varid{exr}{}\<[E]%
\\[\blanklineskip]%
\>[B]{}\Varid{u}\mathbin{\triangledown}\mathrm{0}\mathrel{=}\Varid{inl}\hsdot{\circ }{.}\Varid{u}{}\<[E]%
\\
\>[B]{}\mathrm{0}\mathbin{\triangledown}\Varid{v}\mathrel{=}\Varid{inr}\hsdot{\circ }{.}\Varid{v}{}\<[E]%
\ColumnHook
\end{hscode}\resethooks

\sectionl{Proofs}

\subsection{\corRef{compose}}\proofLabel{corollary:compose}
\begin{hscode}\SaveRestoreHook
\column{B}{@{}>{\hspre}c<{\hspost}@{}}%
\column{BE}{@{}l@{}}%
\column{5}{@{}>{\hspre}l<{\hspost}@{}}%
\column{64}{@{}>{\hspre}l<{\hspost}@{}}%
\column{E}{@{}>{\hspre}l<{\hspost}@{}}%
\>[5]{}\mathcal{D}\!^+\!\;(\Varid{g}\hsdot{\circ }{.}\Varid{f})\;\Varid{a}{}\<[E]%
\\
\>[B]{}\mathrel{=}{}\<[BE]%
\>[5]{}((\Varid{g}\hsdot{\circ }{.}\Varid{f})\;\Varid{a},\mathcal{D}\;(\Varid{g}\hsdot{\circ }{.}\Varid{f})\;\Varid{a}){}\<[64]%
\>[64]{}\mbox{\onelinecomment  definition of \ensuremath{\mathcal{D}\!^+\!}}{}\<[E]%
\\
\>[B]{}\mathrel{=}{}\<[BE]%
\>[5]{}(\Varid{g}\;(\Varid{f}\;\Varid{a}),\mathcal{D}\;(\Varid{g}\hsdot{\circ }{.}\Varid{f})\;\Varid{a}){}\<[64]%
\>[64]{}\mbox{\onelinecomment  definition of \ensuremath{(\hsdot{\circ }{.})}}{}\<[E]%
\\
\>[B]{}\mathrel{=}{}\<[BE]%
\>[5]{}(\Varid{g}\;(\Varid{f}\;\Varid{a}),\mathcal{D}\;\Varid{g}\;(\Varid{f}\;\Varid{a})\hsdot{\circ }{.}\mathcal{D}\;\Varid{f}\;\Varid{a}){}\<[64]%
\>[64]{}\mbox{\onelinecomment  \thmRef{compose}}{}\<[E]%
\\
\>[B]{}\mathrel{=}{}\<[BE]%
\>[5]{}\mathbf{let}\;\Varid{b}\mathrel{=}\Varid{f}\;\Varid{a}\;\mathbf{in}\;(\Varid{g}\;\Varid{b},\mathcal{D}\;\Varid{g}\;\Varid{b}\hsdot{\circ }{.}\mathcal{D}\;\Varid{f}\;\Varid{a}){}\<[64]%
\>[64]{}\mbox{\onelinecomment  refactoring to share \ensuremath{\Varid{f}\;\Varid{a}}}{}\<[E]%
\\
\>[B]{}\mathrel{=}{}\<[BE]%
\>[5]{}\mathbf{let}\;\{\mskip1.5mu (\Varid{b},\Varid{f'})\mathrel{=}\mathcal{D}\!^+\!\;\Varid{f}\;\Varid{a};(\Varid{c},\Varid{g'})\mathrel{=}\mathcal{D}\!^+\!\;\Varid{g}\;\Varid{b}\mskip1.5mu\}\;\mathbf{in}\;(\Varid{c},\Varid{g'}\hsdot{\circ }{.}\Varid{f'}){}\<[64]%
\>[64]{}\mbox{\onelinecomment  refactoring to show compositionality}{}\<[E]%
\ColumnHook
\end{hscode}\resethooks

\subsection{\corRef{cross}}\proofLabel{corollary:cross}
\begin{hscode}\SaveRestoreHook
\column{B}{@{}>{\hspre}c<{\hspost}@{}}%
\column{BE}{@{}l@{}}%
\column{5}{@{}>{\hspre}l<{\hspost}@{}}%
\column{86}{@{}>{\hspre}l<{\hspost}@{}}%
\column{E}{@{}>{\hspre}l<{\hspost}@{}}%
\>[5]{}\mathcal{D}\!^+\!\;(\Varid{f}\times\Varid{g})\;(\Varid{a},\Varid{b}){}\<[E]%
\\
\>[B]{}\mathrel{=}{}\<[BE]%
\>[5]{}((\Varid{f}\times\Varid{g})\;(\Varid{a},\Varid{b}),\mathcal{D}\;(\Varid{f}\times\Varid{g})\;(\Varid{a},\Varid{b})){}\<[86]%
\>[86]{}\mbox{\onelinecomment  definition of \ensuremath{\mathcal{D}\!^+\!}}{}\<[E]%
\\
\>[B]{}\mathrel{=}{}\<[BE]%
\>[5]{}((\Varid{f}\;\Varid{a},\Varid{g}\;\Varid{b}),\mathcal{D}\;(\Varid{f}\times\Varid{g})\;(\Varid{a},\Varid{b})){}\<[86]%
\>[86]{}\mbox{\onelinecomment  definition of \ensuremath{(\times)}}{}\<[E]%
\\
\>[B]{}\mathrel{=}{}\<[BE]%
\>[5]{}((\Varid{f}\;\Varid{a},\Varid{g}\;\Varid{b}),\mathcal{D}\;\Varid{f}\;\Varid{a}\times\mathcal{D}\;\Varid{g}\;\Varid{b}){}\<[86]%
\>[86]{}\mbox{\onelinecomment  \thmRef{cross}}{}\<[E]%
\\
\>[B]{}\mathrel{=}{}\<[BE]%
\>[5]{}\mathbf{let}\;\{\mskip1.5mu (\Varid{c},\Varid{f'})\mathrel{=}(\Varid{f}\;\Varid{a},\mathcal{D}\;\Varid{f}\;\Varid{a});(\Varid{d},\Varid{g'})\mathrel{=}(\Varid{g}\;\Varid{b},\mathcal{D}\;\Varid{g}\;\Varid{b})\mskip1.5mu\}\;\mathbf{in}\;((\Varid{c},\Varid{d}),\Varid{f'}\times\Varid{g'}){}\<[86]%
\>[86]{}\mbox{\onelinecomment  refactoring}{}\<[E]%
\\
\>[B]{}\mathrel{=}{}\<[BE]%
\>[5]{}\mathbf{let}\;\{\mskip1.5mu (\Varid{c},\Varid{f'})\mathrel{=}\mathcal{D}\!^+\!\;\Varid{f}\;\Varid{a};(\Varid{d},\Varid{g'})\mathrel{=}\mathcal{D}\!^+\!\;\Varid{g}\;\Varid{b}\mskip1.5mu\}\;\mathbf{in}\;((\Varid{c},\Varid{d}),\Varid{f'}\times\Varid{g'}){}\<[86]%
\>[86]{}\mbox{\onelinecomment  definition of \ensuremath{\mathcal{D}\!^+\!}}{}\<[E]%
\ColumnHook
\end{hscode}\resethooks

\subsection{\thmRef{cont}}\proofLabel{theorem:cont}

Recall the definition of \ensuremath{\Varid{cont}}:
\begin{hscode}\SaveRestoreHook
\column{B}{@{}>{\hspre}l<{\hspost}@{}}%
\column{E}{@{}>{\hspre}l<{\hspost}@{}}%
\>[B]{}\Varid{cont}\mathbin{::}\Conid{Category}\;\Varid{k}\Rightarrow (\Varid{a}\mathbin{`\Varid{k}`}\Varid{b})\to \Conid{Cont}_{\Varid{k}}^{\Varid{r}}\;\Varid{a}\;\Varid{b}{}\<[E]%
\\
\>[B]{}\Varid{cont}\;\Varid{f}\mathrel{=}\Conid{Cont}\;(\hsdot{\circ }{.}\:{}\Varid{f}){}\<[E]%
\ColumnHook
\end{hscode}\resethooks
To say that \ensuremath{\Varid{cont}} is a functor (\ensuremath{\Conid{Category}} homomorphism) is equivalent to the following two equalities:
\begin{closerCodePars}
\begin{hscode}\SaveRestoreHook
\column{B}{@{}>{\hspre}l<{\hspost}@{}}%
\column{E}{@{}>{\hspre}l<{\hspost}@{}}%
\>[B]{}\Varid{cont}\;\Varid{id}\mathrel{=}\Varid{id}{}\<[E]%
\\[\blanklineskip]%
\>[B]{}\Varid{cont}\;(\Varid{g}\hsdot{\circ }{.}\Varid{f})\mathrel{=}\Varid{cont}\;\Varid{g}\hsdot{\circ }{.}\Varid{cont}\;\Varid{f}{}\<[E]%
\ColumnHook
\end{hscode}\resethooks
\end{closerCodePars}%
Simplify the first homomorphism equation:
\begin{hscode}\SaveRestoreHook
\column{B}{@{}>{\hspre}c<{\hspost}@{}}%
\column{BE}{@{}l@{}}%
\column{5}{@{}>{\hspre}l<{\hspost}@{}}%
\column{27}{@{}>{\hspre}l<{\hspost}@{}}%
\column{E}{@{}>{\hspre}l<{\hspost}@{}}%
\>[5]{}\Varid{cont}\;\Varid{id}{}\<[E]%
\\
\>[B]{}\mathrel{=}{}\<[BE]%
\>[5]{}\Conid{Cont}\;(\hsdot{\circ }{.}\:{}\Varid{id}){}\<[27]%
\>[27]{}\mbox{\onelinecomment  definition of \ensuremath{\Varid{cont}}}{}\<[E]%
\\
\>[B]{}\mathrel{=}{}\<[BE]%
\>[5]{}\Conid{Cont}\;(\lambda \Varid{h}\to \Varid{h}\hsdot{\circ }{.}\Varid{id}){}\<[27]%
\>[27]{}\mbox{\onelinecomment  definition of right section}{}\<[E]%
\\
\>[B]{}\mathrel{=}{}\<[BE]%
\>[5]{}\Conid{Cont}\;(\lambda \Varid{h}\to \Varid{h}){}\<[27]%
\>[27]{}\mbox{\onelinecomment  category law}{}\<[E]%
\\
\>[B]{}\mathrel{=}{}\<[BE]%
\>[5]{}\Conid{Cont}\;\Varid{id}{}\<[27]%
\>[27]{}\mbox{\onelinecomment  definition of \ensuremath{\Varid{id}} for functions}{}\<[E]%
\ColumnHook
\end{hscode}\resethooks
The first homomorphism equation is thus equivalent to \ensuremath{\Varid{id}\mathrel{=}\Conid{Cont}\;\Varid{id}}, which is in solved form.
For the second homomorphism equation, simplify both sides:
\begin{hscode}\SaveRestoreHook
\column{B}{@{}>{\hspre}c<{\hspost}@{}}%
\column{BE}{@{}l@{}}%
\column{5}{@{}>{\hspre}l<{\hspost}@{}}%
\column{44}{@{}>{\hspre}l<{\hspost}@{}}%
\column{E}{@{}>{\hspre}l<{\hspost}@{}}%
\>[5]{}\Varid{cont}\;\Varid{g}\hsdot{\circ }{.}\Varid{cont}\;\Varid{f}{}\<[E]%
\\
\>[B]{}\mathrel{=}{}\<[BE]%
\>[5]{}\Conid{Cont}\;(\hsdot{\circ }{.}\:{}\Varid{g})\hsdot{\circ }{.}\Conid{Cont}\;(\hsdot{\circ }{.}\:{}\Varid{f}){}\<[44]%
\>[44]{}\mbox{\onelinecomment  definition of \ensuremath{\Varid{cont}}}{}\<[E]%
\\[\blanklineskip]%
\>[5]{}\Varid{cont}\;(\Varid{g}\hsdot{\circ }{.}\Varid{f}){}\<[E]%
\\
\>[B]{}\mathrel{=}{}\<[BE]%
\>[5]{}\Varid{cont}\;(\hsdot{\circ }{.}\:{}(\Varid{g}\hsdot{\circ }{.}\Varid{f})){}\<[44]%
\>[44]{}\mbox{\onelinecomment  definition of \ensuremath{\Varid{cont}}}{}\<[E]%
\\
\>[B]{}\mathrel{=}{}\<[BE]%
\>[5]{}\Varid{cont}\;(\lambda \Varid{h}\to \Varid{h}\hsdot{\circ }{.}(\Varid{g}\hsdot{\circ }{.}\Varid{f})){}\<[44]%
\>[44]{}\mbox{\onelinecomment  definition of right section}{}\<[E]%
\\
\>[B]{}\mathrel{=}{}\<[BE]%
\>[5]{}\Varid{cont}\;(\lambda \Varid{h}\to (\Varid{h}\hsdot{\circ }{.}\Varid{g})\hsdot{\circ }{.}\Varid{f}){}\<[44]%
\>[44]{}\mbox{\onelinecomment  category law}{}\<[E]%
\\
\>[B]{}\mathrel{=}{}\<[BE]%
\>[5]{}\Varid{cont}\;(\lambda \Varid{h}\to (\hsdot{\circ }{.}\:{}\Varid{f})\;((\hsdot{\circ }{.}\:{}\Varid{g})\;\Varid{h})){}\<[44]%
\>[44]{}\mbox{\onelinecomment  definition of right section}{}\<[E]%
\\
\>[B]{}\mathrel{=}{}\<[BE]%
\>[5]{}\Conid{Cont}\;((\hsdot{\circ }{.}\:{}\Varid{f})\hsdot{\circ }{.}(\hsdot{\circ }{.}\:{}\Varid{g})){}\<[44]%
\>[44]{}\mbox{\onelinecomment  definition of \ensuremath{(\hsdot{\circ }{.})}}{}\<[E]%
\ColumnHook
\end{hscode}\resethooks
The simplified requirement:
\begin{hscode}\SaveRestoreHook
\column{B}{@{}>{\hspre}l<{\hspost}@{}}%
\column{E}{@{}>{\hspre}l<{\hspost}@{}}%
\>[B]{}\Conid{Cont}\;(\hsdot{\circ }{.}\:{}\Varid{g})\hsdot{\circ }{.}\Conid{Cont}\;(\hsdot{\circ }{.}\:{}\Varid{f})\mathrel{=}\Conid{Cont}\;((\hsdot{\circ }{.}\:{}\Varid{f})\hsdot{\circ }{.}(\hsdot{\circ }{.}\:{}\Varid{g})){}\<[E]%
\ColumnHook
\end{hscode}\resethooks
Generalize to a stronger condition, replacing \ensuremath{(\hsdot{\circ }{.}\:{}\Varid{g})} and \ensuremath{(\hsdot{\circ }{.}\:{}\Varid{f})} with \ensuremath{\Varid{g}} and \ensuremath{\Varid{f}} (appropriately re-typed):
\begin{hscode}\SaveRestoreHook
\column{B}{@{}>{\hspre}l<{\hspost}@{}}%
\column{E}{@{}>{\hspre}l<{\hspost}@{}}%
\>[B]{}\Conid{Cont}\;\Varid{g}\hsdot{\circ }{.}\Conid{Cont}\;\Varid{f}\mathrel{=}\Conid{Cont}\;(\Varid{f}\hsdot{\circ }{.}\Varid{g}){}\<[E]%
\ColumnHook
\end{hscode}\resethooks
This strengthened condition is also in solved form.
Notice the reversal of composition (and, more subtly, of \ensuremath{\Varid{id}}).

The monoidal functor (i.e., a \ensuremath{\Conid{Monoidal}} homomorphism) property:
\begin{hscode}\SaveRestoreHook
\column{B}{@{}>{\hspre}l<{\hspost}@{}}%
\column{E}{@{}>{\hspre}l<{\hspost}@{}}%
\>[B]{}\Varid{cont}\;(\Varid{f}\times\Varid{g})\mathrel{=}\Varid{cont}\;\Varid{f}\times\Varid{cont}\;\Varid{g}{}\<[E]%
\ColumnHook
\end{hscode}\resethooks
Simplify both sides:
\begin{hscode}\SaveRestoreHook
\column{B}{@{}>{\hspre}c<{\hspost}@{}}%
\column{BE}{@{}l@{}}%
\column{5}{@{}>{\hspre}l<{\hspost}@{}}%
\column{72}{@{}>{\hspre}l<{\hspost}@{}}%
\column{E}{@{}>{\hspre}l<{\hspost}@{}}%
\>[5]{}\Varid{cont}\;\Varid{f}\times\Varid{cont}\;\Varid{g}{}\<[E]%
\\
\>[B]{}\mathrel{=}{}\<[BE]%
\>[5]{}\Conid{Cont}\;(\hsdot{\circ }{.}\:{}\Varid{f})\times\Conid{Cont}\;(\hsdot{\circ }{.}\:{}\Varid{g}){}\<[72]%
\>[72]{}\mbox{\onelinecomment  definition of \ensuremath{\Varid{cont}}}{}\<[E]%
\\[\blanklineskip]%
\>[5]{}\Varid{cont}\;(\Varid{f}\times\Varid{g}){}\<[E]%
\\
\>[B]{}\mathrel{=}{}\<[BE]%
\>[5]{}\Conid{Cont}\;(\hsdot{\circ }{.}\:{}(\Varid{f}\times\Varid{g})){}\<[72]%
\>[72]{}\mbox{\onelinecomment  definition of \ensuremath{\Varid{cont}}}{}\<[E]%
\\
\>[B]{}\mathrel{=}{}\<[BE]%
\>[5]{}\Conid{Cont}\;(\lambda \Varid{h}\to \Varid{h}\hsdot{\circ }{.}(\Varid{f}\times\Varid{g})){}\<[72]%
\>[72]{}\mbox{\onelinecomment  definition of right section}{}\<[E]%
\\
\>[B]{}\mathrel{=}{}\<[BE]%
\>[5]{}\Conid{Cont}\;(\lambda \Varid{h}\to \Varid{join}\;(\Varid{unjoin}\;\Varid{h})\hsdot{\circ }{.}(\Varid{f}\times\Varid{g})){}\<[72]%
\>[72]{}\mbox{\onelinecomment  \ensuremath{\Varid{join}\hsdot{\circ }{.}\Varid{unjoin}\mathrel{=}\Varid{id}}}{}\<[E]%
\\
\>[B]{}\mathrel{=}{}\<[BE]%
\>[5]{}\Conid{Cont}\;(\lambda \Varid{h}\to \mathbf{let}\;(\Varid{h}_{\Varid{a}},\Varid{h}_{\Varid{b}})\mathrel{=}\Varid{unjoin}\;\Varid{h}\;\mathbf{in}\;\Varid{join}\;(\Varid{h}_{\Varid{a}},\Varid{h}_{\Varid{b}})\hsdot{\circ }{.}(\Varid{f}\times\Varid{g})){}\<[72]%
\>[72]{}\mbox{\onelinecomment  refactor}{}\<[E]%
\\
\>[B]{}\mathrel{=}{}\<[BE]%
\>[5]{}\Conid{Cont}\;(\lambda \Varid{h}\to \mathbf{let}\mathbin{...}\mathbf{in}\;(\Varid{h}_{\Varid{a}}\mathbin{\triangledown}\Varid{h}_{\Varid{b}})\hsdot{\circ }{.}(\Varid{f}\times\Varid{g})){}\<[72]%
\>[72]{}\mbox{\onelinecomment  definition of \ensuremath{\Varid{join}}}{}\<[E]%
\\
\>[B]{}\mathrel{=}{}\<[BE]%
\>[5]{}\Conid{Cont}\;(\lambda \Varid{h}\to \mathbf{let}\mathbin{...}\mathbf{in}\;(\Varid{h}_{\Varid{a}}\hsdot{\circ }{.}\Varid{f}\mathbin{\triangledown}\Varid{h}_{\Varid{b}}\hsdot{\circ }{.}\Varid{g})){}\<[72]%
\>[72]{}\mbox{\onelinecomment  \citep[Section 1.5.2]{Gibbons2002Calculating}}{}\<[E]%
\\
\>[B]{}\mathrel{=}{}\<[BE]%
\>[5]{}\Conid{Cont}\;(\lambda \Varid{h}\to \mathbf{let}\mathbin{...}\mathbf{in}\;((\hsdot{\circ }{.}\:{}\Varid{f})\;\Varid{h}_{\Varid{a}}\mathbin{\triangledown}(\hsdot{\circ }{.}\:{}\Varid{g})\;\Varid{h}_{\Varid{b}})){}\<[72]%
\>[72]{}\mbox{\onelinecomment  definition of right section}{}\<[E]%
\\
\>[B]{}\mathrel{=}{}\<[BE]%
\>[5]{}\Conid{Cont}\;(\lambda \Varid{h}\to \mathbf{let}\mathbin{...}\mathbf{in}\;\Varid{join}\;((\hsdot{\circ }{.}\:{}\Varid{f})\;\Varid{h}_{\Varid{a}},(\hsdot{\circ }{.}\:{}\Varid{g})\;\Varid{h}_{\Varid{b}})){}\<[72]%
\>[72]{}\mbox{\onelinecomment  definition of \ensuremath{\Varid{join}}}{}\<[E]%
\\
\>[B]{}\mathrel{=}{}\<[BE]%
\>[5]{}\Conid{Cont}\;(\lambda \Varid{h}\to \mathbf{let}\mathbin{...}\mathbf{in}\;\Varid{join}\;(((\hsdot{\circ }{.}\:{}\Varid{f})\times(\hsdot{\circ }{.}\:{}\Varid{g}))\;(\Varid{h}_{\Varid{a}},\Varid{h}_{\Varid{b}}))){}\<[72]%
\>[72]{}\mbox{\onelinecomment  definition of \ensuremath{(\times)}}{}\<[E]%
\\
\>[B]{}\mathrel{=}{}\<[BE]%
\>[5]{}\Conid{Cont}\;(\lambda \Varid{h}\to \Varid{join}\;(((\hsdot{\circ }{.}\:{}\Varid{f})\times(\hsdot{\circ }{.}\:{}\Varid{g}))\;(\Varid{unjoin}\;\Varid{h}))){}\<[72]%
\>[72]{}\mbox{\onelinecomment  eliminate \ensuremath{\mathbf{let}}}{}\<[E]%
\\
\>[B]{}\mathrel{=}{}\<[BE]%
\>[5]{}\Conid{Cont}\;(\Varid{join}\hsdot{\circ }{.}((\hsdot{\circ }{.}\:{}\Varid{f})\times(\hsdot{\circ }{.}\:{}\Varid{g}))\hsdot{\circ }{.}\Varid{unjoin}){}\<[72]%
\>[72]{}\mbox{\onelinecomment  definition of \ensuremath{(\hsdot{\circ }{.})}}{}\<[E]%
\ColumnHook
\end{hscode}\resethooks
The crucial trick here was to note that \ensuremath{\Varid{h}\mathbin{::}(\Varid{a} \times \Varid{b})\mathbin{`\Varid{k}`}\Varid{r}} can be split into two continuations \ensuremath{\Varid{h}_{\Varid{a}}\mathbin{::}\Varid{a}\mathbin{`\Varid{k}`}\Varid{r}} and \ensuremath{\Varid{h}_{\Varid{b}}\mathbin{::}\Varid{b}\mathbin{`\Varid{k}`}\Varid{r}} thanks to \ensuremath{\Varid{join}}/\ensuremath{\Varid{unjoin}} isomorphism from \secref{Derived Operations}.\notefoot{In general, this splitting can lose efficiency, since \ensuremath{\Varid{h}_{\Varid{a}}} and \ensuremath{\Varid{h}_{\Varid{b}}} could duplicate work that was shared in \ensuremath{\Varid{h}}. Investigate this concern.}
Now, strengthen the massaged specification, generalizing from \ensuremath{(\hsdot{\circ }{.}\:{}\Varid{f})} and \ensuremath{(\hsdot{\circ }{.}\:{}\Varid{g})} as usual, resulting in a sufficient condition in solved form:
\begin{hscode}\SaveRestoreHook
\column{B}{@{}>{\hspre}l<{\hspost}@{}}%
\column{E}{@{}>{\hspre}l<{\hspost}@{}}%
\>[B]{}\Conid{Cont}\;\Varid{f}\times\Conid{Cont}\;\Varid{g}\mathrel{=}\Conid{Cont}\;(\Varid{join}\hsdot{\circ }{.}(\Varid{f}\times\Varid{g})\hsdot{\circ }{.}\Varid{unjoin}){}\<[E]%
\ColumnHook
\end{hscode}\resethooks

Next, derive \ensuremath{\Conid{Cartesian}} and \ensuremath{\Conid{Cocartesian}} instances from the specification that \ensuremath{\Varid{cont}} is a cartesian functor and a cocartesian functor (i.e., \ensuremath{\Conid{Cartesian}} and \ensuremath{\Conid{Cocartesian}} homomorphisms), i.e.,\\
{\mathindent2.5em
\begin{minipage}[b]{0.30\textwidth}
\begin{hscode}\SaveRestoreHook
\column{B}{@{}>{\hspre}l<{\hspost}@{}}%
\column{11}{@{}>{\hspre}l<{\hspost}@{}}%
\column{E}{@{}>{\hspre}l<{\hspost}@{}}%
\>[B]{}\Varid{cont}\;\Varid{exl}{}\<[11]%
\>[11]{}\mathrel{=}\Varid{exl}{}\<[E]%
\\
\>[B]{}\Varid{cont}\;\Varid{exr}{}\<[11]%
\>[11]{}\mathrel{=}\Varid{exr}{}\<[E]%
\\
\>[B]{}\Varid{cont}\;\Varid{dup}{}\<[11]%
\>[11]{}\mathrel{=}\Varid{dup}{}\<[E]%
\ColumnHook
\end{hscode}\resethooks
\end{minipage}
\begin{minipage}[b]{0ex}{\rule[3ex]{0.5pt}{0.43in}}\end{minipage}
\begin{minipage}[b]{0.0\textwidth}
\begin{hscode}\SaveRestoreHook
\column{B}{@{}>{\hspre}l<{\hspost}@{}}%
\column{11}{@{}>{\hspre}l<{\hspost}@{}}%
\column{E}{@{}>{\hspre}l<{\hspost}@{}}%
\>[B]{}\Varid{cont}\;\Varid{inl}{}\<[11]%
\>[11]{}\mathrel{=}\Varid{inl}{}\<[E]%
\\
\>[B]{}\Varid{cont}\;\Varid{inr}{}\<[11]%
\>[11]{}\mathrel{=}\Varid{inr}{}\<[E]%
\\
\>[B]{}\Varid{cont}\;\Varid{jam}{}\<[11]%
\>[11]{}\mathrel{=}\Varid{jam}{}\<[E]%
\ColumnHook
\end{hscode}\resethooks
\end{minipage}}\\
Reversing each of these equations puts them in solved form, so they can be used directly as definitions.
\out{\\
\begin{minipage}[b]{0.45\textwidth}
\begin{hscode}\SaveRestoreHook
\column{B}{@{}>{\hspre}l<{\hspost}@{}}%
\column{3}{@{}>{\hspre}l<{\hspost}@{}}%
\column{8}{@{}>{\hspre}l<{\hspost}@{}}%
\column{11}{@{}>{\hspre}l<{\hspost}@{}}%
\column{E}{@{}>{\hspre}l<{\hspost}@{}}%
\>[B]{}\mathbf{instance}\;{}\<[11]%
\>[11]{}\Conid{Cartesian}\;\Varid{k}\Rightarrow {}\<[E]%
\\
\>[11]{}\Conid{Cartesian}\;\Conid{Cont}_{\Varid{k}}^{\Varid{r}}\;\mathbf{where}{}\<[E]%
\\
\>[B]{}\hsindent{3}{}\<[3]%
\>[3]{}\Varid{exl}{}\<[8]%
\>[8]{}\mathrel{=}\Varid{cont}\;\Varid{exl}{}\<[E]%
\\
\>[B]{}\hsindent{3}{}\<[3]%
\>[3]{}\Varid{exr}{}\<[8]%
\>[8]{}\mathrel{=}\Varid{cont}\;\Varid{exr}{}\<[E]%
\\
\>[B]{}\hsindent{3}{}\<[3]%
\>[3]{}\Varid{dup}{}\<[8]%
\>[8]{}\mathrel{=}\Varid{cont}\;\Varid{dup}{}\<[E]%
\ColumnHook
\end{hscode}\resethooks
\end{minipage}
\begin{minipage}[b]{0ex}{\rule[2.5ex]{0.5pt}{0.8in}}\end{minipage}
\begin{minipage}[b]{0.0\textwidth}
\begin{hscode}\SaveRestoreHook
\column{B}{@{}>{\hspre}l<{\hspost}@{}}%
\column{3}{@{}>{\hspre}l<{\hspost}@{}}%
\column{8}{@{}>{\hspre}l<{\hspost}@{}}%
\column{11}{@{}>{\hspre}l<{\hspost}@{}}%
\column{E}{@{}>{\hspre}l<{\hspost}@{}}%
\>[B]{}\mathbf{instance}\;{}\<[11]%
\>[11]{}\Conid{Cocartesian}\;\Varid{k}\Rightarrow {}\<[E]%
\\
\>[11]{}\Conid{Cocartesian}\;\Conid{Cont}_{\Varid{k}}^{\Varid{r}}\;\mathbf{where}{}\<[E]%
\\
\>[B]{}\hsindent{3}{}\<[3]%
\>[3]{}\Varid{inl}{}\<[8]%
\>[8]{}\mathrel{=}\Varid{cont}\;\Varid{inl}{}\<[E]%
\\
\>[B]{}\hsindent{3}{}\<[3]%
\>[3]{}\Varid{inr}{}\<[8]%
\>[8]{}\mathrel{=}\Varid{cont}\;\Varid{inr}{}\<[E]%
\\
\>[B]{}\hsindent{3}{}\<[3]%
\>[3]{}\Varid{jam}{}\<[8]%
\>[8]{}\mathrel{=}\Varid{cont}\;\Varid{jam}{}\<[E]%
\ColumnHook
\end{hscode}\resethooks
\end{minipage}
}
While these definitions are correct, they can be made more efficient.
For instance,
\begin{hscode}\SaveRestoreHook
\column{B}{@{}>{\hspre}c<{\hspost}@{}}%
\column{BE}{@{}l@{}}%
\column{5}{@{}>{\hspre}l<{\hspost}@{}}%
\column{34}{@{}>{\hspre}l<{\hspost}@{}}%
\column{E}{@{}>{\hspre}l<{\hspost}@{}}%
\>[5]{}\Varid{cont}\;\Varid{exl}{}\<[E]%
\\
\>[B]{}\mathrel{=}{}\<[BE]%
\>[5]{}\Conid{Cont}\;(\lambda \Varid{h}\to \Varid{h}\hsdot{\circ }{.}\Varid{exl}){}\<[34]%
\>[34]{}\mbox{\onelinecomment  definition of \ensuremath{\Varid{cont}}}{}\<[E]%
\\
\>[B]{}\mathrel{=}{}\<[BE]%
\>[5]{}\Conid{Cont}\;(\lambda \Varid{h}\to \Varid{h}\mathbin{\triangledown}\mathrm{0}){}\<[34]%
\>[34]{}\mbox{\onelinecomment  \appref{Abelian Categories}}{}\<[E]%
\\
\>[B]{}\mathrel{=}{}\<[BE]%
\>[5]{}\Conid{Cont}\;(\lambda \Varid{h}\to \Varid{join}\;(\Varid{h},\mathrm{0})){}\<[34]%
\>[34]{}\mbox{\onelinecomment  definition of \ensuremath{\Varid{join}}}{}\<[E]%
\\
\>[B]{}\mathrel{=}{}\<[BE]%
\>[5]{}\Conid{Cont}\;(\lambda \Varid{h}\to \Varid{join}\;(\Varid{inl}\;\Varid{h})){}\<[34]%
\>[34]{}\mbox{\onelinecomment  definition of \ensuremath{\Varid{inl}} for functions}{}\<[E]%
\\
\>[B]{}\mathrel{=}{}\<[BE]%
\>[5]{}\Conid{Cont}\;(\Varid{join}\hsdot{\circ }{.}\Varid{inl}){}\<[34]%
\>[34]{}\mbox{\onelinecomment  definition of \ensuremath{(\hsdot{\circ }{.})} for functions}{}\<[E]%
\ColumnHook
\end{hscode}\resethooks
Similarly, \ensuremath{\Varid{cont}\;\Varid{exr}\mathrel{=}\Conid{Cont}\;(\Varid{join}\hsdot{\circ }{.}\Varid{inr})}.
For \ensuremath{\Varid{dup}\mathbin{::}\Varid{a}\mathbin{`\Varid{k}`}(\Varid{a} \times \Varid{a})}, we'll have \ensuremath{\Varid{h}\mathbin{::}(\Varid{a} \times \Varid{a}) \leadsto \Varid{r}}, so we can split \ensuremath{\Varid{h}} with \ensuremath{\Varid{unjoin}}:
\begin{hscode}\SaveRestoreHook
\column{B}{@{}>{\hspre}c<{\hspost}@{}}%
\column{BE}{@{}l@{}}%
\column{5}{@{}>{\hspre}l<{\hspost}@{}}%
\column{64}{@{}>{\hspre}l<{\hspost}@{}}%
\column{E}{@{}>{\hspre}l<{\hspost}@{}}%
\>[5]{}\Varid{cont}\;\Varid{dup}{}\<[E]%
\\
\>[B]{}\mathrel{=}{}\<[BE]%
\>[5]{}\Conid{Cont}\;(\lambda \Varid{h}\to \Varid{h}\hsdot{\circ }{.}\Varid{dup}){}\<[64]%
\>[64]{}\mbox{\onelinecomment  definition of \ensuremath{\Varid{cont}}}{}\<[E]%
\\
\>[B]{}\mathrel{=}{}\<[BE]%
\>[5]{}\Conid{Cont}\;(\lambda \Varid{h}\to \Varid{join}\;(\Varid{unjoin}\;\Varid{h})\hsdot{\circ }{.}\Varid{dup}){}\<[64]%
\>[64]{}\mbox{\onelinecomment  \ensuremath{\Varid{join}\hsdot{\circ }{.}\Varid{unjoin}\mathrel{=}\Varid{id}}}{}\<[E]%
\\
\>[B]{}\mathrel{=}{}\<[BE]%
\>[5]{}\Conid{Cont}\;(\lambda \Varid{h}\to \mathbf{let}\;(\Varid{h}_{\Varid{a}},\Varid{h}_{\Varid{b}})\mathrel{=}\Varid{unjoin}\;\Varid{h}\;\mathbf{in}\;(\Varid{h}_{\Varid{a}}\mathbin{\triangledown}\Varid{h}_{\Varid{b}})\hsdot{\circ }{.}\Varid{dup}){}\<[64]%
\>[64]{}\mbox{\onelinecomment  refactor; definition of \ensuremath{\Varid{join}}}{}\<[E]%
\\
\>[B]{}\mathrel{=}{}\<[BE]%
\>[5]{}\Conid{Cont}\;(\lambda \Varid{h}\to \mathbf{let}\;(\Varid{h}_{\Varid{a}},\Varid{h}_{\Varid{b}})\mathrel{=}\Varid{unjoin}\;\Varid{h}\;\mathbf{in}\;\Varid{h}_{\Varid{a}}\mathbin{+}\Varid{h}_{\Varid{b}}){}\<[64]%
\>[64]{}\mbox{\onelinecomment  \appref{Abelian Categories}}{}\<[E]%
\\
\>[B]{}\mathrel{=}{}\<[BE]%
\>[5]{}\Conid{Cont}\;(\lambda \Varid{h}\to \mathbf{let}\;(\Varid{h}_{\Varid{a}},\Varid{h}_{\Varid{b}})\mathrel{=}\Varid{unjoin}\;\Varid{h}\;\mathbf{in}\;\Varid{jam}\;(\Varid{h}_{\Varid{a}},\Varid{h}_{\Varid{b}})){}\<[64]%
\>[64]{}\mbox{\onelinecomment  definition of \ensuremath{\Varid{jam}} for functions}{}\<[E]%
\\
\>[B]{}\mathrel{=}{}\<[BE]%
\>[5]{}\Conid{Cont}\;(\lambda \Varid{h}\to \Varid{jam}\;(\Varid{unjoin}\;\Varid{h})){}\<[64]%
\>[64]{}\mbox{\onelinecomment  eliminate the \ensuremath{\mathbf{let}}}{}\<[E]%
\\
\>[B]{}\mathrel{=}{}\<[BE]%
\>[5]{}\Conid{Cont}\;(\Varid{jam}\hsdot{\circ }{.}\Varid{unjoin}){}\<[64]%
\>[64]{}\mbox{\onelinecomment  definition of \ensuremath{(\hsdot{\circ }{.})} on functions}{}\<[E]%
\ColumnHook
\end{hscode}\resethooks

For \ensuremath{\Conid{Cocartesian}}, we reason dually:
\begin{hscode}\SaveRestoreHook
\column{B}{@{}>{\hspre}c<{\hspost}@{}}%
\column{BE}{@{}l@{}}%
\column{5}{@{}>{\hspre}l<{\hspost}@{}}%
\column{64}{@{}>{\hspre}l<{\hspost}@{}}%
\column{E}{@{}>{\hspre}l<{\hspost}@{}}%
\>[5]{}\Varid{cont}\;\Varid{inl}{}\<[E]%
\\
\>[B]{}\mathrel{=}{}\<[BE]%
\>[5]{}\Conid{Cont}\;(\lambda \Varid{h}\to \Varid{h}\hsdot{\circ }{.}\Varid{inl}){}\<[64]%
\>[64]{}\mbox{\onelinecomment  definition of \ensuremath{\Varid{inl}}}{}\<[E]%
\\
\>[B]{}\mathrel{=}{}\<[BE]%
\>[5]{}\Conid{Cont}\;(\lambda \Varid{h}\to \Varid{join}\;(\Varid{unjoin}\;\Varid{h})\hsdot{\circ }{.}\Varid{inl}){}\<[64]%
\>[64]{}\mbox{\onelinecomment  \ensuremath{\Varid{join}\hsdot{\circ }{.}\Varid{unjoin}\mathrel{=}\Varid{id}}}{}\<[E]%
\\
\>[B]{}\mathrel{=}{}\<[BE]%
\>[5]{}\Conid{Cont}\;(\lambda \Varid{h}\to \mathbf{let}\;(\Varid{h}_{\Varid{a}},\Varid{h}_{\Varid{b}})\mathrel{=}\Varid{unjoin}\;\Varid{h}\;\mathbf{in}\;(\Varid{h}_{\Varid{a}}\mathbin{\triangledown}\Varid{h}_{\Varid{b}})\hsdot{\circ }{.}\Varid{inl}){}\<[64]%
\>[64]{}\mbox{\onelinecomment  definition of \ensuremath{\Varid{join}}}{}\<[E]%
\\
\>[B]{}\mathrel{=}{}\<[BE]%
\>[5]{}\Conid{Cont}\;(\lambda \Varid{h}\to \mathbf{let}\;(\Varid{h}_{\Varid{a}},\Varid{h}_{\Varid{b}})\mathrel{=}\Varid{unjoin}\;\Varid{h}\;\mathbf{in}\;\Varid{h}_{\Varid{a}}){}\<[64]%
\>[64]{}\mbox{\onelinecomment  \citep[Section 1.5.2]{Gibbons2002Calculating}}{}\<[E]%
\\
\>[B]{}\mathrel{=}{}\<[BE]%
\>[5]{}\Conid{Cont}\;(\lambda \Varid{h}\to \Varid{exl}\;(\Varid{unjoin}\;\Varid{h})){}\<[64]%
\>[64]{}\mbox{\onelinecomment  definition of \ensuremath{\Varid{exl}} for functions}{}\<[E]%
\\
\>[B]{}\mathrel{=}{}\<[BE]%
\>[5]{}\Conid{Cont}\;(\Varid{exl}\hsdot{\circ }{.}\Varid{unjoin}){}\<[64]%
\>[64]{}\mbox{\onelinecomment  definition of \ensuremath{(\hsdot{\circ }{.})} for functions}{}\<[E]%
\ColumnHook
\end{hscode}\resethooks
Similarly, \ensuremath{\Varid{cont}\;\Varid{inr}\mathrel{=}\Conid{Cont}\;(\Varid{exr}\hsdot{\circ }{.}\Varid{unjoin})}.
Next,
\begin{hscode}\SaveRestoreHook
\column{B}{@{}>{\hspre}c<{\hspost}@{}}%
\column{BE}{@{}l@{}}%
\column{5}{@{}>{\hspre}l<{\hspost}@{}}%
\column{38}{@{}>{\hspre}l<{\hspost}@{}}%
\column{E}{@{}>{\hspre}l<{\hspost}@{}}%
\>[5]{}\Varid{cont}\;\Varid{jam}{}\<[E]%
\\
\>[B]{}\mathrel{=}{}\<[BE]%
\>[5]{}\Conid{Cont}\;(\lambda \Varid{h}\to \Varid{h}\hsdot{\circ }{.}\Varid{jam}){}\<[38]%
\>[38]{}\mbox{\onelinecomment  definition of \ensuremath{\Varid{cont}}}{}\<[E]%
\\
\>[B]{}\mathrel{=}{}\<[BE]%
\>[5]{}\Conid{Cont}\;(\lambda \Varid{h}\to \Varid{h}\hsdot{\circ }{.}(\Varid{id}\mathbin{\triangledown}\Varid{id})){}\<[38]%
\>[38]{}\mbox{\onelinecomment  a law for \ensuremath{\Varid{jam}} and \ensuremath{(\mathbin{\triangledown})}}{}\<[E]%
\\
\>[B]{}\mathrel{=}{}\<[BE]%
\>[5]{}\Conid{Cont}\;(\lambda \Varid{h}\to \Varid{h}\hsdot{\circ }{.}\Varid{id}\mathbin{\triangledown}\Varid{h}\hsdot{\circ }{.}\Varid{id}){}\<[38]%
\>[38]{}\mbox{\onelinecomment  \citep[Section 1.5.2]{Gibbons2002Calculating}}{}\<[E]%
\\
\>[B]{}\mathrel{=}{}\<[BE]%
\>[5]{}\Conid{Cont}\;(\lambda \Varid{h}\to \Varid{h}\mathbin{\triangledown}\Varid{h}){}\<[38]%
\>[38]{}\mbox{\onelinecomment  category law}{}\<[E]%
\\
\>[B]{}\mathrel{=}{}\<[BE]%
\>[5]{}\Conid{Cont}\;(\lambda \Varid{h}\to \Varid{join}\;(\Varid{h},\Varid{h})){}\<[38]%
\>[38]{}\mbox{\onelinecomment  definition of \ensuremath{\Varid{join}}}{}\<[E]%
\\
\>[B]{}\mathrel{=}{}\<[BE]%
\>[5]{}\Conid{Cont}\;(\Varid{join}\hsdot{\circ }{.}\Varid{dup}){}\<[38]%
\>[38]{}\mbox{\onelinecomment  definition of \ensuremath{\Varid{dup}} on functions}{}\<[E]%
\ColumnHook
\end{hscode}\resethooks

The final element of our linear vocabulary is scalar multiplication:\notefoot{Is there a more general argument to make? I haven't wanted to say that \ensuremath{\Varid{h}} is linear.}
\begin{hscode}\SaveRestoreHook
\column{B}{@{}>{\hspre}c<{\hspost}@{}}%
\column{BE}{@{}l@{}}%
\column{5}{@{}>{\hspre}l<{\hspost}@{}}%
\column{32}{@{}>{\hspre}l<{\hspost}@{}}%
\column{E}{@{}>{\hspre}l<{\hspost}@{}}%
\>[5]{}\Varid{cont}\;(\Varid{scale}\;\Varid{s}){}\<[E]%
\\
\>[B]{}\mathrel{=}{}\<[BE]%
\>[5]{}\Conid{Cont}\;(\lambda \Varid{h}\to \Varid{h}\hsdot{\circ }{.}\Varid{scale}\;\Varid{s}){}\<[32]%
\>[32]{}\mbox{\onelinecomment  definition of \ensuremath{\Varid{cont}}}{}\<[E]%
\\
\>[B]{}\mathrel{=}{}\<[BE]%
\>[5]{}\Conid{Cont}\;(\lambda \Varid{h}\to \Varid{scale}\;\Varid{s}\hsdot{\circ }{.}\Varid{h}){}\<[32]%
\>[32]{}\mbox{\onelinecomment  linearity of \ensuremath{\Varid{h}}}{}\<[E]%
\\
\>[B]{}\mathrel{=}{}\<[BE]%
\>[5]{}\Conid{Cont}\;(\lambda \Varid{h}\to \Varid{scale}\;\Varid{s}\;\Varid{h}){}\<[32]%
\>[32]{}\mbox{\onelinecomment  definition of \ensuremath{\Varid{scale}} for functions/maps}{}\<[E]%
\\
\>[B]{}\mathrel{=}{}\<[BE]%
\>[5]{}\Conid{Cont}\;(\Varid{scale}\;\Varid{s}){}\<[32]%
\>[32]{}\mbox{\onelinecomment  $\eta$-reduction}{}\<[E]%
\ColumnHook
\end{hscode}\resethooks
These optimized solved forms match the definitions in \figref{cont}.

\subsection{\thmRef{asDual}}\proofLabel{theorem:asDual}

\nc\lemDot[1]{\lemRef{dot-properties}, part \ref{#1}}
\nc\lemDotTwo[2]{Lemma \ref{lemma:dot-properties}, parts \ref{#1} \& \ref{#2}}

To derive instances for \ensuremath{\Conid{Dual}_{\Varid{k}}}, we'll need some properties.
\begin{lemma} \lemLabel{dot-properties}
The following properties hold:
% https://tex.stackexchange.com/questions/38260/non-italic-text-in-theorems-definitions-examples
\normalfont
\begin{enumerate}
\item \ensuremath{\Varid{dot}} is linear. \label{dot-linear}
\item \ensuremath{\Varid{dot}^{-1}} is linear. \label{unDot-linear}
\item \ensuremath{\Varid{unjoin}\hsdot{\circ }{.}\Varid{dot}\mathrel{=}\Varid{dot}\times\Varid{dot}} \label{unjoin-dot}
\item \ensuremath{\Varid{dot}^{-1}\hsdot{\circ }{.}\Varid{join}\mathrel{=}\Varid{dot}^{-1}\times\Varid{dot}^{-1}} \label{unDot-join}
\item \ensuremath{\Varid{dot}\;\Varid{u}\mathbin{\triangledown}\Varid{dot}\;\Varid{v}\mathrel{=}\Varid{dot}\;(\Varid{u},\Varid{v})} \label{dot-dot-join}
\item \ensuremath{\Varid{dot}\;\mathrm{0}\mathrel{=}\mathrm{0}} (zero vector vs zero morphism) \label{dot-zeroV}
\end{enumerate}
\end{lemma}
\emph{Proof:}
\begin{enumerate}
\item Follows from the bilinearity of uncurried dot product:\notefoot{I'm treating linear maps here as functions. Revisit.}
\begin{hscode}\SaveRestoreHook
\column{B}{@{}>{\hspre}c<{\hspost}@{}}%
\column{BE}{@{}l@{}}%
\column{5}{@{}>{\hspre}l<{\hspost}@{}}%
\column{31}{@{}>{\hspre}l<{\hspost}@{}}%
\column{E}{@{}>{\hspre}l<{\hspost}@{}}%
\>[5]{}\Varid{dot}\;(\Varid{u}\mathbin{+}\Varid{v}){}\<[E]%
\\
\>[B]{}\mathrel{=}{}\<[BE]%
\>[5]{}\lambda \Varid{w}\to \Varid{dot}\;(\Varid{u}\mathbin{+}\Varid{v})\;\Varid{w}{}\<[31]%
\>[31]{}\mbox{\onelinecomment  $\eta$-expansion}{}\<[E]%
\\
\>[B]{}\mathrel{=}{}\<[BE]%
\>[5]{}\lambda \Varid{w}\to \Varid{dot}\;\Varid{u}\;\Varid{w}\mathbin{+}\Varid{dot}\;\Varid{v}\;\Varid{w}{}\<[31]%
\>[31]{}\mbox{\onelinecomment  bilinearity of uncurried dot product}{}\<[E]%
\\
\>[B]{}\mathrel{=}{}\<[BE]%
\>[5]{}\Varid{dot}\;\Varid{u}\mathbin{+}\Varid{dot}\;\Varid{v}{}\<[31]%
\>[31]{}\mbox{\onelinecomment  definition of \ensuremath{(\mathbin{+})} of functions}{}\<[E]%
\\[\blanklineskip]%
\>[5]{}\Varid{dot}\;(\Varid{s} \cdot \Varid{u}){}\<[E]%
\\
\>[B]{}\mathrel{=}{}\<[BE]%
\>[5]{}\lambda \Varid{w}\to \Varid{dot}\;(\Varid{s} \cdot \Varid{u})\;\Varid{w}{}\<[31]%
\>[31]{}\mbox{\onelinecomment  $\eta$-expansion}{}\<[E]%
\\
\>[B]{}\mathrel{=}{}\<[BE]%
\>[5]{}\lambda \Varid{w}\to \Varid{s} \cdot \Varid{dot}\;\Varid{u}\;\Varid{w}{}\<[31]%
\>[31]{}\mbox{\onelinecomment  bilinearity of uncurried dot product}{}\<[E]%
\\
\>[B]{}\mathrel{=}{}\<[BE]%
\>[5]{}\Varid{s} \cdot \Varid{dot}\;\Varid{u}{}\<[31]%
\>[31]{}\mbox{\onelinecomment  definition of \ensuremath{( \cdot )} on functions}{}\<[E]%
\ColumnHook
\end{hscode}\resethooks
\item Invertible linear functions have linear inverses. In particular,
\begin{hscode}\SaveRestoreHook
\column{B}{@{}>{\hspre}c<{\hspost}@{}}%
\column{BE}{@{}l@{}}%
\column{5}{@{}>{\hspre}l<{\hspost}@{}}%
\column{44}{@{}>{\hspre}l<{\hspost}@{}}%
\column{E}{@{}>{\hspre}l<{\hspost}@{}}%
\>[5]{}\Varid{dot}^{-1}\;(\Varid{u}\mathbin{+}\Varid{v}){}\<[E]%
\\
\>[B]{}\mathrel{=}{}\<[BE]%
\>[5]{}\Varid{dot}^{-1}\;(\Varid{dot}\;(\Varid{dot}^{-1}\;\Varid{u})\mathbin{+}\Varid{dot}\;(\Varid{dot}^{-1}\;\Varid{v})){}\<[44]%
\>[44]{}\mbox{\onelinecomment  \ensuremath{\Varid{dot}\hsdot{\circ }{.}\Varid{dot}^{-1}\mathrel{=}\Varid{id}}}{}\<[E]%
\\
\>[B]{}\mathrel{=}{}\<[BE]%
\>[5]{}\Varid{dot}^{-1}\;(\Varid{dot}\;(\Varid{dot}^{-1}\;\Varid{u}\mathbin{+}\Varid{dot}^{-1}\;\Varid{v})){}\<[44]%
\>[44]{}\mbox{\onelinecomment  linearity of \ensuremath{\Varid{dot}}}{}\<[E]%
\\
\>[B]{}\mathrel{=}{}\<[BE]%
\>[5]{}\Varid{dot}^{-1}\;\Varid{u}\mathbin{+}\Varid{dot}^{-1}\;\Varid{v}{}\<[44]%
\>[44]{}\mbox{\onelinecomment  \ensuremath{\Varid{dot}^{-1}\hsdot{\circ }{.}\Varid{dot}\mathrel{=}\Varid{id}}}{}\<[E]%
\\[\blanklineskip]%
\>[5]{}\Varid{dot}^{-1}\;(\Varid{s} \cdot \Varid{u}){}\<[E]%
\\
\>[B]{}\mathrel{=}{}\<[BE]%
\>[5]{}\Varid{dot}^{-1}\;(\Varid{s} \cdot \Varid{dot}\;(\Varid{dot}^{-1}\;\Varid{u})){}\<[44]%
\>[44]{}\mbox{\onelinecomment  \ensuremath{\Varid{dot}\hsdot{\circ }{.}\Varid{dot}^{-1}\mathrel{=}\Varid{id}}}{}\<[E]%
\\
\>[B]{}\mathrel{=}{}\<[BE]%
\>[5]{}\Varid{dot}^{-1}\;(\Varid{dot}\;(\Varid{s} \cdot \Varid{dot}^{-1}\;\Varid{u})){}\<[44]%
\>[44]{}\mbox{\onelinecomment  linearity of \ensuremath{\Varid{dot}} }{}\<[E]%
\\
\>[B]{}\mathrel{=}{}\<[BE]%
\>[5]{}\Varid{s} \cdot \Varid{dot}^{-1}\;\Varid{u}{}\<[44]%
\>[44]{}\mbox{\onelinecomment  \ensuremath{\Varid{dot}^{-1}\hsdot{\circ }{.}\Varid{dot}\mathrel{=}\Varid{id}}}{}\<[E]%
\ColumnHook
\end{hscode}\resethooks
\item Noting that the argument of both sides is a pair,
\begin{hscode}\SaveRestoreHook
\column{B}{@{}>{\hspre}c<{\hspost}@{}}%
\column{BE}{@{}l@{}}%
\column{5}{@{}>{\hspre}l<{\hspost}@{}}%
\column{72}{@{}>{\hspre}l<{\hspost}@{}}%
\column{E}{@{}>{\hspre}l<{\hspost}@{}}%
\>[5]{}\Varid{unjoin}\hsdot{\circ }{.}\Varid{dot}{}\<[E]%
\\
\>[B]{}\mathrel{=}{}\<[BE]%
\>[5]{}\lambda (\Varid{u},\Varid{v})\to \Varid{unjoin}\;(\Varid{dot}\;(\Varid{u},\Varid{v})){}\<[72]%
\>[72]{}\mbox{\onelinecomment  $\eta$-expansion}{}\<[E]%
\\
\>[B]{}\mathrel{=}{}\<[BE]%
\>[5]{}\lambda (\Varid{u},\Varid{v})\to (\Varid{dot}\;(\Varid{u},\Varid{v})\hsdot{\circ }{.}\Varid{inl},\Varid{dot}\;(\Varid{u},\Varid{v})\hsdot{\circ }{.}\Varid{inr}){}\<[72]%
\>[72]{}\mbox{\onelinecomment  definition of \ensuremath{\Varid{unjoin}}}{}\<[E]%
\\
\>[B]{}\mathrel{=}{}\<[BE]%
\>[5]{}\lambda (\Varid{u},\Varid{v})\to (\lambda \Varid{x}\to \Varid{dot}\;(\Varid{u},\Varid{v})\;(\Varid{inl}\;\Varid{x}),\lambda \Varid{y}\to \Varid{dot}\;(\Varid{u},\Varid{v})\;(\Varid{inr}\;\Varid{y})){}\<[72]%
\>[72]{}\mbox{\onelinecomment  def'n of \ensuremath{(\hsdot{\circ }{.})} for \ensuremath{(\to )}}{}\<[E]%
\\
\>[B]{}\mathrel{=}{}\<[BE]%
\>[5]{}\lambda (\Varid{u},\Varid{v})\to (\lambda \Varid{x}\to \Varid{dot}\;(\Varid{u},\Varid{v})\;(\Varid{x},\mathrm{0}),\lambda \Varid{y}\to \Varid{dot}\;(\Varid{u},\Varid{v})\;(\mathrm{0},\Varid{y})){}\<[72]%
\>[72]{}\mbox{\onelinecomment  def'n of \ensuremath{\Varid{inl}} for \ensuremath{(\multimap)}}{}\<[E]%
\\
\>[B]{}\mathrel{=}{}\<[BE]%
\>[5]{}\lambda (\Varid{u},\Varid{v})\to (\lambda \Varid{x}\to \Varid{dot}\;\Varid{u}\;\Varid{x}\mathbin{+}\Varid{dot}\;\Varid{v}\;\mathrm{0},\lambda \Varid{y}\to \Varid{dot}\;\Varid{u}\;\mathrm{0}\mathbin{+}\Varid{dot}\;\Varid{v}\;\Varid{y}){}\<[72]%
\>[72]{}\mbox{\onelinecomment  def'n of \ensuremath{\Varid{dot}} for pairs}{}\<[E]%
\\
\>[B]{}\mathrel{=}{}\<[BE]%
\>[5]{}\lambda (\Varid{u},\Varid{v})\to (\lambda \Varid{x}\to \Varid{dot}\;\Varid{u}\;\Varid{x},\lambda \Varid{y}\to \Varid{dot}\;\Varid{v}\;\Varid{y}){}\<[72]%
\>[72]{}\mbox{\onelinecomment  linearity of \ensuremath{\Varid{dot}}}{}\<[E]%
\\
\>[B]{}\mathrel{=}{}\<[BE]%
\>[5]{}\lambda (\Varid{u},\Varid{v})\to (\Varid{dot}\;\Varid{u},\Varid{dot}\;\Varid{v}){}\<[72]%
\>[72]{}\mbox{\onelinecomment  $\eta$-reduction}{}\<[E]%
\\
\>[B]{}\mathrel{=}{}\<[BE]%
\>[5]{}\Varid{dot}\times\Varid{dot}{}\<[72]%
\>[72]{}\mbox{\onelinecomment  def'n of \ensuremath{(\times)} for \ensuremath{(\to )}}{}\<[E]%
\ColumnHook
\end{hscode}\resethooks
\item Follows from inverting each side of part \ref{unjoin-dot}.
\item Noting again that the argument of both sides is a pair,
\begin{hscode}\SaveRestoreHook
\column{B}{@{}>{\hspre}c<{\hspost}@{}}%
\column{BE}{@{}l@{}}%
\column{5}{@{}>{\hspre}l<{\hspost}@{}}%
\column{48}{@{}>{\hspre}l<{\hspost}@{}}%
\column{E}{@{}>{\hspre}l<{\hspost}@{}}%
\>[5]{}\Varid{dot}\;\Varid{u}\mathbin{\triangledown}\Varid{dot}\;\Varid{v}{}\<[E]%
\\
\>[B]{}\mathrel{=}{}\<[BE]%
\>[5]{}\Varid{jam}\hsdot{\circ }{.}(\Varid{dot}\;\Varid{u}\times\Varid{dot}\;\Varid{v}){}\<[48]%
\>[48]{}\mbox{\onelinecomment  definition of \ensuremath{(\mathbin{\triangledown})}}{}\<[E]%
\\
\>[B]{}\mathrel{=}{}\<[BE]%
\>[5]{}\lambda (\Varid{x},\Varid{y})\to \Varid{jam}\;((\Varid{dot}\;\Varid{u}\times\Varid{dot}\;\Varid{v})\;(\Varid{x},\Varid{y})){}\<[48]%
\>[48]{}\mbox{\onelinecomment  definition of \ensuremath{(\hsdot{\circ }{.})} for functions}{}\<[E]%
\\
\>[B]{}\mathrel{=}{}\<[BE]%
\>[5]{}\lambda (\Varid{x},\Varid{y})\to \Varid{jam}\;(\Varid{dot}\;\Varid{u}\;\Varid{x},\Varid{dot}\;\Varid{v}\;\Varid{y}){}\<[48]%
\>[48]{}\mbox{\onelinecomment  definition of \ensuremath{(\times)} for functions}{}\<[E]%
\\
\>[B]{}\mathrel{=}{}\<[BE]%
\>[5]{}\lambda (\Varid{x},\Varid{y})\to \Varid{dot}\;\Varid{u}\;\Varid{x}\mathbin{+}\Varid{dot}\;\Varid{v}\;\Varid{y}{}\<[48]%
\>[48]{}\mbox{\onelinecomment  definition of \ensuremath{\Varid{jam}} for functions}{}\<[E]%
\\
\>[B]{}\mathrel{=}{}\<[BE]%
\>[5]{}\lambda (\Varid{x},\Varid{y})\to \Varid{dot}\;(\Varid{u},\Varid{v})\;(\Varid{x},\Varid{y}){}\<[48]%
\>[48]{}\mbox{\onelinecomment  definition of \ensuremath{\Varid{dot}} for pairs}{}\<[E]%
\\
\>[B]{}\mathrel{=}{}\<[BE]%
\>[5]{}\Varid{dot}\;(\Varid{u},\Varid{v}){}\<[48]%
\>[48]{}\mbox{\onelinecomment  $\eta$-reduction}{}\<[E]%
\ColumnHook
\end{hscode}\resethooks
\item Immediate from linearity and the definition of \ensuremath{\mathrm{0}} for functions.
\end{enumerate}
\emph{End of proof of \lemRef{dot-properties}}.\\

Recall the definition of \ensuremath{\Varid{asDual}} from \secref{Gradients and Duality}:
\begin{hscode}\SaveRestoreHook
\column{B}{@{}>{\hspre}l<{\hspost}@{}}%
\column{E}{@{}>{\hspre}l<{\hspost}@{}}%
\>[B]{}\Varid{asDual}\mathbin{::}(\Conid{HasDot}^{\Varid{s}}\;\Varid{a},\Conid{HasDot}^{\Varid{s}}\;\Varid{b})\Rightarrow \Conid{Cont}_{\Varid{k}}^{\Varid{s}}\;\Varid{a}\;\Varid{b}\to \Conid{Dual}_{\Varid{k}}\;\Varid{a}\;\Varid{b}{}\<[E]%
\\
\>[B]{}\Varid{asDual}\;(\Conid{Cont}\;\Varid{f})\mathrel{=}\Conid{Dual}\;(\Varid{onDot}\;\Varid{f}){}\<[E]%
\ColumnHook
\end{hscode}\resethooks
where
\begin{hscode}\SaveRestoreHook
\column{B}{@{}>{\hspre}l<{\hspost}@{}}%
\column{E}{@{}>{\hspre}l<{\hspost}@{}}%
\>[B]{}\Varid{onDot}\mathbin{::}(\Conid{HasDot}^{\Varid{s}}\;\Varid{a},\Conid{HasDot}^{\Varid{s}}\;\Varid{b})\Rightarrow ((\Varid{b}\multimap\Varid{s})\to (\Varid{a}\multimap\Varid{s}))\to (\Varid{b}\multimap\Varid{a}){}\<[E]%
\\
\>[B]{}\Varid{onDot}\;\Varid{f}\mathrel{=}\Varid{dot}^{-1}\hsdot{\circ }{.}\Varid{f}\hsdot{\circ }{.}\Varid{dot}{}\<[E]%
\ColumnHook
\end{hscode}\resethooks
For the \ensuremath{\Conid{Category}} instance of \ensuremath{\Conid{Dual}_{\Varid{k}}}, we'll need that \ensuremath{\Varid{id}\mathrel{=}\Varid{asDual}\;\Varid{id}}.
Simplifying the RHS,
\begin{hscode}\SaveRestoreHook
\column{B}{@{}>{\hspre}c<{\hspost}@{}}%
\column{BE}{@{}l@{}}%
\column{5}{@{}>{\hspre}l<{\hspost}@{}}%
\column{30}{@{}>{\hspre}l<{\hspost}@{}}%
\column{E}{@{}>{\hspre}l<{\hspost}@{}}%
\>[5]{}\Varid{asDual}\;\Varid{id}{}\<[E]%
\\
\>[B]{}\mathrel{=}{}\<[BE]%
\>[5]{}\Varid{asDual}\;(\Conid{Cont}\;\Varid{id}){}\<[30]%
\>[30]{}\mbox{\onelinecomment  definition of \ensuremath{\Varid{id}} for \ensuremath{\Conid{Cont}_{\Varid{k}}^{\Varid{r}}} (\figref{cont})}{}\<[E]%
\\
\>[B]{}\mathrel{=}{}\<[BE]%
\>[5]{}\Conid{Dual}\;(\Varid{dot}^{-1}\hsdot{\circ }{.}\Varid{id}\hsdot{\circ }{.}\Varid{dot}){}\<[30]%
\>[30]{}\mbox{\onelinecomment  definition of \ensuremath{\Varid{asDual}}}{}\<[E]%
\\
\>[B]{}\mathrel{=}{}\<[BE]%
\>[5]{}\Conid{Dual}\;(\Varid{dot}^{-1}\hsdot{\circ }{.}\Varid{dot}){}\<[30]%
\>[30]{}\mbox{\onelinecomment  \ensuremath{\Conid{Category}} law for \ensuremath{\Varid{id}}/\ensuremath{(\hsdot{\circ }{.})}}{}\<[E]%
\\
\>[B]{}\mathrel{=}{}\<[BE]%
\>[5]{}\Conid{Dual}\;\Varid{id}{}\<[30]%
\>[30]{}\mbox{\onelinecomment  \ensuremath{\Varid{dot}^{-1}\hsdot{\circ }{.}\Varid{dot}\mathrel{=}\Varid{id}}}{}\<[E]%
\ColumnHook
\end{hscode}\resethooks
We also need \ensuremath{\Varid{asDual}\;(\Varid{g}\hsdot{\circ }{.}\Varid{f})\mathrel{=}\Varid{asDual}\;\Varid{g}\hsdot{\circ }{.}\Varid{asDual}\;\Varid{f}}, or (without loss of generality)
\begin{hscode}\SaveRestoreHook
\column{B}{@{}>{\hspre}l<{\hspost}@{}}%
\column{E}{@{}>{\hspre}l<{\hspost}@{}}%
\>[B]{}\Varid{asDual}\;(\Conid{Cont}\;\Varid{g}\hsdot{\circ }{.}\Conid{Cont}\;\Varid{f})\mathrel{=}\Varid{asDual}\;(\Conid{Cont}\;\Varid{g})\hsdot{\circ }{.}\Varid{asDual}\;(\Conid{Cont}\;\Varid{f}){}\<[E]%
\ColumnHook
\end{hscode}\resethooks
Simplifying both sides,
\begin{hscode}\SaveRestoreHook
\column{B}{@{}>{\hspre}c<{\hspost}@{}}%
\column{BE}{@{}l@{}}%
\column{5}{@{}>{\hspre}l<{\hspost}@{}}%
\column{47}{@{}>{\hspre}l<{\hspost}@{}}%
\column{E}{@{}>{\hspre}l<{\hspost}@{}}%
\>[5]{}\Varid{asDual}\;(\Conid{Cont}\;\Varid{g}\hsdot{\circ }{.}\Conid{Cont}\;\Varid{f}){}\<[E]%
\\
\>[B]{}\mathrel{=}{}\<[BE]%
\>[5]{}\Varid{asDual}\;(\Conid{Cont}\;(\Varid{f}\hsdot{\circ }{.}\Varid{g})){}\<[47]%
\>[47]{}\mbox{\onelinecomment  definition of \ensuremath{(\hsdot{\circ }{.})} for \ensuremath{\Conid{Cont}_{\Varid{k}}^{\Varid{r}}}}{}\<[E]%
\\
\>[B]{}\mathrel{=}{}\<[BE]%
\>[5]{}\Conid{Dual}\;(\Varid{dot}^{-1}\hsdot{\circ }{.}\Varid{f}\hsdot{\circ }{.}\Varid{g}\hsdot{\circ }{.}\Varid{dot}){}\<[47]%
\>[47]{}\mbox{\onelinecomment  definition of \ensuremath{\Varid{asDual}}}{}\<[E]%
\\
\>[B]{}\mathrel{=}{}\<[BE]%
\>[5]{}\Conid{Dual}\;(\Varid{dot}^{-1}\hsdot{\circ }{.}\Varid{f}\hsdot{\circ }{.}\Varid{dot}\hsdot{\circ }{.}\Varid{dot}^{-1}\hsdot{\circ }{.}\Varid{g}\hsdot{\circ }{.}\Varid{dot}){}\<[47]%
\>[47]{}\mbox{\onelinecomment  \ensuremath{\Varid{dot}\hsdot{\circ }{.}\Varid{dot}^{-1}\mathrel{=}\Varid{id}}}{}\<[E]%
\\
\>[B]{}\mathrel{=}{}\<[BE]%
\>[5]{}\Conid{Dual}\;(\Varid{onDot}\;\Varid{f}\hsdot{\circ }{.}\Varid{onDot}\;\Varid{g}){}\<[47]%
\>[47]{}\mbox{\onelinecomment  definition of \ensuremath{\Varid{onDot}}}{}\<[E]%
\\[\blanklineskip]%
\>[5]{}\Varid{asDual}\;(\Conid{Cont}\;\Varid{g})\hsdot{\circ }{.}\Varid{asDual}\;(\Conid{Cont}\;\Varid{f}){}\<[E]%
\\
\>[B]{}\mathrel{=}{}\<[BE]%
\>[5]{}\Conid{Dual}\;(\Varid{onDot}\;\Varid{g})\hsdot{\circ }{.}\Varid{asDual}\;(\Varid{onDot}\;\Varid{f}){}\<[47]%
\>[47]{}\mbox{\onelinecomment  definition of \ensuremath{\Varid{asDual}}}{}\<[E]%
\ColumnHook
\end{hscode}\resethooks
As usual, strengthen this equality by replacing \ensuremath{\Varid{onDot}\;\Varid{g}} and \ensuremath{\Varid{onDot}\;\Varid{f}} by re-typed \ensuremath{\Varid{g}} and \ensuremath{\Varid{f}}, and read off a sufficient definition:
\begin{hscode}\SaveRestoreHook
\column{B}{@{}>{\hspre}l<{\hspost}@{}}%
\column{E}{@{}>{\hspre}l<{\hspost}@{}}%
\>[B]{}\Conid{Dual}\;(\Varid{f}\hsdot{\circ }{.}\Varid{g})\mathrel{=}\Conid{Dual}\;\Varid{g}\hsdot{\circ }{.}\Varid{asDual}\;\Varid{f}{}\<[E]%
\ColumnHook
\end{hscode}\resethooks

For \ensuremath{\Conid{Monoidal}}, the homomorphism condition is \ensuremath{\Varid{asDual}\;(\Varid{f}\times\Varid{g})\mathrel{=}\Varid{asDual}\;\Varid{f}\times\Varid{asDual}\;\Varid{g}}.
Simplify both sides:
\begin{hscode}\SaveRestoreHook
\column{B}{@{}>{\hspre}c<{\hspost}@{}}%
\column{BE}{@{}l@{}}%
\column{5}{@{}>{\hspre}l<{\hspost}@{}}%
\column{59}{@{}>{\hspre}l<{\hspost}@{}}%
\column{E}{@{}>{\hspre}l<{\hspost}@{}}%
\>[5]{}\Varid{asDual}\;(\Conid{Cont}\;\Varid{f})\times\Varid{asDual}\;(\Conid{Cont}\;\Varid{g}){}\<[E]%
\\
\>[B]{}\mathrel{=}{}\<[BE]%
\>[5]{}\Conid{Dual}\;(\Varid{onDot}\;\Varid{f})\times\Conid{Dual}\;(\Varid{onDot}\;\Varid{g}){}\<[59]%
\>[59]{}\mbox{\onelinecomment  definition of \ensuremath{\Varid{asDual}}}{}\<[E]%
\\[\blanklineskip]%
\>[5]{}\Varid{asDual}\;(\Conid{Cont}\;\Varid{f}\times\Conid{Cont}\;\Varid{g}){}\<[E]%
\\
\>[B]{}\mathrel{=}{}\<[BE]%
\>[5]{}\Varid{asDual}\;(\Conid{Cont}\;(\Varid{join}\hsdot{\circ }{.}(\Varid{f}\times\Varid{g})\hsdot{\circ }{.}\Varid{unjoin})){}\<[59]%
\>[59]{}\mbox{\onelinecomment  definition of \ensuremath{(\times)} on \ensuremath{\Conid{Cont}}}{}\<[E]%
\\
\>[B]{}\mathrel{=}{}\<[BE]%
\>[5]{}\Conid{Dual}\;(\Varid{onDot}\;(\Varid{join}\hsdot{\circ }{.}(\Varid{f}\times\Varid{g})\hsdot{\circ }{.}\Varid{unjoin})){}\<[59]%
\>[59]{}\mbox{\onelinecomment  definition of \ensuremath{\Varid{asDual}}}{}\<[E]%
\\
\>[B]{}\mathrel{=}{}\<[BE]%
\>[5]{}\Conid{Dual}\;(\Varid{dot}^{-1}\hsdot{\circ }{.}\Varid{join}\hsdot{\circ }{.}(\Varid{f}\times\Varid{g})\hsdot{\circ }{.}\Varid{unjoin}\hsdot{\circ }{.}\Varid{dot}){}\<[59]%
\>[59]{}\mbox{\onelinecomment  definition of \ensuremath{\Varid{onDot}}}{}\<[E]%
\\
\>[B]{}\mathrel{=}{}\<[BE]%
\>[5]{}\Conid{Dual}\;((\Varid{dot}^{-1}\times\Varid{dot}^{-1})\hsdot{\circ }{.}(\Varid{f}\times\Varid{g})\hsdot{\circ }{.}(\Varid{dot}\times\Varid{dot})){}\<[59]%
\>[59]{}\mbox{\onelinecomment  \lemDotTwo{unjoin-dot}{unDot-join} }{}\<[E]%
\\
\>[B]{}\mathrel{=}{}\<[BE]%
\>[5]{}\Conid{Dual}\;(\Varid{dot}^{-1}\hsdot{\circ }{.}\Varid{f}\hsdot{\circ }{.}\Varid{dot}\times\Varid{dot}^{-1}\hsdot{\circ }{.}\Varid{g}\hsdot{\circ }{.}\Varid{dot}^{-1}){}\<[59]%
\>[59]{}\mbox{\onelinecomment  law about \ensuremath{(\times)}/\ensuremath{(\hsdot{\circ }{.})}}{}\<[E]%
\\
\>[B]{}\mathrel{=}{}\<[BE]%
\>[5]{}\Conid{Dual}\;(\Varid{onDot}\;\Varid{f}\times\Varid{onDot}\;\Varid{g}){}\<[59]%
\>[59]{}\mbox{\onelinecomment  definition of \ensuremath{\Varid{onDot}}}{}\<[E]%
\ColumnHook
\end{hscode}\resethooks
Strengthening from \ensuremath{\Varid{onDot}\;\Varid{f}} and \ensuremath{\Varid{onDot}\;\Varid{g}} gives a simple sufficient condition:
\begin{hscode}\SaveRestoreHook
\column{B}{@{}>{\hspre}l<{\hspost}@{}}%
\column{E}{@{}>{\hspre}l<{\hspost}@{}}%
\>[B]{}\Conid{Dual}\;\Varid{f}\times\Conid{Dual}\;\Varid{g}\mathrel{=}\Conid{Dual}\;(\Varid{f}\times\Varid{g}){}\<[E]%
\ColumnHook
\end{hscode}\resethooks

For \ensuremath{\Conid{Cartesian}},
\begin{hscode}\SaveRestoreHook
\column{B}{@{}>{\hspre}c<{\hspost}@{}}%
\column{BE}{@{}l@{}}%
\column{5}{@{}>{\hspre}l<{\hspost}@{}}%
\column{49}{@{}>{\hspre}l<{\hspost}@{}}%
\column{58}{@{}>{\hspre}l<{\hspost}@{}}%
\column{E}{@{}>{\hspre}l<{\hspost}@{}}%
\>[5]{}\Varid{exl}{}\<[E]%
\\
\>[B]{}\mathrel{=}{}\<[BE]%
\>[5]{}\Varid{asDual}\;\Varid{exl}{}\<[49]%
\>[49]{}\mbox{\onelinecomment  specification}{}\<[E]%
\\
\>[B]{}\mathrel{=}{}\<[BE]%
\>[5]{}\Varid{asDual}\;(\Conid{Cont}\;(\Varid{join}\hsdot{\circ }{.}\Varid{inl})){}\<[49]%
\>[49]{}\mbox{\onelinecomment  definition of \ensuremath{\Varid{exl}} for \ensuremath{\Conid{Cont}_{\Varid{k}}^{\Varid{r}}}}{}\<[E]%
\\
\>[B]{}\mathrel{=}{}\<[BE]%
\>[5]{}\Conid{Dual}\;(\Varid{onDot}\;(\Varid{join}\hsdot{\circ }{.}\Varid{inl})){}\<[49]%
\>[49]{}\mbox{\onelinecomment  definition of \ensuremath{\Varid{asDual}}}{}\<[E]%
\\
\>[B]{}\mathrel{=}{}\<[BE]%
\>[5]{}\Conid{Dual}\;(\Varid{dot}^{-1}\hsdot{\circ }{.}\Varid{join}\hsdot{\circ }{.}\Varid{inl}\hsdot{\circ }{.}\Varid{dot}){}\<[49]%
\>[49]{}\mbox{\onelinecomment  definition of \ensuremath{\Varid{onDot}}, and associativity of \ensuremath{(\hsdot{\circ }{.})}}{}\<[E]%
\\
\>[B]{}\mathrel{=}{}\<[BE]%
\>[5]{}\Conid{Dual}\;(\lambda \Varid{u}\to \Varid{dot}^{-1}\;(\Varid{join}\;(\Varid{inl}\;(\Varid{dot}\;\Varid{u})))){}\<[49]%
\>[49]{}\mbox{\onelinecomment  definition of \ensuremath{(\hsdot{\circ }{.})} for functions}{}\<[E]%
\\
\>[B]{}\mathrel{=}{}\<[BE]%
\>[5]{}\Conid{Dual}\;(\lambda \Varid{u}\to \Varid{dot}^{-1}\;(\Varid{join}\;(\Varid{dot}\;\Varid{u},\mathrm{0}))){}\<[49]%
\>[49]{}\mbox{\onelinecomment  definition of \ensuremath{\Varid{inl}} for functions}{}\<[E]%
\\
\>[B]{}\mathrel{=}{}\<[BE]%
\>[5]{}\Conid{Dual}\;(\lambda \Varid{u}\to \Varid{dot}^{-1}\;(\Varid{dot}\;\Varid{u}\mathbin{\triangledown}\mathrm{0})){}\<[49]%
\>[49]{}\mbox{\onelinecomment  definition of \ensuremath{\Varid{join}}}{}\<[E]%
\\
\>[B]{}\mathrel{=}{}\<[BE]%
\>[5]{}\Conid{Dual}\;(\lambda \Varid{u}\to \Varid{dot}^{-1}\;(\Varid{dot}\;\Varid{u}\mathbin{\triangledown}\Varid{dot}\;\mathrm{0})){}\<[49]%
\>[49]{}\mbox{\onelinecomment  \lemDot{dot-zeroV}}{}\<[E]%
\\
\>[B]{}\mathrel{=}{}\<[BE]%
\>[5]{}\Conid{Dual}\;(\lambda \Varid{u}\to \Varid{dot}^{-1}\;(\Varid{dot}\;(\Varid{u},\mathrm{0}))){}\<[49]%
\>[49]{}\mbox{\onelinecomment  \lemDot{dot-dot-join}}{}\<[E]%
\\
\>[B]{}\mathrel{=}{}\<[BE]%
\>[5]{}\Conid{Dual}\;(\lambda \Varid{u}\to (\Varid{u},\mathrm{0})){}\<[49]%
\>[49]{}\mbox{\onelinecomment  \ensuremath{\Varid{dot}^{-1}\hsdot{\circ }{.}\Varid{dot}\mathrel{=}\Varid{id}}}{}\<[E]%
\\
\>[B]{}\mathrel{=}{}\<[BE]%
\>[5]{}\Conid{Dual}\;(\lambda \Varid{u}\to \Varid{inl}\;\Varid{u}){}\<[49]%
\>[49]{}\mbox{\onelinecomment  definition of \ensuremath{\Varid{inl}} for functions}{}\<[E]%
\\
\>[B]{}\mathrel{=}{}\<[BE]%
\>[5]{}\Conid{Dual}\;\Varid{inl}{}\<[49]%
\>[49]{}\mbox{\onelinecomment  $\eta$-reduction}{}\<[E]%
\\[\blanklineskip]%
\>[5]{}\Varid{exrP}{}\<[E]%
\\
\>[B]{}\mathrel{=}{}\<[BE]%
\>[5]{}\Conid{Dual}\;\Varid{inr}{}\<[49]%
\>[49]{}\mbox{\onelinecomment  as with \ensuremath{\Varid{exlP}}}{}\<[E]%
\\[\blanklineskip]%
\>[5]{}\Varid{dup}{}\<[E]%
\\
\>[B]{}\mathrel{=}{}\<[BE]%
\>[5]{}\Varid{asDual}\;\Varid{dup}{}\<[58]%
\>[58]{}\mbox{\onelinecomment  specification}{}\<[E]%
\\
\>[B]{}\mathrel{=}{}\<[BE]%
\>[5]{}\Varid{asDual}\;(\Conid{Cont}\;(\Varid{jam}\hsdot{\circ }{.}\Varid{unjoin})){}\<[58]%
\>[58]{}\mbox{\onelinecomment  definition of \ensuremath{\Varid{dup}} for \ensuremath{\Conid{Cont}_{\Varid{k}}^{\Varid{r}}}}{}\<[E]%
\\
\>[B]{}\mathrel{=}{}\<[BE]%
\>[5]{}\Conid{Dual}\;(\Varid{onDot}\;(\Varid{jam}\hsdot{\circ }{.}\Varid{unjoin})){}\<[58]%
\>[58]{}\mbox{\onelinecomment  definition of \ensuremath{\Varid{asDual}}}{}\<[E]%
\\
\>[B]{}\mathrel{=}{}\<[BE]%
\>[5]{}\Conid{Dual}\;(\Varid{dot}^{-1}\hsdot{\circ }{.}\Varid{jam}\hsdot{\circ }{.}\Varid{unjoin}\hsdot{\circ }{.}\Varid{dot}){}\<[58]%
\>[58]{}\mbox{\onelinecomment  definition of \ensuremath{\Varid{onDot}}}{}\<[E]%
\\
\>[B]{}\mathrel{=}{}\<[BE]%
\>[5]{}\Conid{Dual}\;(\lambda (\Varid{u},\Varid{v})\to \Varid{dot}^{-1}\;(\Varid{jam}\;(\Varid{unjoin}\;(\Varid{dot}\;(\Varid{u},\Varid{v}))))){}\<[58]%
\>[58]{}\mbox{\onelinecomment  definition of \ensuremath{(\hsdot{\circ }{.})} for functions}{}\<[E]%
\\
\>[B]{}\mathrel{=}{}\<[BE]%
\>[5]{}\Conid{Dual}\;(\lambda (\Varid{u},\Varid{v})\to \Varid{dot}^{-1}\;(\Varid{jam}\;(\Varid{dot}\;\Varid{u},\Varid{dot}\;\Varid{v}))){}\<[58]%
\>[58]{}\mbox{\onelinecomment  \lemDot{unjoin-dot}}{}\<[E]%
\\
\>[B]{}\mathrel{=}{}\<[BE]%
\>[5]{}\Conid{Dual}\;(\lambda (\Varid{u},\Varid{v})\to \Varid{dot}^{-1}\;(\Varid{dot}\;\Varid{u}\mathbin{+}\Varid{dot}\;\Varid{v})){}\<[58]%
\>[58]{}\mbox{\onelinecomment  definition of \ensuremath{\Varid{jam}} for functions}{}\<[E]%
\\
\>[B]{}\mathrel{=}{}\<[BE]%
\>[5]{}\Conid{Dual}\;(\lambda (\Varid{u},\Varid{v})\to \Varid{dot}^{-1}\;(\Varid{dot}\;\Varid{u})\mathbin{+}\Varid{dot}^{-1}\;(\Varid{dot}\;\Varid{v})){}\<[58]%
\>[58]{}\mbox{\onelinecomment  \lemDot{unDot-linear}}{}\<[E]%
\\
\>[B]{}\mathrel{=}{}\<[BE]%
\>[5]{}\Conid{Dual}\;(\lambda (\Varid{u},\Varid{v})\to \Varid{u}\mathbin{+}\Varid{v}){}\<[58]%
\>[58]{}\mbox{\onelinecomment  \ensuremath{\Varid{dot}^{-1}\hsdot{\circ }{.}\Varid{dot}\mathrel{=}\Varid{id}}}{}\<[E]%
\\
\>[B]{}\mathrel{=}{}\<[BE]%
\>[5]{}\Conid{Dual}\;\Varid{jam}{}\<[58]%
\>[58]{}\mbox{\onelinecomment  definition of \ensuremath{\Varid{jam}} for functions}{}\<[E]%
\ColumnHook
\end{hscode}\resethooks
The \ensuremath{\Conid{Cocartesian}} instance comes out similarly:
\begin{hscode}\SaveRestoreHook
\column{B}{@{}>{\hspre}c<{\hspost}@{}}%
\column{BE}{@{}l@{}}%
\column{5}{@{}>{\hspre}l<{\hspost}@{}}%
\column{49}{@{}>{\hspre}l<{\hspost}@{}}%
\column{58}{@{}>{\hspre}l<{\hspost}@{}}%
\column{E}{@{}>{\hspre}l<{\hspost}@{}}%
\>[5]{}\Varid{inl}{}\<[E]%
\\
\>[B]{}\mathrel{=}{}\<[BE]%
\>[5]{}\Varid{asDual}\;\Varid{inl}{}\<[58]%
\>[58]{}\mbox{\onelinecomment  specification}{}\<[E]%
\\
\>[B]{}\mathrel{=}{}\<[BE]%
\>[5]{}\Varid{asDual}\;(\Conid{Cont}\;(\Varid{exl}\hsdot{\circ }{.}\Varid{unjoin})){}\<[58]%
\>[58]{}\mbox{\onelinecomment  definition of \ensuremath{\Varid{inl}} for \ensuremath{\Conid{Cont}_{\Varid{k}}^{\Varid{r}}}}{}\<[E]%
\\
\>[B]{}\mathrel{=}{}\<[BE]%
\>[5]{}\Conid{Dual}\;(\Varid{onDot}\;(\Varid{exl}\hsdot{\circ }{.}\Varid{unjoin})){}\<[58]%
\>[58]{}\mbox{\onelinecomment  definition of \ensuremath{\Varid{asDual}}}{}\<[E]%
\\
\>[B]{}\mathrel{=}{}\<[BE]%
\>[5]{}\Conid{Dual}\;(\Varid{dot}^{-1}\hsdot{\circ }{.}\Varid{exl}\hsdot{\circ }{.}\Varid{unjoin}\hsdot{\circ }{.}\Varid{dot}){}\<[58]%
\>[58]{}\mbox{\onelinecomment  definition of \ensuremath{\Varid{onDot}}}{}\<[E]%
\\
\>[B]{}\mathrel{=}{}\<[BE]%
\>[5]{}\Conid{Dual}\;(\lambda (\Varid{u},\Varid{v})\to \Varid{dot}^{-1}\;(\Varid{exl}\;(\Varid{unjoin}\;(\Varid{dot}\;(\Varid{u},\Varid{v}))))){}\<[58]%
\>[58]{}\mbox{\onelinecomment  definition of \ensuremath{(\hsdot{\circ }{.})} for functions}{}\<[E]%
\\
\>[B]{}\mathrel{=}{}\<[BE]%
\>[5]{}\Conid{Dual}\;(\lambda (\Varid{u},\Varid{v})\to \Varid{dot}^{-1}\;(\Varid{exl}\;(\Varid{dot}\;\Varid{u},\Varid{dot}\;\Varid{v}))){}\<[58]%
\>[58]{}\mbox{\onelinecomment  \lemDot{unjoin-dot}}{}\<[E]%
\\
\>[B]{}\mathrel{=}{}\<[BE]%
\>[5]{}\Conid{Dual}\;(\lambda (\Varid{u},\Varid{v})\to \Varid{dot}^{-1}\;(\Varid{dot}\;\Varid{u})){}\<[58]%
\>[58]{}\mbox{\onelinecomment  definition of \ensuremath{\Varid{exl}} on functions}{}\<[E]%
\\
\>[B]{}\mathrel{=}{}\<[BE]%
\>[5]{}\Conid{Dual}\;(\lambda (\Varid{u},\Varid{v})\to \Varid{u}){}\<[58]%
\>[58]{}\mbox{\onelinecomment  \ensuremath{\Varid{dot}^{-1}\hsdot{\circ }{.}\Varid{dot}\mathrel{=}\Varid{id}}}{}\<[E]%
\\
\>[B]{}\mathrel{=}{}\<[BE]%
\>[5]{}\Conid{Dual}\;\Varid{exl}{}\<[58]%
\>[58]{}\mbox{\onelinecomment  definition of \ensuremath{\Varid{exl}} for functions}{}\<[E]%
\\[\blanklineskip]%
\>[5]{}\Varid{inr}{}\<[E]%
\\
\>[B]{}\mathrel{=}{}\<[BE]%
\>[5]{}\Conid{Dual}\;\Varid{exr}{}\<[58]%
\>[58]{}\mbox{\onelinecomment  \ldots{} as with \ensuremath{\Varid{inl}} \ldots}{}\<[E]%
\\[\blanklineskip]%
\>[5]{}\Varid{jam}{}\<[E]%
\\
\>[B]{}\mathrel{=}{}\<[BE]%
\>[5]{}\Varid{asDual}\;\Varid{jam}{}\<[49]%
\>[49]{}\mbox{\onelinecomment  specification}{}\<[E]%
\\
\>[B]{}\mathrel{=}{}\<[BE]%
\>[5]{}\Varid{asDual}\;(\Conid{Cont}\;(\Varid{join}\hsdot{\circ }{.}\Varid{dup})){}\<[49]%
\>[49]{}\mbox{\onelinecomment  definition of \ensuremath{\Varid{jam}} on \ensuremath{\Conid{Cont}}}{}\<[E]%
\\
\>[B]{}\mathrel{=}{}\<[BE]%
\>[5]{}\Conid{Dual}\;(\Varid{onDot}\;(\Varid{join}\hsdot{\circ }{.}\Varid{dup})){}\<[49]%
\>[49]{}\mbox{\onelinecomment  definition of \ensuremath{\Varid{asDual}}}{}\<[E]%
\\
\>[B]{}\mathrel{=}{}\<[BE]%
\>[5]{}\Conid{Dual}\;(\Varid{dot}^{-1}\hsdot{\circ }{.}\Varid{join}\hsdot{\circ }{.}\Varid{dup}\hsdot{\circ }{.}\Varid{dot}){}\<[49]%
\>[49]{}\mbox{\onelinecomment  definition of \ensuremath{\Varid{onDot}}}{}\<[E]%
\\
\>[B]{}\mathrel{=}{}\<[BE]%
\>[5]{}\Conid{Dual}\;(\lambda \Varid{u}\to \Varid{dot}^{-1}\;(\Varid{join}\;(\Varid{dup}\;(\Varid{dot}\;\Varid{u})))){}\<[49]%
\>[49]{}\mbox{\onelinecomment  definition of \ensuremath{(\hsdot{\circ }{.})} on functions}{}\<[E]%
\\
\>[B]{}\mathrel{=}{}\<[BE]%
\>[5]{}\Conid{Dual}\;(\lambda \Varid{u}\to \Varid{dot}^{-1}\;(\Varid{join}\;(\Varid{dot}\;\Varid{u},\Varid{dot}\;\Varid{u}))){}\<[49]%
\>[49]{}\mbox{\onelinecomment  definition of \ensuremath{\Varid{dup}} for functions}{}\<[E]%
\\
\>[B]{}\mathrel{=}{}\<[BE]%
\>[5]{}\Conid{Dual}\;(\lambda \Varid{u}\to \Varid{dot}^{-1}\;(\Varid{dot}\;\Varid{u}\mathbin{\triangledown}\Varid{dot}\;\Varid{u})){}\<[49]%
\>[49]{}\mbox{\onelinecomment  definition of \ensuremath{\Varid{join}}}{}\<[E]%
\\
\>[B]{}\mathrel{=}{}\<[BE]%
\>[5]{}\Conid{Dual}\;(\lambda \Varid{u}\to \Varid{dot}^{-1}\;(\Varid{dot}\;(\Varid{u},\Varid{u}))){}\<[49]%
\>[49]{}\mbox{\onelinecomment  \lemDot{dot-dot-join}}{}\<[E]%
\\
\>[B]{}\mathrel{=}{}\<[BE]%
\>[5]{}\Conid{Dual}\;(\lambda \Varid{u}\to (\Varid{u},\Varid{u})){}\<[49]%
\>[49]{}\mbox{\onelinecomment  \ensuremath{\Varid{dot}^{-1}\hsdot{\circ }{.}\Varid{dot}\mathrel{=}\Varid{id}}}{}\<[E]%
\\
\>[B]{}\mathrel{=}{}\<[BE]%
\>[5]{}\Conid{Dual}\;(\lambda \Varid{u}\to \Varid{dup}\;\Varid{u}){}\<[49]%
\>[49]{}\mbox{\onelinecomment  definition of \ensuremath{\Varid{dup}} on functions}{}\<[E]%
\\
\>[B]{}\mathrel{=}{}\<[BE]%
\>[5]{}\Conid{Dual}\;\Varid{dup}{}\<[49]%
\>[49]{}\mbox{\onelinecomment  $\eta$-reduction}{}\<[E]%
\ColumnHook
\end{hscode}\resethooks

Finally, scaling:
\begin{hscode}\SaveRestoreHook
\column{B}{@{}>{\hspre}c<{\hspost}@{}}%
\column{BE}{@{}l@{}}%
\column{5}{@{}>{\hspre}l<{\hspost}@{}}%
\column{35}{@{}>{\hspre}l<{\hspost}@{}}%
\column{E}{@{}>{\hspre}l<{\hspost}@{}}%
\>[5]{}\Varid{scale}\;\Varid{s}{}\<[E]%
\\
\>[B]{}\mathrel{=}{}\<[BE]%
\>[5]{}\Varid{asDual}\;(\Varid{scale}\;\Varid{s}){}\<[35]%
\>[35]{}\mbox{\onelinecomment  specification}{}\<[E]%
\\
\>[B]{}\mathrel{=}{}\<[BE]%
\>[5]{}\Varid{asDual}\;(\Conid{Cont}\;(\Varid{scale}\;\Varid{s})){}\<[35]%
\>[35]{}\mbox{\onelinecomment  definition of \ensuremath{\Varid{scale}} for \ensuremath{\Conid{Cont}_{\Varid{k}}^{\Varid{r}}}}{}\<[E]%
\\
\>[B]{}\mathrel{=}{}\<[BE]%
\>[5]{}\Conid{Dual}\;(\Varid{onDot}\;(\Varid{scale}\;\Varid{s})){}\<[35]%
\>[35]{}\mbox{\onelinecomment  definition of \ensuremath{\Varid{asDual}}}{}\<[E]%
\\
\>[B]{}\mathrel{=}{}\<[BE]%
\>[5]{}\Conid{Dual}\;(\Varid{dot}^{-1}\hsdot{\circ }{.}\Varid{scale}\;\Varid{s}\hsdot{\circ }{.}\Varid{dot}){}\<[35]%
\>[35]{}\mbox{\onelinecomment  definition of \ensuremath{\Varid{onDot}}}{}\<[E]%
\\
\>[B]{}\mathrel{=}{}\<[BE]%
\>[5]{}\Conid{Dual}\;(\Varid{scale}\;\Varid{s}\hsdot{\circ }{.}\Varid{dot}^{-1}\hsdot{\circ }{.}\Varid{dot}){}\<[35]%
\>[35]{}\mbox{\onelinecomment  \lemDot{unDot-linear}}{}\<[E]%
\\
\>[B]{}\mathrel{=}{}\<[BE]%
\>[5]{}\Conid{Dual}\;(\Varid{scale}\;\Varid{s}){}\<[35]%
\>[35]{}\mbox{\onelinecomment  \ensuremath{\Varid{dot}^{-1}\hsdot{\circ }{.}\Varid{dot}\mathrel{=}\Varid{id}}}{}\<[E]%
\ColumnHook
\end{hscode}\resethooks

\subsection{\corRef{dual-derived}}\proofLabel{corollary:dual-derived}
Given the definitions in \figref{asDual},
\begin{hscode}\SaveRestoreHook
\column{B}{@{}>{\hspre}c<{\hspost}@{}}%
\column{BE}{@{}l@{}}%
\column{5}{@{}>{\hspre}l<{\hspost}@{}}%
\column{33}{@{}>{\hspre}l<{\hspost}@{}}%
\column{E}{@{}>{\hspre}l<{\hspost}@{}}%
\>[5]{}\Conid{Dual}\;\Varid{f}\mathbin{\vartriangle}\Conid{Dual}\;\Varid{g}{}\<[E]%
\\
\>[B]{}\mathrel{=}{}\<[BE]%
\>[5]{}(\Conid{Dual}\;\Varid{f}\times\Conid{Dual}\;\Varid{g})\hsdot{\circ }{.}\Varid{dup}{}\<[33]%
\>[33]{}\mbox{\onelinecomment  definition of \ensuremath{(\mathbin{\vartriangle})}}{}\<[E]%
\\
\>[B]{}\mathrel{=}{}\<[BE]%
\>[5]{}\Conid{Dual}\;(\Varid{f}\times\Varid{g})\hsdot{\circ }{.}\Varid{dup}{}\<[33]%
\>[33]{}\mbox{\onelinecomment  definition of \ensuremath{(\times)} for \ensuremath{\Conid{Dual}_{\Varid{k}}}}{}\<[E]%
\\
\>[B]{}\mathrel{=}{}\<[BE]%
\>[5]{}\Conid{Dual}\;(\Varid{f}\times\Varid{g})\hsdot{\circ }{.}\Conid{Dual}\;\Varid{jam}{}\<[33]%
\>[33]{}\mbox{\onelinecomment  definition of \ensuremath{\Varid{dup}} for \ensuremath{\Conid{Dual}_{\Varid{k}}}}{}\<[E]%
\\
\>[B]{}\mathrel{=}{}\<[BE]%
\>[5]{}\Conid{Dual}\;(\Varid{jam}\hsdot{\circ }{.}(\Varid{f}\times\Varid{g})){}\<[33]%
\>[33]{}\mbox{\onelinecomment  definition of \ensuremath{(\hsdot{\circ }{.})} for \ensuremath{\Conid{Dual}_{\Varid{k}}}}{}\<[E]%
\\
\>[B]{}\mathrel{=}{}\<[BE]%
\>[5]{}\Conid{Dual}\;(\Varid{f}\mathbin{\triangledown}\Varid{g}){}\<[33]%
\>[33]{}\mbox{\onelinecomment  definition of \ensuremath{(\mathbin{\triangledown})}}{}\<[E]%
\\[\blanklineskip]%
\>[5]{}\Conid{Dual}\;\Varid{f}\mathbin{\triangledown}\Conid{Dual}\;\Varid{g}{}\<[E]%
\\
\>[B]{}\mathrel{=}{}\<[BE]%
\>[5]{}\Varid{jam}\hsdot{\circ }{.}(\Conid{Dual}\;\Varid{f}\times\Conid{Dual}\;\Varid{g}){}\<[33]%
\>[33]{}\mbox{\onelinecomment  definition of \ensuremath{(\mathbin{\triangledown})}}{}\<[E]%
\\
\>[B]{}\mathrel{=}{}\<[BE]%
\>[5]{}\Varid{jam}\hsdot{\circ }{.}\Conid{Dual}\;(\Varid{f}\times\Varid{g}){}\<[33]%
\>[33]{}\mbox{\onelinecomment  definition of \ensuremath{(\times)} for \ensuremath{\Conid{Dual}_{\Varid{k}}}}{}\<[E]%
\\
\>[B]{}\mathrel{=}{}\<[BE]%
\>[5]{}\Conid{Dual}\;\Varid{dup}\hsdot{\circ }{.}\Conid{Dual}\;(\Varid{f}\times\Varid{g}){}\<[33]%
\>[33]{}\mbox{\onelinecomment  definition of \ensuremath{\Varid{jam}} for \ensuremath{\Conid{Dual}_{\Varid{k}}}}{}\<[E]%
\\
\>[B]{}\mathrel{=}{}\<[BE]%
\>[5]{}\Conid{Dual}\;((\Varid{f}\times\Varid{g})\hsdot{\circ }{.}\Varid{dup}){}\<[33]%
\>[33]{}\mbox{\onelinecomment  definition of \ensuremath{(\hsdot{\circ }{.})} for \ensuremath{\Conid{Dual}_{\Varid{k}}}}{}\<[E]%
\\
\>[B]{}\mathrel{=}{}\<[BE]%
\>[5]{}\Conid{Dual}\;(\Varid{f}\mathbin{\vartriangle}\Varid{g}){}\<[33]%
\>[33]{}\mbox{\onelinecomment  definition of \ensuremath{(\mathbin{\vartriangle})}}{}\<[E]%
\ColumnHook
\end{hscode}\resethooks

\subsection{\thmRef{indexed}}\proofLabel{theorem:indexed}

We will need an indexed counterpart to \thmRef{cross}, which says
$$\ensuremath{\mathcal{D}\;(\Varid{f}\times\Varid{g})\;(\Varid{a},\Varid{b})\mathrel{=}\mathcal{D}\;\Varid{f}\;\Varid{a}\times\mathcal{D}\;\Varid{g}\;\Varid{b}}$$
Letting \ensuremath{\Varid{cross}\mathrel{=}\Varid{uncurry}\;(\times)}, we can rephrase this theorem:
\begin{hscode}\SaveRestoreHook
\column{B}{@{}>{\hspre}c<{\hspost}@{}}%
\column{BE}{@{}l@{}}%
\column{5}{@{}>{\hspre}l<{\hspost}@{}}%
\column{49}{@{}>{\hspre}l<{\hspost}@{}}%
\column{E}{@{}>{\hspre}l<{\hspost}@{}}%
\>[5]{}\mathcal{D}\;(\Varid{f}\times\Varid{g}){}\<[E]%
\\
\>[B]{}\mathrel{=}{}\<[BE]%
\>[5]{}\lambda (\Varid{a},\Varid{b})\to \mathcal{D}\;\Varid{f}\;\Varid{a}\times\mathcal{D}\;\Varid{g}\;\Varid{b}{}\<[49]%
\>[49]{}\mbox{\onelinecomment  \thmRef{cross}}{}\<[E]%
\\
\>[B]{}\mathrel{=}{}\<[BE]%
\>[5]{}\lambda (\Varid{a},\Varid{b})\to \Varid{cross}\;(\mathcal{D}\;\Varid{f}\;\Varid{a},\mathcal{D}\;\Varid{g}\;\Varid{b}){}\<[49]%
\>[49]{}\mbox{\onelinecomment  definition of \ensuremath{\Varid{cross}}}{}\<[E]%
\\
\>[B]{}\mathrel{=}{}\<[BE]%
\>[5]{}\lambda (\Varid{a},\Varid{b})\to \Varid{cross}\;((\mathcal{D}\;\Varid{f}\times\mathcal{D}\;\Varid{g})\;(\Varid{a},\Varid{b})){}\<[49]%
\>[49]{}\mbox{\onelinecomment  definition of \ensuremath{(\times)} on functions}{}\<[E]%
\\
\>[B]{}\mathrel{=}{}\<[BE]%
\>[5]{}\Varid{cross}\hsdot{\circ }{.}(\mathcal{D}\;\Varid{f}\times\mathcal{D}\;\Varid{g}){}\<[49]%
\>[49]{}\mbox{\onelinecomment  definition of \ensuremath{(\hsdot{\circ }{.})} on functions}{}\<[E]%
\ColumnHook
\end{hscode}\resethooks
Likewise, extend from binary to $n$-ary:
\begin{theorem}[indexed cross rule] \thmLabel{crossF}
$$\ensuremath{\mathcal{D}\;(\Varid{crossI}\;\Varid{fs})\mathrel{=}\Varid{crossI}\hsdot{\circ }{.}\Varid{crossI}\;(\Varid{fmap}\;\mathcal{D}\;\Varid{fs})}$$
\end{theorem}
If \ensuremath{\Varid{fs}\mathbin{::}\Varid{h}\;(\Varid{a}\to \Varid{b})}, then both sides of this equation have type \ensuremath{\Varid{h}\;\Varid{a}\to (\Varid{h}\;\Varid{a}\multimap\Varid{h}\;\Varid{b})}.
The proof is similar to \thmRef{cross} \citep[variant of Theorem 2-3 (3)]{Spivak65}.

\thmRef{crossF} gives us what we need to construct \ensuremath{\mathcal{D}\!^+\!\;(\Varid{crossI}\;\Varid{fs})} compositionally:
\begin{corollary} \corLabel{crossF}
\ensuremath{\mathcal{D}\!^+\!} is compositional with respect to \ensuremath{\Varid{crossI}}. Specifically,
$$\ensuremath{\mathcal{D}\!^+\!\;(\Varid{crossI}\;\Varid{fs})\mathrel{=}\Varid{second}\;\Varid{crossI}\hsdot{\circ }{.}\Varid{unzip}\hsdot{\circ }{.}\Varid{crossI}\;(\Varid{fmap}\;\mathcal{D}\!^+\!\;\Varid{fs})}$$
\end{corollary}
The proof is analogous to that of \corRef{cross}:
\begin{hscode}\SaveRestoreHook
\column{B}{@{}>{\hspre}c<{\hspost}@{}}%
\column{BE}{@{}l@{}}%
\column{5}{@{}>{\hspre}l<{\hspost}@{}}%
\column{75}{@{}>{\hspre}l<{\hspost}@{}}%
\column{E}{@{}>{\hspre}l<{\hspost}@{}}%
\>[5]{}\mathcal{D}\!^+\!\;(\Varid{crossI}\;\Varid{fs})\;\Varid{as}{}\<[E]%
\\
\>[B]{}\mathrel{=}{}\<[BE]%
\>[5]{}(\Varid{crossI}\;\Varid{fs}\;\Varid{as},\mathcal{D}\;(\Varid{crossI}\;\Varid{fs})\;\Varid{as}){}\<[75]%
\>[75]{}\mbox{\onelinecomment  definition of \ensuremath{\mathcal{D}\!^+\!}}{}\<[E]%
\\
\>[B]{}\mathrel{=}{}\<[BE]%
\>[5]{}(\Varid{crossI}\;\Varid{fs}\;\Varid{as},\Varid{crossI}\;(\Varid{crossI}\;(\Varid{fmap}\;\mathcal{D}\;\Varid{fs})\;\Varid{as})){}\<[75]%
\>[75]{}\mbox{\onelinecomment  \thmRef{crossF}}{}\<[E]%
\\
\>[B]{}\mathrel{=}{}\<[BE]%
\>[5]{}\Varid{second}\;\Varid{crossI}\;(\Varid{crossI}\;\Varid{fs}\;\Varid{as},\Varid{crossI}\;(\Varid{fmap}\;\mathcal{D}\;\Varid{fs})\;\Varid{as}){}\<[75]%
\>[75]{}\mbox{\onelinecomment  definition of \ensuremath{\Varid{second}} (\figref{indexed})}{}\<[E]%
\\
\>[B]{}\mathrel{=}{}\<[BE]%
\>[5]{}\Varid{second}\;\Varid{crossI}\;((\Varid{crossI}\;\Varid{fs}\mathbin{\vartriangle}\Varid{crossI}\;(\Varid{fmap}\;\mathcal{D}\;\Varid{fs}))\;\Varid{as}){}\<[75]%
\>[75]{}\mbox{\onelinecomment  definition of \ensuremath{(\mathbin{\vartriangle})} on functions}{}\<[E]%
\\
\>[B]{}\mathrel{=}{}\<[BE]%
\>[5]{}(\Varid{second}\;\Varid{crossI}\hsdot{\circ }{.}(\Varid{crossI}\;\Varid{fs}\mathbin{\vartriangle}\Varid{crossI}\;(\Varid{fmap}\;\mathcal{D}\;\Varid{fs})))\;\Varid{as}{}\<[75]%
\>[75]{}\mbox{\onelinecomment  definition of \ensuremath{(\hsdot{\circ }{.})} on functions}{}\<[E]%
\\
\>[B]{}\mathrel{=}{}\<[BE]%
\>[5]{}(\Varid{second}\;\Varid{crossI}\hsdot{\circ }{.}\Varid{unzip}\hsdot{\circ }{.}\Varid{crossI}\;(\Varid{zipWith}\;(\mathbin{\vartriangle})\;\Varid{fs}\;(\Varid{fmap}\;\mathcal{D}\;\Varid{fs})))\;\Varid{as}{}\<[75]%
\>[75]{}\mbox{\onelinecomment  \lemRef{crossZip} below}{}\<[E]%
\\
\>[B]{}\mathrel{=}{}\<[BE]%
\>[5]{}(\Varid{second}\;\Varid{crossI}\hsdot{\circ }{.}\Varid{unzip}\hsdot{\circ }{.}\Varid{crossI}\;(\Varid{fmap}\;\mathcal{D}\!^+\!\;\Varid{fs}))\;\Varid{as}{}\<[75]%
\>[75]{}\mbox{\onelinecomment  definition of \ensuremath{\mathcal{D}\!^+\!}}{}\<[E]%
\ColumnHook
\end{hscode}\resethooks
For the second-to-last step,
\begin{lemma}\lemLabel{crossZip}
\ensuremath{\Varid{crossI}\;\Varid{fs}\mathbin{\vartriangle}\Varid{crossI}\;\Varid{gs}\mathrel{=}\Varid{unzip}\hsdot{\circ }{.}\Varid{crossI}\;(\Varid{zipWith}\;(\mathbin{\vartriangle})\;\Varid{fs}\;\Varid{gs})}.
\end{lemma}
For now, let's prove just the binary version of this lemma, namely
$$ \ensuremath{(\Varid{f}\times\Varid{f'})\mathbin{\vartriangle}(\Varid{g}\times\Varid{g'})\mathrel{=}\Varid{transpose}\hsdot{\circ }{.}((\Varid{f}\mathbin{\vartriangle}\Varid{g})\times(\Varid{g'}\mathbin{\vartriangle}\Varid{g'}))} $$
where
\begin{hscode}\SaveRestoreHook
\column{B}{@{}>{\hspre}l<{\hspost}@{}}%
\column{E}{@{}>{\hspre}l<{\hspost}@{}}%
\>[B]{}\Varid{transpose}\mathbin{::}((\Varid{a} \times \Varid{b}) \times (\Varid{c} \times \Varid{d}))\to ((\Varid{a} \times \Varid{c}) \times (\Varid{b} \times \Varid{d})){}\<[E]%
\\
\>[B]{}\Varid{transpose}\;((\Varid{a},\Varid{b}),(\Varid{c},\Varid{d}))\mathrel{=}((\Varid{a},\Varid{c}),(\Varid{b},\Varid{d})){}\<[E]%
\ColumnHook
\end{hscode}\resethooks
\out{For general cartesian categories, \ensuremath{\Varid{transpose}\mathrel{=}(\Varid{exl}\hsdot{\circ }{.}\Varid{exl}\mathbin{\vartriangle}\Varid{exl}\hsdot{\circ }{.}\Varid{exr})\mathbin{\vartriangle}(\Varid{exr}\hsdot{\circ }{.}\Varid{exl}\mathbin{\vartriangle}\Varid{exr}\hsdot{\circ }{.}\Varid{exr})}.}
Proof:
\begin{hscode}\SaveRestoreHook
\column{B}{@{}>{\hspre}c<{\hspost}@{}}%
\column{BE}{@{}l@{}}%
\column{5}{@{}>{\hspre}l<{\hspost}@{}}%
\column{69}{@{}>{\hspre}l<{\hspost}@{}}%
\column{E}{@{}>{\hspre}l<{\hspost}@{}}%
\>[5]{}(\Varid{f}\times\Varid{f'})\mathbin{\vartriangle}(\Varid{g}\times\Varid{g'}){}\<[E]%
\\
\>[B]{}\mathrel{=}{}\<[BE]%
\>[5]{}(\Varid{inl}\hsdot{\circ }{.}\Varid{f}\mathbin{\triangledown}\Varid{inr}\hsdot{\circ }{.}\Varid{f'})\mathbin{\vartriangle}(\Varid{inl}\hsdot{\circ }{.}\Varid{g}\mathbin{\triangledown}\Varid{inr}\hsdot{\circ }{.}\Varid{g'}){}\<[69]%
\>[69]{}\mbox{\onelinecomment  \citep[Equation (17)]{MacedoOliveira2013Typing}}{}\<[E]%
\\
\>[B]{}\mathrel{=}{}\<[BE]%
\>[5]{}(\Varid{inl}\hsdot{\circ }{.}\Varid{f}\mathbin{\vartriangle}\Varid{inl}\hsdot{\circ }{.}\Varid{g})\mathbin{\triangledown}(\Varid{inr}\hsdot{\circ }{.}\Varid{f'}\mathbin{\vartriangle}\Varid{inr}\hsdot{\circ }{.}\Varid{g'}){}\<[69]%
\>[69]{}\mbox{\onelinecomment  exchange law \citep[Section 1.5.4]{Gibbons2002Calculating}}{}\<[E]%
\\
\>[B]{}\mathrel{=}{}\<[BE]%
\>[5]{}\Varid{transpose}\hsdot{\circ }{.}\Varid{inl}\hsdot{\circ }{.}(\Varid{f}\mathbin{\vartriangle}\Varid{g})\mathbin{\triangledown}\Varid{transpose}\hsdot{\circ }{.}\Varid{inr}\hsdot{\circ }{.}(\Varid{f'}\mathbin{\vartriangle}\Varid{g'}){}\<[69]%
\>[69]{}\mbox{\onelinecomment  \lemRef{inlFork} below}{}\<[E]%
\\
\>[B]{}\mathrel{=}{}\<[BE]%
\>[5]{}\Varid{transpose}\hsdot{\circ }{.}(\Varid{inl}\hsdot{\circ }{.}(\Varid{f}\mathbin{\vartriangle}\Varid{g})\mathbin{\triangledown}\Varid{inr}\hsdot{\circ }{.}(\Varid{f'}\mathbin{\vartriangle}\Varid{g'})){}\<[69]%
\>[69]{}\mbox{\onelinecomment  \citep[Section 1.5.2]{Gibbons2002Calculating}}{}\<[E]%
\\
\>[B]{}\mathrel{=}{}\<[BE]%
\>[5]{}\Varid{transpose}\hsdot{\circ }{.}((\Varid{f}\mathbin{\vartriangle}\Varid{g})\times(\Varid{f'}\mathbin{\vartriangle}\Varid{g'})){}\<[69]%
\>[69]{}\mbox{\onelinecomment  \citep[Equation (17)]{MacedoOliveira2013Typing}}{}\<[E]%
\ColumnHook
\end{hscode}\resethooks

For the third step, we need two more properties.
\begin{lemma}\lemLabel{inlFork}
$$ \ensuremath{\Varid{inl}\hsdot{\circ }{.}\Varid{f}\mathbin{\vartriangle}\Varid{inl}\hsdot{\circ }{.}\Varid{g}\mathrel{=}\Varid{transpose}\hsdot{\circ }{.}\Varid{inl}\hsdot{\circ }{.}(\Varid{f}\mathbin{\vartriangle}\Varid{g})} $$
$$ \ensuremath{\Varid{inr}\hsdot{\circ }{.}\Varid{f}\mathbin{\vartriangle}\Varid{inr}\hsdot{\circ }{.}\Varid{g}\mathrel{=}\Varid{transpose}\hsdot{\circ }{.}\Varid{inr}\hsdot{\circ }{.}(\Varid{f}\mathbin{\vartriangle}\Varid{g})} $$
\end{lemma}
Below is a proof in the \ensuremath{(\to )} category, which suffice for our purpose.
(To do: does the property hold for general biproduct categories?)
\begin{hscode}\SaveRestoreHook
\column{B}{@{}>{\hspre}c<{\hspost}@{}}%
\column{BE}{@{}l@{}}%
\column{5}{@{}>{\hspre}l<{\hspost}@{}}%
\column{50}{@{}>{\hspre}l<{\hspost}@{}}%
\column{E}{@{}>{\hspre}l<{\hspost}@{}}%
\>[5]{}\Varid{inl}\hsdot{\circ }{.}\Varid{f}\mathbin{\vartriangle}\Varid{inl}\hsdot{\circ }{.}\Varid{g}{}\<[E]%
\\
\>[B]{}\mathrel{=}{}\<[BE]%
\>[5]{}\lambda \Varid{a}\to (\Varid{inl}\hsdot{\circ }{.}\Varid{f}\mathbin{\vartriangle}\Varid{inl}\hsdot{\circ }{.}\Varid{g})\;\Varid{a}{}\<[50]%
\>[50]{}\mbox{\onelinecomment  $\eta$-expand}{}\<[E]%
\\
\>[B]{}\mathrel{=}{}\<[BE]%
\>[5]{}\lambda \Varid{a}\to (\Varid{inl}\;(\Varid{f}\;\Varid{a}),\Varid{inl}\;(\Varid{g}\;\Varid{a})){}\<[50]%
\>[50]{}\mbox{\onelinecomment  definition of \ensuremath{(\mathbin{\vartriangle})} for functions}{}\<[E]%
\\
\>[B]{}\mathrel{=}{}\<[BE]%
\>[5]{}\lambda \Varid{a}\to ((\Varid{f}\;\Varid{a},\mathrm{0}),(\Varid{g}\;\Varid{a},\mathrm{0})){}\<[50]%
\>[50]{}\mbox{\onelinecomment  definition of \ensuremath{\Varid{inl}} for functions}{}\<[E]%
\\
\>[B]{}\mathrel{=}{}\<[BE]%
\>[5]{}\lambda \Varid{a}\to \Varid{transpose}\;((\Varid{f}\;\Varid{a},\Varid{g}\;\Varid{a}),(\mathrm{0},\mathrm{0})){}\<[50]%
\>[50]{}\mbox{\onelinecomment  definition of \ensuremath{\Varid{transpose}}}{}\<[E]%
\\
\>[B]{}\mathrel{=}{}\<[BE]%
\>[5]{}\lambda \Varid{a}\to \Varid{transpose}\;((\Varid{f}\;\Varid{a},\Varid{g}\;\Varid{a}),\mathrm{0}){}\<[50]%
\>[50]{}\mbox{\onelinecomment  definition of \ensuremath{\mathrm{0}} for pairs}{}\<[E]%
\\
\>[B]{}\mathrel{=}{}\<[BE]%
\>[5]{}\lambda \Varid{a}\to \Varid{transpose}\;(\Varid{inl}\;(\Varid{f}\;\Varid{a},\Varid{g}\;\Varid{a})){}\<[50]%
\>[50]{}\mbox{\onelinecomment  definition of \ensuremath{\Varid{inl}} for functions}{}\<[E]%
\\
\>[B]{}\mathrel{=}{}\<[BE]%
\>[5]{}\Varid{transpose}\hsdot{\circ }{.}\Varid{inl}\hsdot{\circ }{.}(\Varid{f}\mathbin{\vartriangle}\Varid{g}){}\<[50]%
\>[50]{}\mbox{\onelinecomment  definition of \ensuremath{(\hsdot{\circ }{.})} for functions}{}\<[E]%
\ColumnHook
\end{hscode}\resethooks
Similarly for the second property (with \ensuremath{\Varid{inr}}), noting that \ensuremath{((\mathrm{0},\Varid{f}\;\Varid{a}),(\mathrm{0},\Varid{g}\;\Varid{a}))\mathrel{=}\Varid{transpose}\;((\mathrm{0},\mathrm{0}),(\Varid{f}\;\Varid{a},\Varid{g}\;\Varid{a}))}.

The \ensuremath{\Conid{CartesianI}} and \ensuremath{\Conid{CocartesianI}} instances follow from linearity (\thmRef{linear}).

\bibliography{bib}

\end{document}